\documentclass[prd,preprint,preprintnumbers,nofootinbib,eqsecnum,superscriptaddress]{revtex4}

 \usepackage[dvips,final]{graphicx}
  \usepackage{amssymb}
   \usepackage{amsmath}
    \usepackage{amsfonts}
     \usepackage{epsfig}
      \usepackage{bm}

\usepackage{mathpazo}

\usepackage[section]{placeins}

\input epsf.tex
\def\desepsf(#1 width #2){\epsfxsize=#2 \epsfbox{#1}}

\usepackage{multirow}
\usepackage{ctable}
\usepackage{booktabs}
\usepackage{array}
\usepackage{tabularx}
\usepackage{xcolor}
\usepackage{pstricks}
\definecolor{fred}{rgb}{0.90053, 0.00369, 0.00159}  

\newcommand{\be}{\begin{eqnarray}}
\newcommand{\ee}{\end{eqnarray}}

\begin{document}

\author{Rafa{\l} Maciu{\l}a}
\email{rafal.maciula@ifj.edu.pl}
\affiliation{Institute of Nuclear
Physics, Polish Academy of Sciences, ul. Radzikowskiego 152, PL-31-342 Krak{\'o}w, Poland}

\author{Antoni Szczurek\footnote{also at University of Rzesz\'ow, PL-35-959 Rzesz\'ow, Poland}}
\email{antoni.szczurek@ifj.edu.pl} 
\affiliation{Institute of Nuclear
Physics, Polish Academy of Sciences, ul. Radzikowskiego 152, PL-31-342 Krak{\'o}w, Poland}


\title{Intrinsic charm in the nucleon
and charm production at large rapidities
in collinear, hybrid and \bm{$k_T$}-factorization approaches}

\begin{abstract}
We discuss the role of intrinsic charm (IC) in the nucleon 
for forward production of $c$-quark (or $\bar c$-antiquark) in proton-proton collisions 
for low and high energies.
The calculations are performed in collinear-factorization approach with on-shell partons,
$k_T$-factorization approach with off-shell partons as well as in a hybrid approach 
using collinear charm distributions and unintegrated (transverse momentum dependent) gluon distributions.
For the collinear-factorization approach we use matrix elements
for both massless and massive charm quarks/antiquarks.
The distributions in rapidity and transverse momentum of charm quark/antiquark 
are shown for a few different models of IC. Forward charm production is dominated by 
$gc$-fusion processes. The IC contribution dominates over the standard pQCD (extrinsic)
$gg$-fusion mechanism of $c\bar c$-pair production at large rapidities or Feynman-$x_F$.
We perform similar calculations within leading-order and next-to-leading
order $k_T$-factorization approach. The $k_T$-factorization approach
leads to much larger cross sections than the LO collinear approach.
At high energies and large rapidities of $c$-quark or $\bar c$-antiquark one tests 
gluon distributions at extremely small $x$.
The IC contribution has important consequences for 
high-energy neutrino production in the Ice-Cube experiment and can be,
to some extent, tested at the LHC by the SHIP and FASER experiments
by studies of the $\nu_{\tau}$ neutrino production.  
\end{abstract} 

\maketitle

\section{Introduction}

The text-book proton consists of $u u d$ valence quarks.
This picture is by far too simplified. In fact there is strong evidence
for internal strangeness and somewhat smaller for internal charm content
of the nonperturbative proton. Different pictures of nonperturbative
$c \bar c$ content were proposed in the past. A first example is relatively old 
BHPS model \cite{BHPS1980} which assumes $u u d c \bar c$ 5-parton Fock configurations (see also Refs.~\cite{BHK1982,VB1996}).
Another picture proposed in the literature is a meson cloud model (MCM)
\cite{NNNT1996,MT1997,SMT1999,CDNN2001,HLM2014},
where the $p \to {\bar D}^0 \Lambda_c$ or $D \Sigma_c$ fluctuations of the proton
are considered. 
While in the first model $c(x) = {\bar c}(x)$ in the MCM 
$c(x) \ne {\bar c}(x)$.
The models do not allow to predict precisely the absolute probability 
for the $c$-quark or $\bar c$-antiquark content of the proton. Experimental data put 
only loose constraints on the charm content:
\begin{equation}
\int_0^1 c(x) \; dx = \int_0^1 {\bar c}(x) \; dx < 0.01 \; .
\label{charm_probability}
\end{equation}
It is rather upper limit but this value depends somewhat on
the model of charm content of a proton. In general, for sea-like models the
probability can be slightly larger than for the BHPS one. In the sea-like case the charm
is concentrated at lower values of $x$.
Very recent lattice study of charm quark electromagnetic form factors
suggested asymmetry of $c$-quark and $\bar c$-antiquark distributions \cite{latticeQCD}.

Recently there is a renewed interest in the intrinsic charm (IC) which
is related to experiments being performed at the LHC 
\cite{BBLS2015,RKAG2016,LLSB2016,BBLLMST2018}.
The intrinsic charm is often included in global parton analyses
of world experimental data \cite{BKLSSV2015,Ball:2014uwa,Hou:2017khm}.

The highly energetic neutrino experiments, such as IceCube, could put
further constraints on the intrinsic charm \cite{ERS2008,LB2016,GGN2018}.
Here, however, the IC contribution may compete with a concept of the subleading
fragmentation \cite{Maciula:2017wov}. Similarly, future LHC high and low energy forward experiments like FASER and SHIP
could also be very helpful in this context (see e.g. Ref.~\cite{Bai:2020ukz} and Ref.~\cite{Bai:2018xum}, respectively). Also the LHCb experiment 
in its fixed-target mode could be sensitive to the contributions coming from intrinsic charm in a proton, especially in the case of open charm production \cite{Aaij:2018ogq}, where some problems with a satisfactory theoretical description of the experimental data were reported (see also discussion in Ref.~\cite{Maciula:2020cfy}). 

In this paper we concentrate therefore on forward production of
charm quarks/antiquarks.
There were already some studies performed with color glass condensate
approach and compared to the dipole approach at 
forward directions \cite{GNU2010,CGGN2017}.
In our approach we will use instead collinear, hybrid and
$k_T$-factorization approach. The latter two were not studied so far
in the context of IC and forward production of charm.
 
\section{Models of intrinsic charm in a nucleon}

In the five-quark Fock component $u u d c \bar c$ heavy quark/antiquark carries
rather large fraction of the mother proton. In the BHPS model,
after some approximations the probability to find $c$ or $\bar c$
(the same for both) can be expressed via a simple formula: 
\begin{equation}
\frac{dP}{dx} = c(x) = {\bar c}(x) =
A x^2 \left( 6 x (1+x) ln(x) + (1-x)(1+10x+x^2) \right) \; .
\end{equation}
The normalization constant $A$ depends on integrated probability for
$c \bar c$ component and is 6 for 1 \% probability.
Please note that the quark mass is not explicit in this simplified formula.

In the meson cloud models $c$ is in the baryon-like object and
$\bar c$ in the meson-like object.
Then the probabilistic distribution can be obtained as
\begin{eqnarray}
\frac{dP_c}{dx} &=& \int_x^1 \frac{dy}{y} f_B(y) f_{c/B}(x/y) \; , \\
\frac{dP_{\bar c}}{dx} &=& \int_x^1 \frac{dy}{y} f_M(y) f_{\bar c/M}(x/y) \; .
\label{MCM}
\end{eqnarray}
The $f_B$ and $f_M$ functions, the probability to find meson or baryon
in proton, can be calculated from corresponding Lagrangians
supplemented by a somewhat arbitrary and poorly known vertex form factors
and can be found \textit{e.g.} in Ref.~\cite{HLM2014}. In general, such an approach leads
to $c(x) \ne {\bar c}(x)$.

In practice both models give rather similar distributions as will be
shown in the following, so using one of them as 
an example is representative and sufficient. 
These are models of large-$x$ components of IC.
In principle, the IC may have also small-$x$ component known under 
the name of sea-like, however, only simple \textit{ad hoc} parametrizations 
were used in the literature. 

\begin{figure}[!h]
\begin{minipage}{0.35\textwidth}
  \centerline{\includegraphics[width=1.0\textwidth]{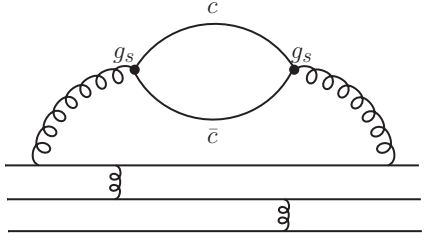}}
\end{minipage}
  \caption{A dynamical process leading to sea-like IC.
\small   
}
\label{fig:sea-like}
\end{figure}

\begin{figure}[h]
\begin{minipage}{0.47\textwidth}
  \centerline{\includegraphics[width=1.0\textwidth]{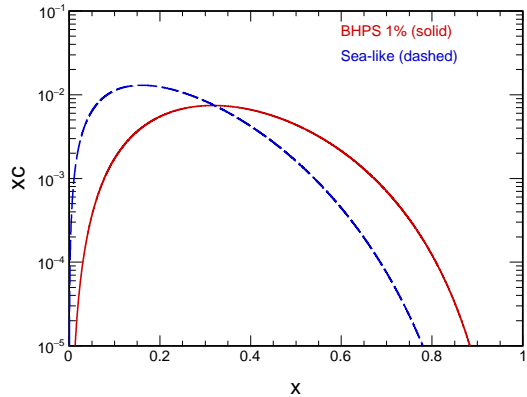}}
\end{minipage}
\caption{Charm quark/antiquark distribution for the
two different models of IC. The solid line represents the BHPS model
while the dashed line is for sea-like glue as obtained in a way
described above. In this calculation BHPS model with 1\% probability was used for
illustration.
}
\label{fig:xcharm_IC}
\end{figure}

There is another category of processes leading to sea-like IC
(see Fig.~\ref{fig:sea-like} where an example of corresponding dynamical
processes is shown). Using intrinsic glue in the nucleon (see \textit{e.g.} Ref.~\cite{EI1998}) one can 
generate intrinsic charm sea. The intrinsic gluon distribution fulfil by construction the relation:
\begin{equation}
\int_0^1 \left( x u_v(x) + x d_v(x) + x g(x) \right) dx = 1 \; .
\end{equation}
For massless charm the intrinsic charm can be calculated
as the convolution with initial (intrinsic) glue
\begin{equation}
c(x) = {\bar c}(x) = \alpha_s(4 m_c^2) / (2 \pi) \int_x^1 dy 
\left( \frac{1}{y} \right) 
P_{q g}\left( \frac{x}{y} \right) g(y) \; ,
\end{equation}
where $g$ is the intrinsic gluon distribution. With the model
from Ref.~\cite{EI1998} the $c$ and $\bar c$ distributions are integrable,
concentrated at $x \sim$ 0.1-0.2
and corresponding probability is 2.7 \%.
It should be less for massive quarks/antiquarks.
In Fig.~\ref{fig:xcharm_IC} we show $x$-distribution of the IC for the BHPS and for the
sea-like model described above.

In the GRV approach \cite{Gluck:1994uf} the charm contribution is calculated
fully radiatively as the convolution of gluon distribution with
appropriate mass-dependent splitting function:
\begin{equation}
x c(x,Q^2) = \frac{\alpha_s(\mu'^2)}{2 \pi}
\int_{a x}^1 dy \left( \frac{x}{y} \right) C_{g,2}^c \left(\frac{x}{y}, \
\frac{m_c^2}{Q^2} \right) g(y,\mu'^2) \; .
\label{GRV_charm}
\end{equation}
The explicit formula for $C_{g,2}^c$ and $a$, including mass of
quarks/antiquarks, can be found in Ref.~\cite{Gluck:1994uf}.
In the following calculations we will use more modern gluon distributions.

In this paper we concentrate on large-$x$
component and completely ignore the sea-like component(s).
Charm can also be generated by evolution equations via $g \to c \bar c$
transition (splitting).
Often it was included in the evolution as a massless parton with
zero as initial condition at the starting scale $\mu^2 \sim m_c^2$.
In a dedicated fits, the intrinsic charm distribution is used as initial condition
for DGLAP evolved charm distributions (see \textit{e.g.} Ref.~\cite{PLT2007}). In the right panel
Fig.~\ref{fig:xc} we show charm distribution in a proton 
without (dashed line) and with (solid line) the IC distribution taken as initial condition of the evolution.


\section{Cross section for associated charm production}

\subsection{The collinear approach}

In the present study we discuss production of the final states with one charm quark or charm antiquark. In the collinear approach \cite{Collins:1989gx} the final state charm must be associated with at least one additional gluon or (light) quark. 
Typical leading-order mechanisms for charm production initiated by charm quark in a initial state are shown in Fig.~\ref{fig1}. The diagrams correspond to the $gc \to gc$ (or $g \bar c \to g\bar c$) subprocesses that are expected to be dominant at high energies, however, the $q c \to q c$ and $\bar q c \to \bar q c$ (or $q \bar c \to q \bar c$ and $\bar q \bar c \to \bar q \bar c$) mechanisms with $q = u,d,s$ are also possible and will be taken into account in the following numerical calculations. 

\begin{figure}[!h]
\begin{minipage}{0.28\textwidth}
  \centerline{\includegraphics[width=1.0\textwidth]{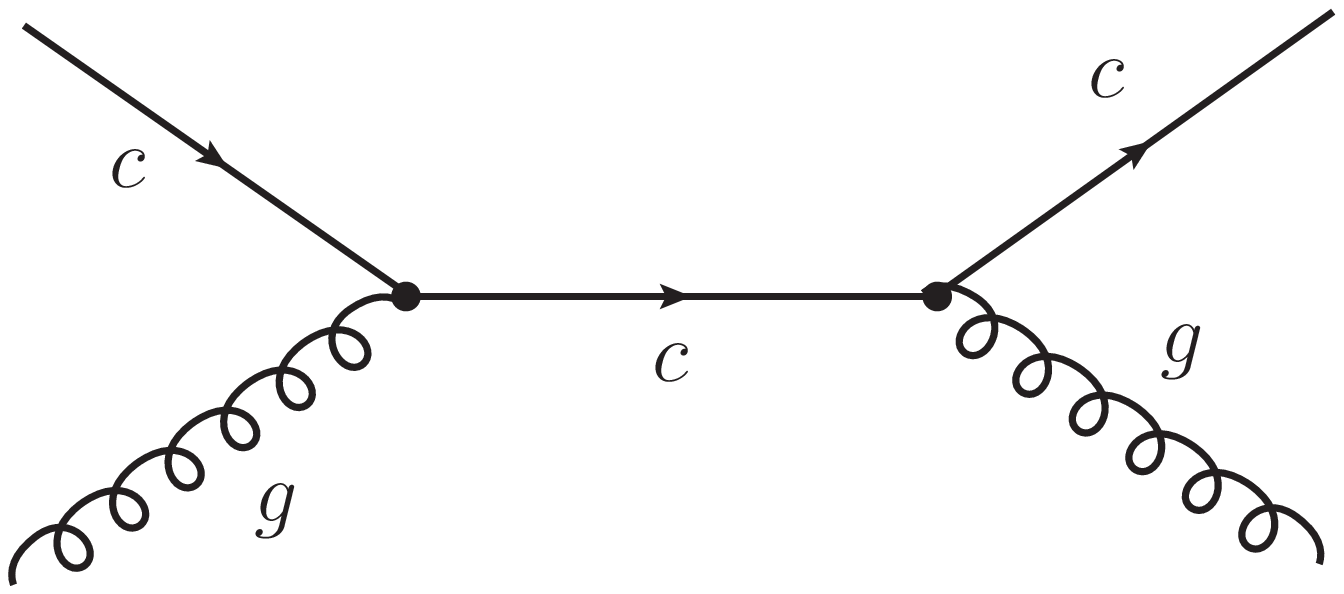}}
\end{minipage}
\hspace{+5mm}
\begin{minipage}{0.28\textwidth}
  \centerline{\includegraphics[width=1.0\textwidth]{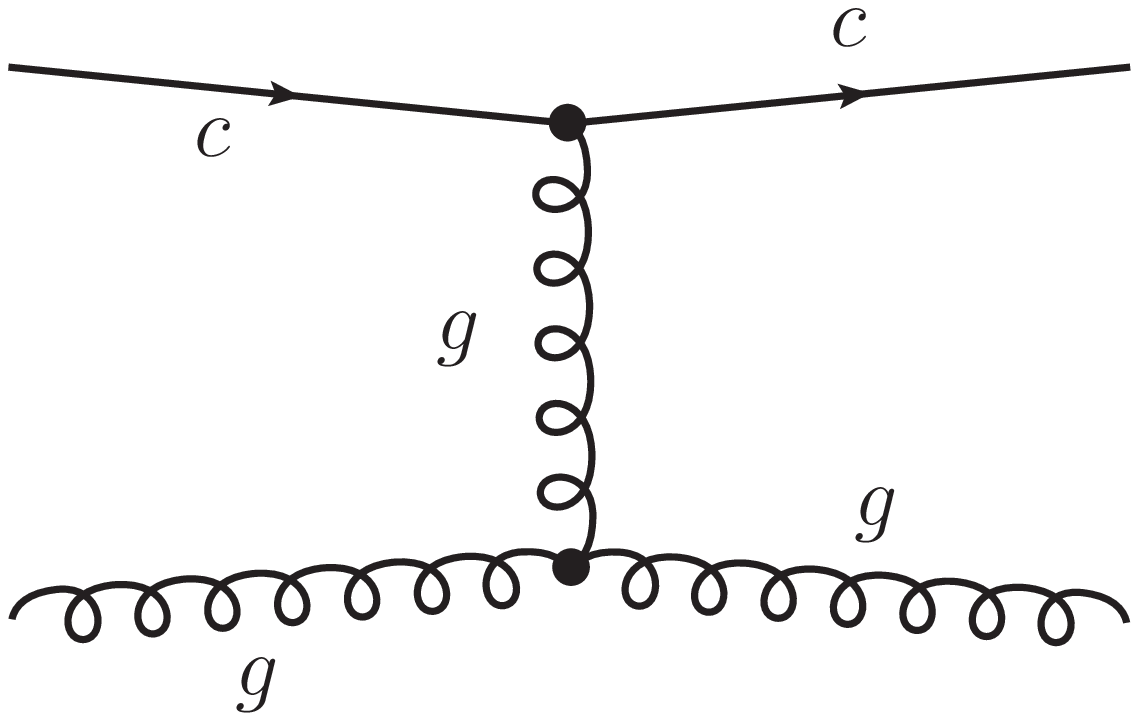}}
\end{minipage}
\hspace{+5mm}
\begin{minipage}{0.28\textwidth}
  \centerline{\includegraphics[width=1.0\textwidth]{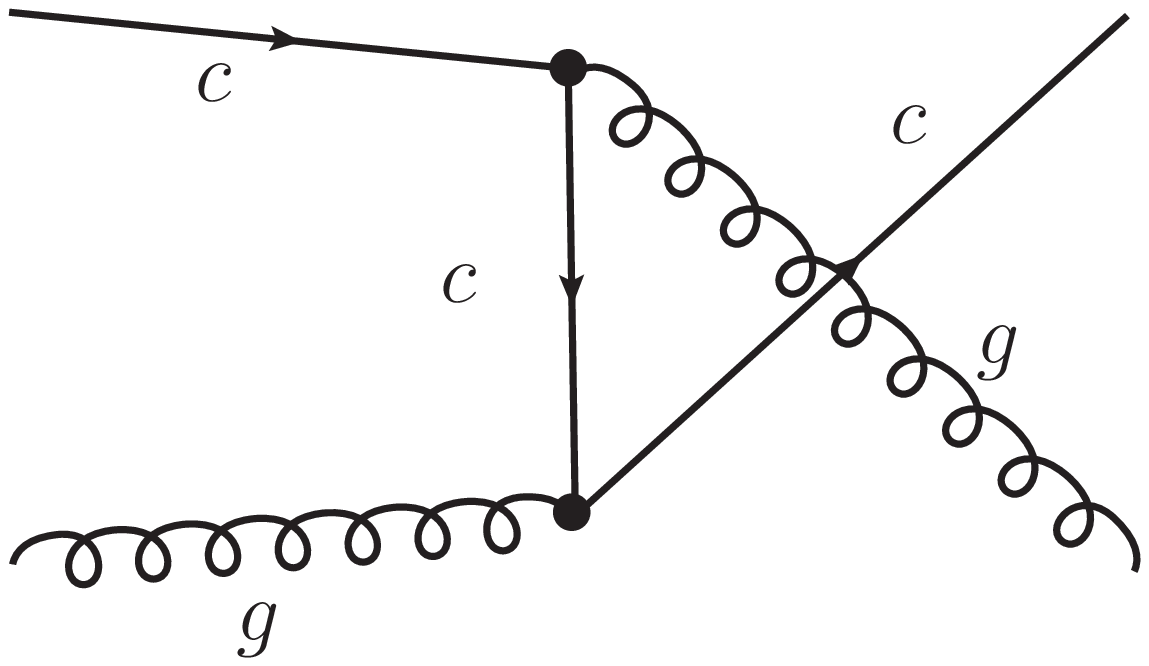}}
\end{minipage}
  \caption{
\small Typical leading-order (2 $\to$ 2) mechanisms of production
of $c$ quarks or $\bar c$ antiquarks in the collinear parton model.   
}
\label{fig1}
\end{figure}

In the collinear approach the differential cross section for forward charm production within the $gc \to gc$ mechanism\footnote{Here and in the following we concentrate only on the forward production mechanisms (with charm quark having positive-rapidity), but, the formalism for symmetric backward configuration is the same.} can be calculated as
\begin{eqnarray}
\frac{d \sigma}{d y_1 d y_2 d^2 p_t} = \frac{1}{16 \pi {\hat s}^2} 
\overline{| {\cal M}_{g c \to g c} |^2} x_{1} g(x_1, \mu^2) x_{2} c(x_2, \mu^2)
\; ,
\end{eqnarray}
where ${\cal M}_{g c \to g c}$ is the on-shell matrix element for $gc \to gc$ subprocesses and 
$g(x_1, \mu^2)$ and $c(x_2, \mu^2)$ are the collinear gluon and charm quark PDFs evaluated at longitudinal momentum fractions $x$ and factorization scale $\mu^{2}$.

Including the mass of charm quark the on-shell matrix element takes the following form:
\begin{eqnarray}
\overline{|{\cal M}_{gc \to gc}|^2} &=&  g_s^4  \left[ \left( - m_c^4 ( 3\hat{s}^2 + 14 \hat{s}\hat{u} + 3\hat{u}^2 )
+ m_c^2 ( \hat{s}^3 + 7 \hat{s}^2\hat{u} + 7 \hat{s} \hat{u}^2 + \hat{u}^3) \right. \right.  \nonumber \\
&& \left. \left. + 6m_c^8-\hat{s}\hat{u} ( \hat{s}^2+\hat{u}^2 ) \right)  \left( -18m_{c}^2 (\hat{s}+\hat{u}) +18m_c^4+9\hat{s}^2+9\hat{u}^2-\hat{t}^2 \right) \right] \nonumber \\
&& / \left( 18\hat{t}^2 ( \hat{u}-m_c^2)^2 ( \hat{s}-m_c^2) \right)^2 ,
\end{eqnarray} 
where $g_s^2 = 4 \pi \alpha_{s}(\mu)$. In the massless limit $m_c \to 0$ one recovers the known textbook formula:
\begin{equation}
\overline{|{\cal M}_{gc \to gc}|^2} =  g_s^4 
\left( -\frac{4}{9} 
\left( \frac{{\hat u}^2 + {\hat s}^2}{{\hat u}{\hat s}} \right)
+ \left(  \frac{{\hat u}^2+{\hat s}^2}{{\hat t}^2} \right)
\right) \; .
\end{equation}
A role of the charm quark mass in the matrix element will be discussed when presenting numerical results.

\begin{figure}[!h]
\begin{minipage}{0.47\textwidth}
  \centerline{\includegraphics[width=1.0\textwidth]{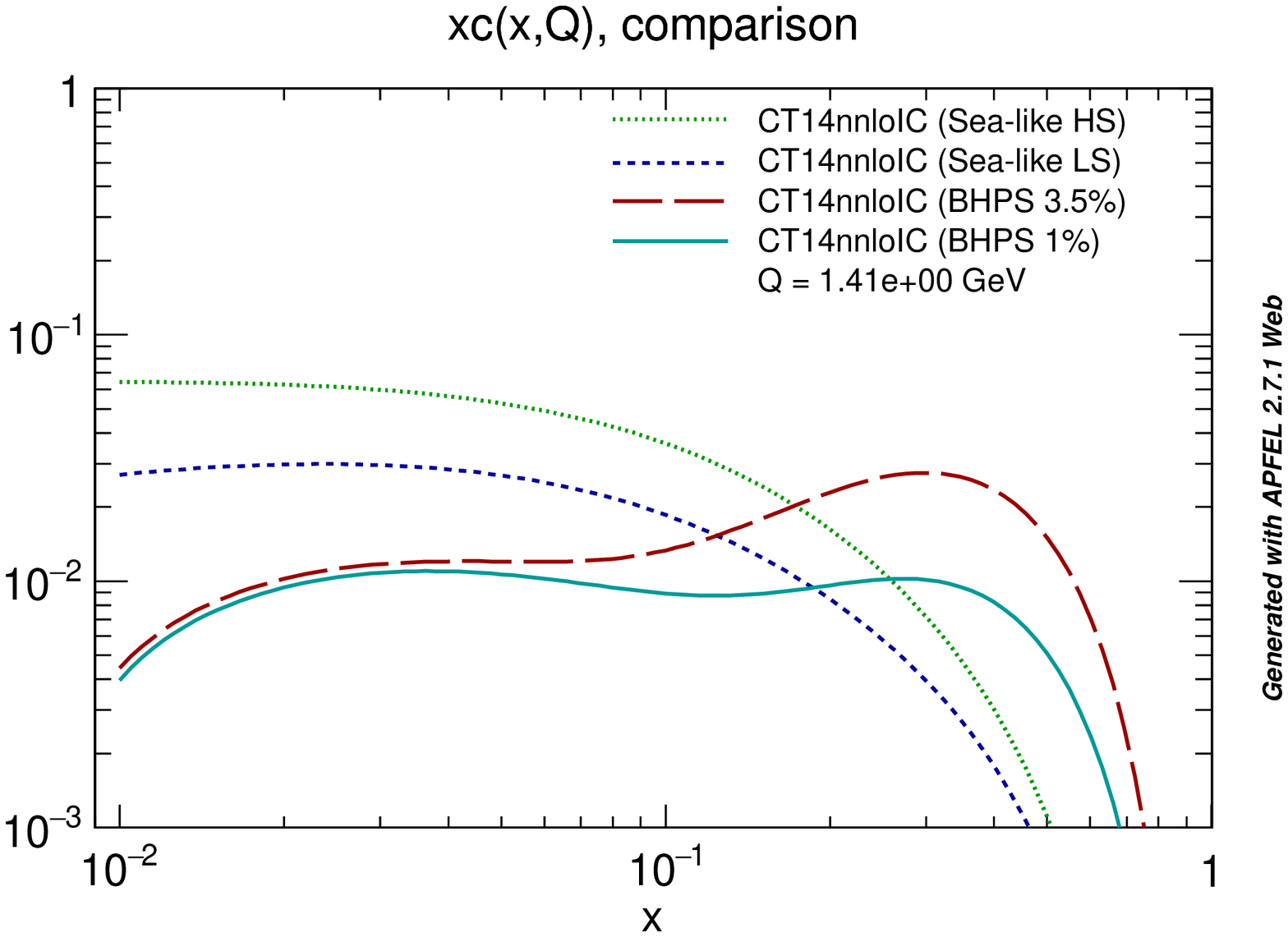}}
\end{minipage}
\begin{minipage}{0.47\textwidth}
  \centerline{\includegraphics[width=1.0\textwidth]{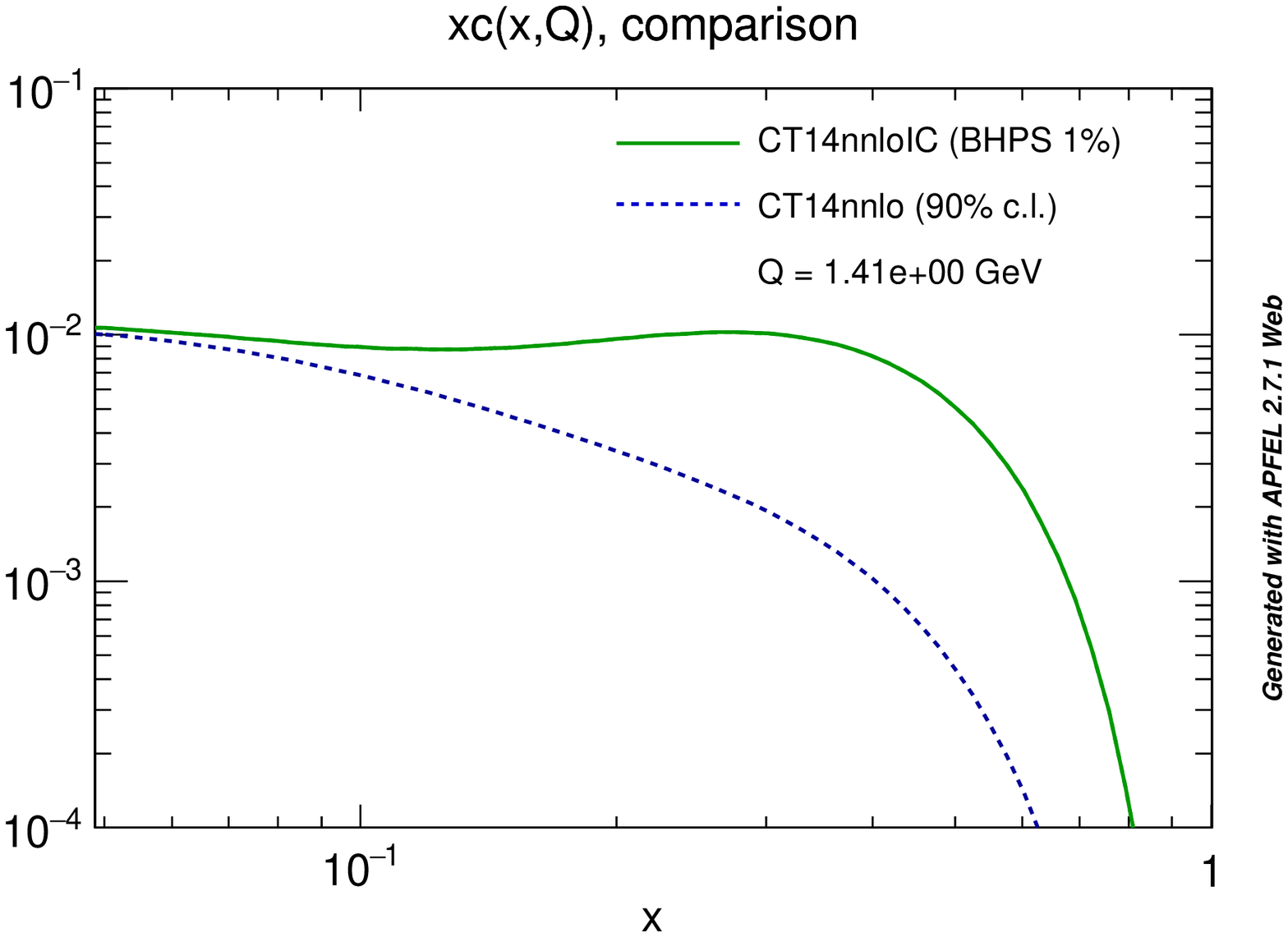}}
\end{minipage}
  \caption{
\small Charm quark distributions in a proton as a function of
longitudinal momentum fraction $x$. Here different models for initial 
intrinsic charm quark distributions are shown (left panel) and 
a comparison between charm quark distributions obtained with and 
without concept of intrinsic charm in the proton.  
}
\label{fig:xc}
\end{figure}

In the numerical calculations below the intrinsic charm PDFs are taken at the initial scale $m_{c} = 1.3$ GeV, so the perturbative charm contribution is intentionally not taken into account. We apply four different grids of the intrinsic charm distributions from the CT14nnloIC PDF \cite{Hou:2017khm} that correspond to 
the BHPS 1\% and BHPS 3.5\% as well as the sea-like LS (low-strength) and sea-like HS (high-strength) models for initial intrinsic charm distribution.    
The distributions are compared with each other in the left panel of Fig.~\ref{fig:xc}. In the right panel we present in addition the difference between the CT14nnloIC charm PDF obtained with and without intrinsic-charm concept.  

On the other hand the collinear gluon PDFs $g(x, \mu^2)$ are taken at the running factorization scale related to the averaged transverse momentum of the outgoing particles, i.e. $\mu = \sqrt{\frac{p_{t1}^2 + p_{t2}^2}{2} + m_c^2}$. The charm quark mass $m_{c} = 1.3$ GeV plays here a role of the minimal scale
and ensures that we are not going beyond the fitted PDF grids where unconstrained extrapolation procedures are applied. We keep the charm quark mass here even when the massless matrix element and/or kinematics are used.     

As will be shown later, the numerical results strongly depend on how the longitudinal momentum fractions $x_1$ and $x_2$ (arguments
of parton distributions) are calculated. In the massive scheme of the calculations the quantities are defined as follows: 
\begin{eqnarray}
x_1 &=& \frac{p_{t1}}{\sqrt{s}} \exp(+y_1) + \frac{m_{t2}}{\sqrt{s}} \exp(+y_2)
\; , \nonumber \\
x_2 &=& \frac{p_{t1}}{\sqrt{s}} \exp(-y_1) + \frac{m_{t2}}{\sqrt{s}} \exp(-y_2)
\; . 
\end{eqnarray}
In this equations $p_{t1}$ is transverse momentum of the outgoing gluon (or light quark/antiquark) and the $m_{t2}$ is $c$ quark ($\bar c$ antiquark) transverse
mass defined as $m_{t} = \sqrt{p_t^2+m_c^2}$. As will be discussed further it is crucial to include in kinematics the mass of the final charm,
while the initial charm can be considered massless. In the following numerical studies all the calculations in the massless limit with massless matrix elements will be done within the kinematics corrected in the above manner. The effect of the correction will be also explicitly shown.

\begin{figure}[!h]
\begin{minipage}{0.47\textwidth}
  \centerline{\includegraphics[width=1.0\textwidth]{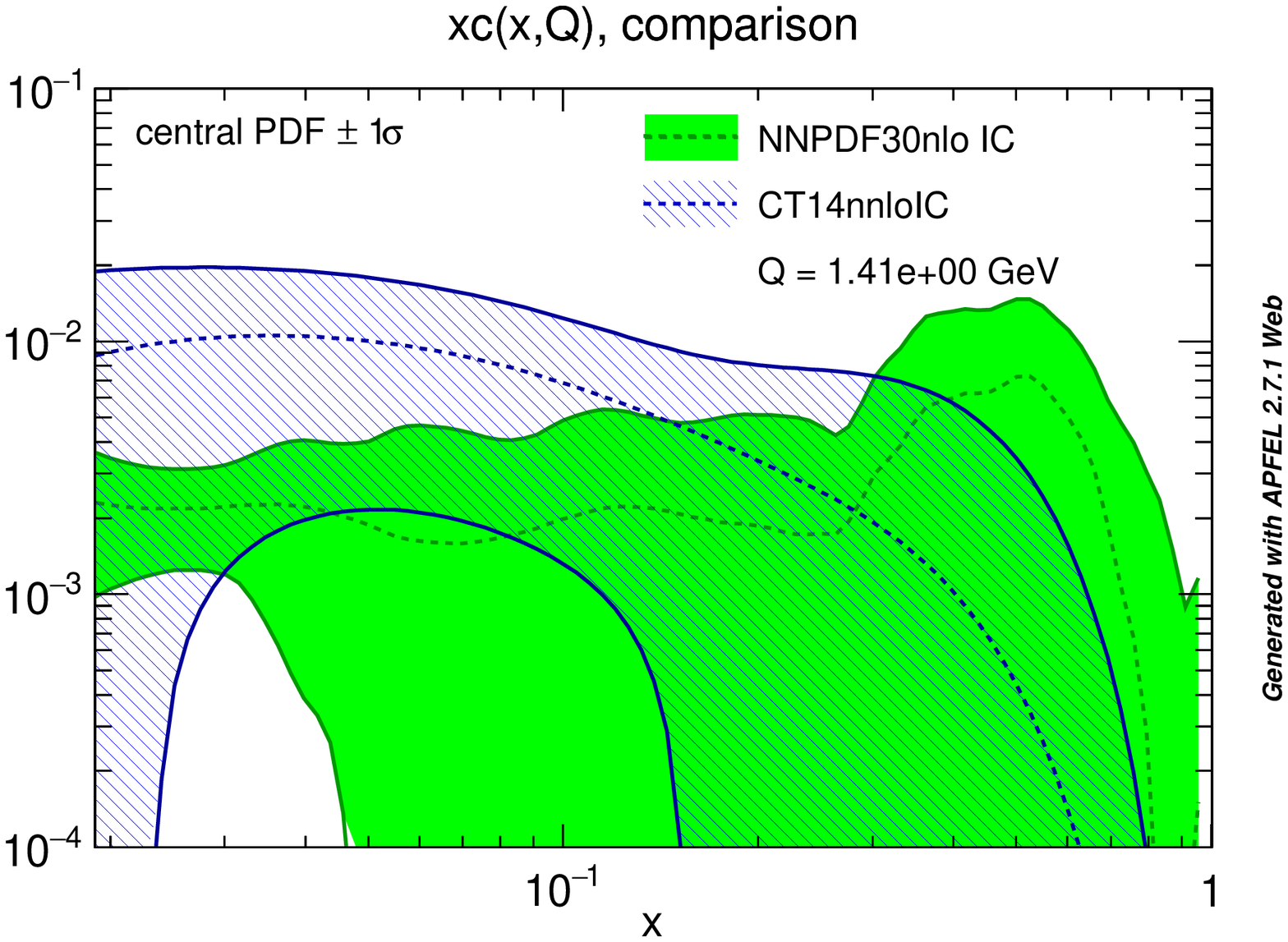}}
\end{minipage}
\begin{minipage}{0.47\textwidth}
  \centerline{\includegraphics[width=1.0\textwidth]{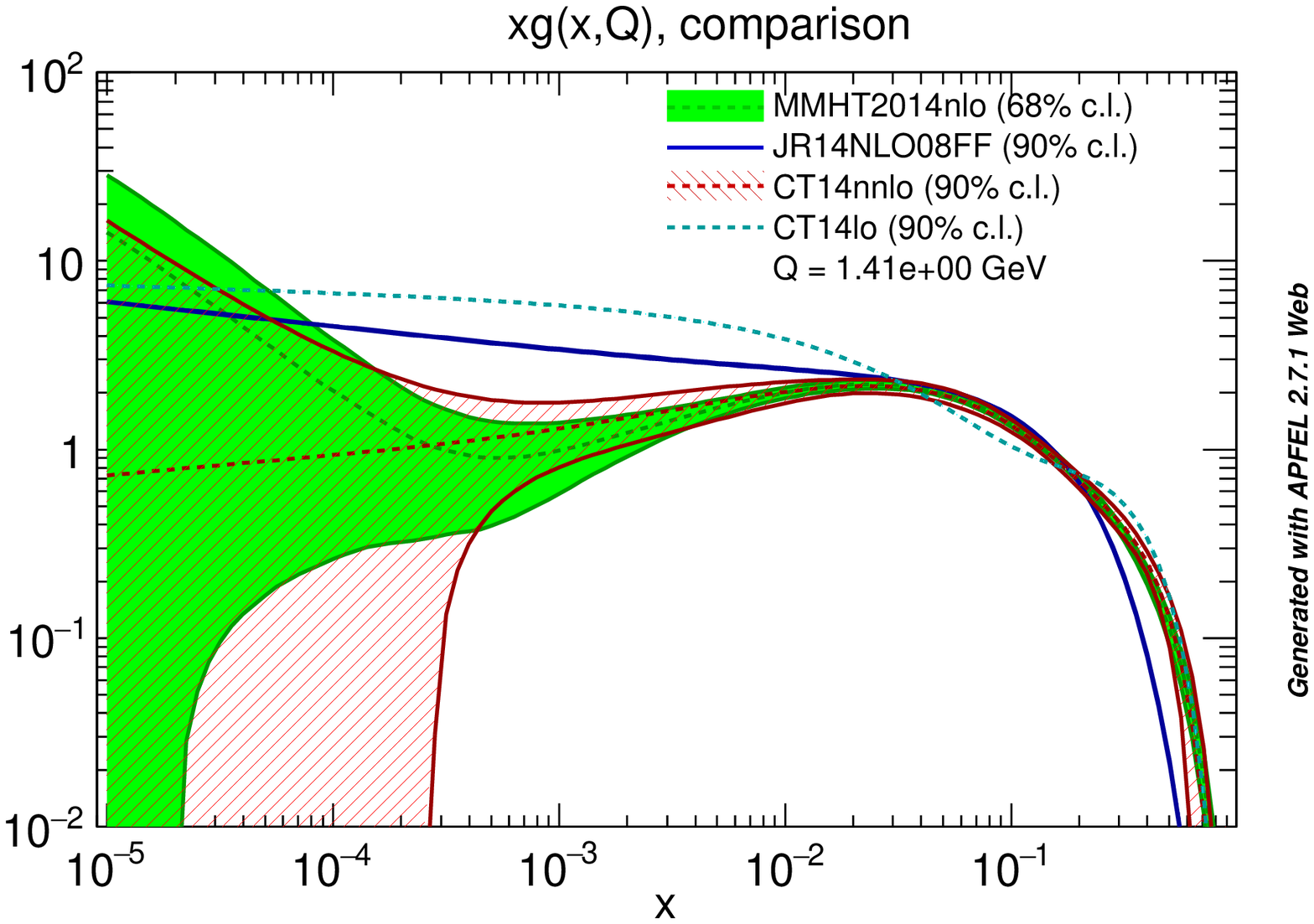}}
\end{minipage}
  \caption{
\small The intrinsic charm (left panel) and gluon (right panel)  distributions in a proton as a function of longitudinal momentum fraction $x$. Here different sets of collinear PDFs are shown including uncertainties.   
}
\label{fig:xcxg}
\end{figure}

Considering forward production of charm at the LHC energies one is exploring asymmetric kinematical regions where $x_1$ is very small (down to $10^{-5}$) and 
$x_2$ is rather large (about $10^{-1}$). Thus in this reaction small-$x$ gluon PDF and intrinsic large-$x$ charm content of the proton are probed simultaneously.
As it is shown in Fig.~\ref{fig:xcxg} both distributions are not well
constrained by the global experimental data. In the left panel we show
the central fits for intrinsic charm distribution from the CT14nnloIC
and the NNPDF30nloIC PDF sets \cite{Ball:2016neh} together with $1\sigma$ standard
deviation. In the right panel we compare gluon PDF fits from different
collaborations, including MMHT2014nlo \cite{Harland-Lang:2014zoa}, JR14NLO08FF \cite{Jimenez-Delgado:2014twa} and CT14lo/nnlo
sets. Clearly the current level of knowledge of both distributions is
rather limited and the large uncertainties prevent definite conclusions. 
In principle, a study of far-forward production of charm may improve the situation by exploring unconstrained areas.

\begin{figure}[!h]
\begin{minipage}{0.28\textwidth}
  \centerline{\includegraphics[width=1.0\textwidth]{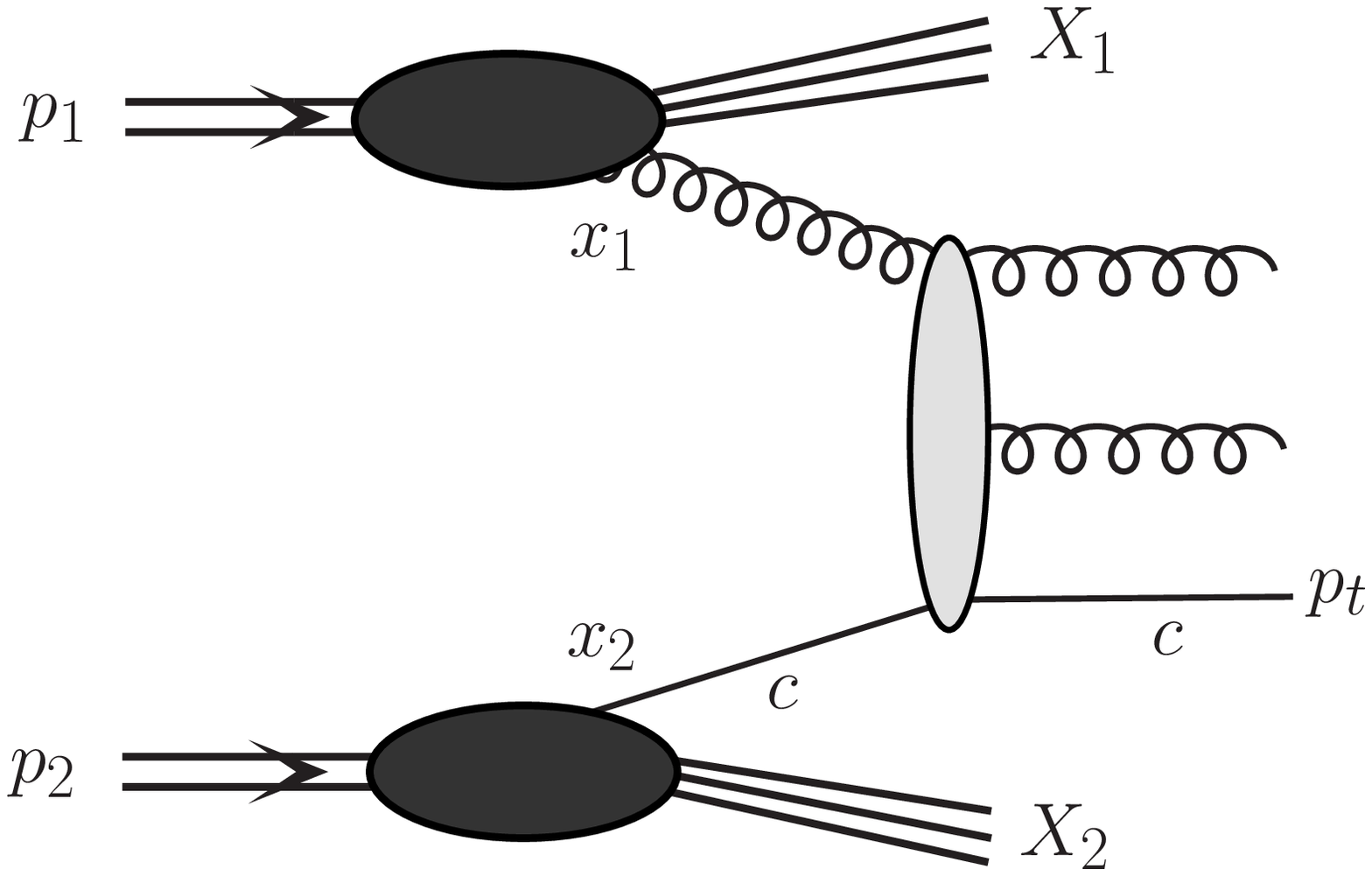}}
\end{minipage}
\hspace{+5mm}
\begin{minipage}{0.28\textwidth}
  \centerline{\includegraphics[width=1.0\textwidth]{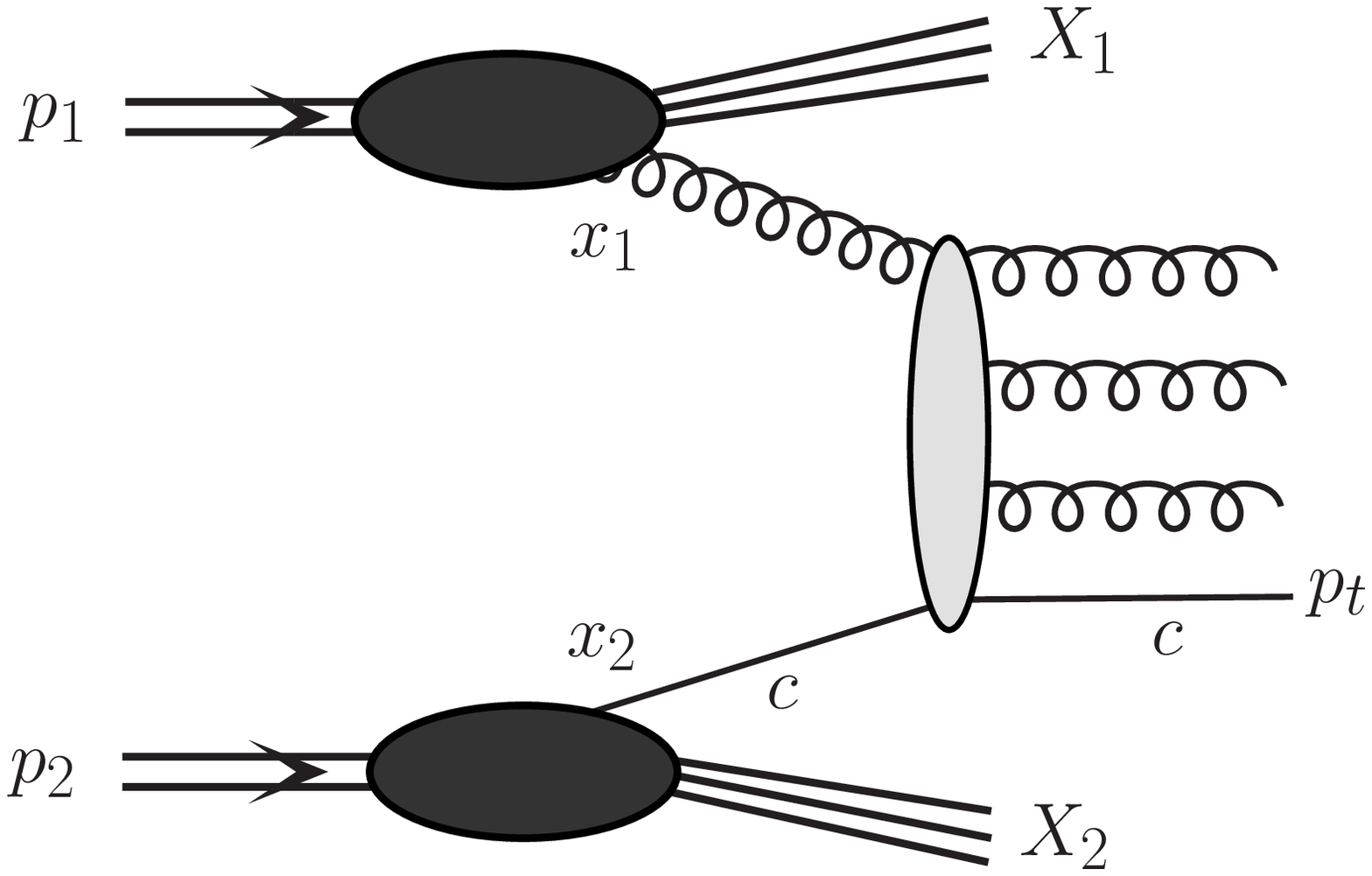}}
\end{minipage}
  \caption{
\small Examples of the $2\to3$ (left panel) and the $2\to4$ (right panel) mechanisms of production
of $c$ quarks or $\bar c$ antiquarks in the collinear parton model.   
}
\label{figHO}
\end{figure}

In the present study we go beyond the leading-order mechanisms and
include also higher-order processes that are expected to play important
role. We take into account all $2\to 3$ and $2\to 4$ processes at
tree-level that lead to a production of charm
quark or antiquark and are driven by the $gc$ and $qc$ (or $\bar q c$)
initial state interactions. Examples of the diagrams corresponding to
the processes are shown in Fig.~\ref{figHO}. 
The relevant cross sections are calculated with the help of 
the \textsc{KaTie} Monte Carlo generator \cite{vanHameren:2016kkz}.  

Having massless partons (minijets) in the final states considered in the present work
it is necessary to regularize the cross section that has a singularity
in the $p_{t} \to 0$ limit. 
We follow here the known prescription 
adopted in \textsc{Pythia} where a special suppression factor is
introduced at the cross section level \cite{Sjostrand:2014zea}:
\begin{equation}
F(p_t) = \frac{p_t^2}{ p_{t0}^2 + p_t^2 } \; 
\label{Phytia_formfactor}
\end{equation}
for each of the outgoing massless partons with transverse momentum $p_t$, where $p_{t0}$ is a free parameter of the form factor.

\subsection{The hybrid model}

Within the asymmetric kinematic situation $x_1 \ll x_2$ described above
the cross section for the processes under consideration can be also
expressed in the so-called hybrid factorization model motivated by the
works in Refs.~\cite{Deak:2009xt,Kutak:2012rf}. 
In this framework the small-$x$ gluon is taken to be off mass shell and 
the differential cross section e.g. for $pp \to g c X$ via $g^* c \to g c$ mechanism reads:
\begin{eqnarray}
d \sigma_{pp \to gc X} = \int d^ 2 k_{t} \int \frac{dx_1}{x_1} \int dx_2 \;
{\cal F}_{g^{*}}(x_1, k_{t}^{2}, \mu^2) \; c(x_2, \mu^2) \; d\hat{\sigma}_{g^{*}c \to gc} \; ,
\end{eqnarray}
where ${\cal F}_{g^{*}}(x_1, k_{t}^{2}, \mu^2)$ is the unintegrated gluon distribution in one proton and $c(x_2, \mu^2)$ a collinear PDF in the second one. The $d\hat{\sigma}_{g^{*}c \to gc}$ is the hard partonic cross section obtained from a gauge invariant tree-level off-shell amplitude. In the present paper we shall not discuss the validity of the hybrid model on the theoretical level and concentrate only on its phenomenological application in forward production. A derivation of the hybrid factorization from the dilute limit of the Color Glass Condensate approach can be found in Ref.~\cite{Kotko:2015ura}.

The gluon uPDF depends on gluon longitudinal momentum fraction $x$, transverse momentum
squared $k_t^2$ of the gluons entering the hard process, and in general also on a (factorization) scale of the hard process $\mu^2$. In the numerical calculations we take different models of unintegrated parton densities from the literature: the JH-2013-set2 \cite{Hautmann:2013tba} model obtained from the CCFM evolution equations, the Kutak-Sapeta (KS) \cite{Kutak:2014wga} model being a solution of linear and non-linear BK evolution, the DGLAP-based PB-NLO-set1 \cite{Martinez:2018jxt} model from the parton-branching (PB) method and the Kimber-Martin-Ryskin (KMR) prescription \cite{Watt:2003mx}.

\begin{figure}[!h]
\begin{minipage}{0.55\textwidth}
  \centerline{\includegraphics[width=1.0\textwidth]{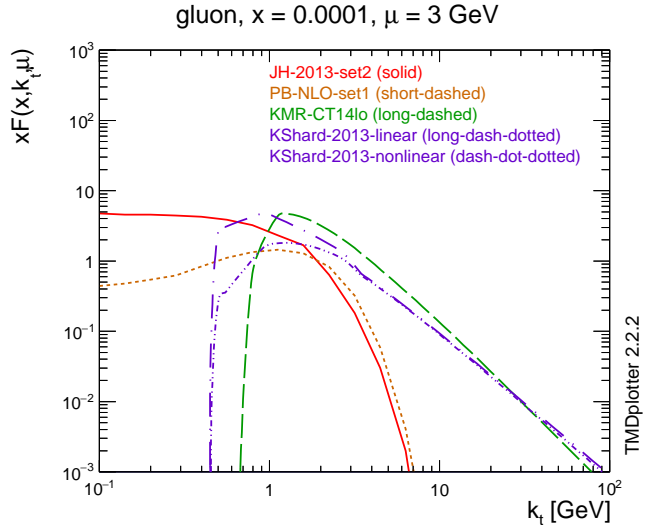}}
\end{minipage}
  \caption{
\small The ingoing gluon transverse momentum distributions from the different models of unintegrated gluon densities in a proton.   
}
\label{fig-uPDFs}
\end{figure}

All of the models, except the PB-NLO-set1, are constructed in the way that allows for resummation of extra hard emissions from the uPDFs. It means that in the hybrid model already at leading-order some part of radiative higher-order corrections can be effectively included via uPDFs. However, it is true only for those uPDF models in which extra emissions of soft and even hard partons are encoded, including $k_{t}^{2} > \mu^{2}$ configurations. Then, when calculating the charm production cross section via \textit{e.g.} the $g^* c \to g c$ mechanism one could expect to effectively include contributions related to an additional extra partonic emission (\textit{i.e.} $g^* c \to g g c$) which in some sense plays a role of the initial state parton shower.  
In Fig.~\ref{fig-uPDFs} we plot the gluon transverse momentum dependence of the different gluon uPDFs from the literature. At the small $x$-values and low scales the differences between the model are quite significant.

There are ongoing intensive works on construction of the full NLO Monte Carlo generator for off-shell initial state partons that are expected to be finished in near future \cite{private-Hameren}. This framework seems to be necessary in phenomenological studies that are based on the PB uPDFs \cite{Maciula:2019izq}. The extra hard emissions from the DGLAP-based uPDFs are usually strongly suppressed which leaves a room for higher-order terms. Therefore, in this case one needs to include usual leading order subprocesses properly matched with a number of additional higher-order radiative corrections at the level of hard matrix elements. In the moment, it can be done only at tree-level.

In consequence, the numerical calculations with the PB-NLO-set1 uPDFs, are done including in addition all $2\to 3$ and $2\to 4$ channels of partonic subprocesses that lead to a production of charm quark or antiquark and are driven by the $g^*c$ and $q^*c$ (or $\bar q^* c$) initial state interactions, similarly as in the collinear case. Here we follow a dedicated matching procedure to avoid double-counting as introduced in Ref.~\cite{Maciula:2019izq}, and further used in Refs.~\cite{Lipatov:2019izq,Maciula:2020cfy}.

\subsection{The $\bm{k_{T}}$-factorization}

Another possible theoretical approach to perform the calculations for the processes considered here is the $k_{T}$-factorization \cite{kTfactorization}. This framework extends the hybrid model formalism and includes in addition effects related to off-shellness of the initial state charm quark. In principle, it allows to study intrinsic charm contribution to charm production via mechanisms where both incident partons are off mass shell.

A topology of possible diagrams present in the $k_{T}$-factorization in the case of intrinsic charm studies is not the same as in the collinear case. Here one can follow two different ways of calculation and consider:
\begin{itemize}
\item $g^* c^* \to gc$ (and/or $q^* c^* \to qc$, $\bar{q}^* c^* \to \bar qc$ ) mechanism,
\item $g^* c^* \to c$ mechanism.
\end{itemize}
The second one is not present in other approaches and can be treated as leading-order.
The first mechanism directly corresponds to the scheme of the calculations applied in the hybrid model and can be classified as higher-order. However, their mutual coincidence is not clear and strongly depends on the model of unintegrated PDFs used in the numerical calculations. 

\subsubsection{The $2 \to 2$ partonic mechanism}

The $k_{T}$-factorization cross section for the $pp \to gc X$ reaction driven by the typical $2\to2$ mechanisms,
e.g. like the $g^* c^* \to gc$, can be expressed as follows:
\begin{eqnarray}\label{LO_kt-factorization} 
\frac{d \sigma(p p \to g c \, X)}{d y_1 d y_2 d^2p_{1,t} d^2p_{2,t}} &=&
\int \frac{d^2 k_{1,t}}{\pi} \frac{d^2 k_{2,t}}{\pi}
\frac{1}{16 \pi^2 (x_1 x_2 s)^2} \; \overline{ | {\cal M}^{\mathrm{off-shell}}_{g^* c^* \to g  c} |^2}
 \\  
&& \times  \; \delta^{2} \left( \vec{k}_{1,t} + \vec{k}_{2,t} 
                 - \vec{p}_{1,t} - \vec{p}_{2,t} \right) \;
{\cal F}_g(x_1,k_{1,t}^2,\mu^2) \; {\cal F}_c(x_2,k_{2,t}^2,\mu^2) \; \nonumber ,   
\end{eqnarray}
where ${\cal F}_g(x_1,k_{1,t}^2,\mu^2)$ and ${\cal F}_c(x_2,k_{2,t}^2,\mu^2)$
are the gluon and intrinsic charm quark uPDFs, respectively, 
for both colliding hadrons and 
${\cal M}^{\mathrm{off-shell}}_{g^* c^* \to g c}$ 
is the off-shell matrix element for the hard subprocess.
Here the Feynmann diagrams are the same as shown in Fig.~\ref{fig1}. 
The extra integration is over transverse momenta of the initial
partons. Here, one keeps exact kinematics from the very beginning 
and additional hard dynamics coming from transverse momenta of incident
partons. Explicit treatment of the transverse momenta makes 
the approach very efficient in studies of correlation observables.

Considering forward production of charm one should not expect that the
initial state (intrinsic) charm quark could have large transverse
momenta. Rather small deviations from the collinear limit are more
physically motivated here. Therefore, for the unintegrated charm
distribution ${\cal F}_c(x,k_{t}^2,\mu^2)$ we will assume Gaussian distributions with 
rather small smearing parameter $\sigma_{0}$. 
The unintegrated $c$ ($\bar c$) distributions are constructed as:
\begin{equation}
{\cal F}_c(x,k_t^2) = \pi \; G(k_t^2) \cdot x c ( x,\mu^2 ),
\label{UPDF_c}
\end{equation}
where 
\begin{equation}
G(k_t^2) = \frac{1}{2 \pi \sigma_0^2} 
\exp\left( \frac{-k_t^2}{2 \sigma_0^2} \right)
\end{equation}
is a standard two-dimensional Gaussian distribution and $\sigma_0$
is in principle a free parameter which governs the nonperturbative
effects in the proton wave function.
The factor $\pi$ is because of our normalization of unintegrated
parton distributions:
\begin{equation}
\int d k_t^2 {\cal F}_c(x,k_t^2) = x c(x) \; .
\label{normalization_of_UPDFs}
\end{equation}

The hard off-shell matrix element ${\cal M}^{\mathrm{off-shell}}_{g^* c^* \to g c}$ is known only in the massless limit.
Within this limit the relevant calculations can be done in the \textsc{KaTie} Monte Carlo code, where the matrix element is computed numerically. Its analytic form can be obtained according to Parton-Reggeization-Approach (PRA) and was published in Ref.~\cite{Nefedov:2013ywa}. For the higher-order tree-level diagrams with off-shell initial state partons and with extra partonic legs in the final state one can also use \textsc{KaTie}, which is very efficient in this type of calculations and which was very recently equipped with tools that allow in addition for generation of analytic form of matrix-elements for a given hard multileg processes \cite{vanHameren:2016kkz}. 

\subsubsection{The $2\to 1$ partonic mechanism}

\begin{figure}[!h]
\begin{minipage}{0.35\textwidth}
  \centerline{\includegraphics[width=1.0\textwidth]{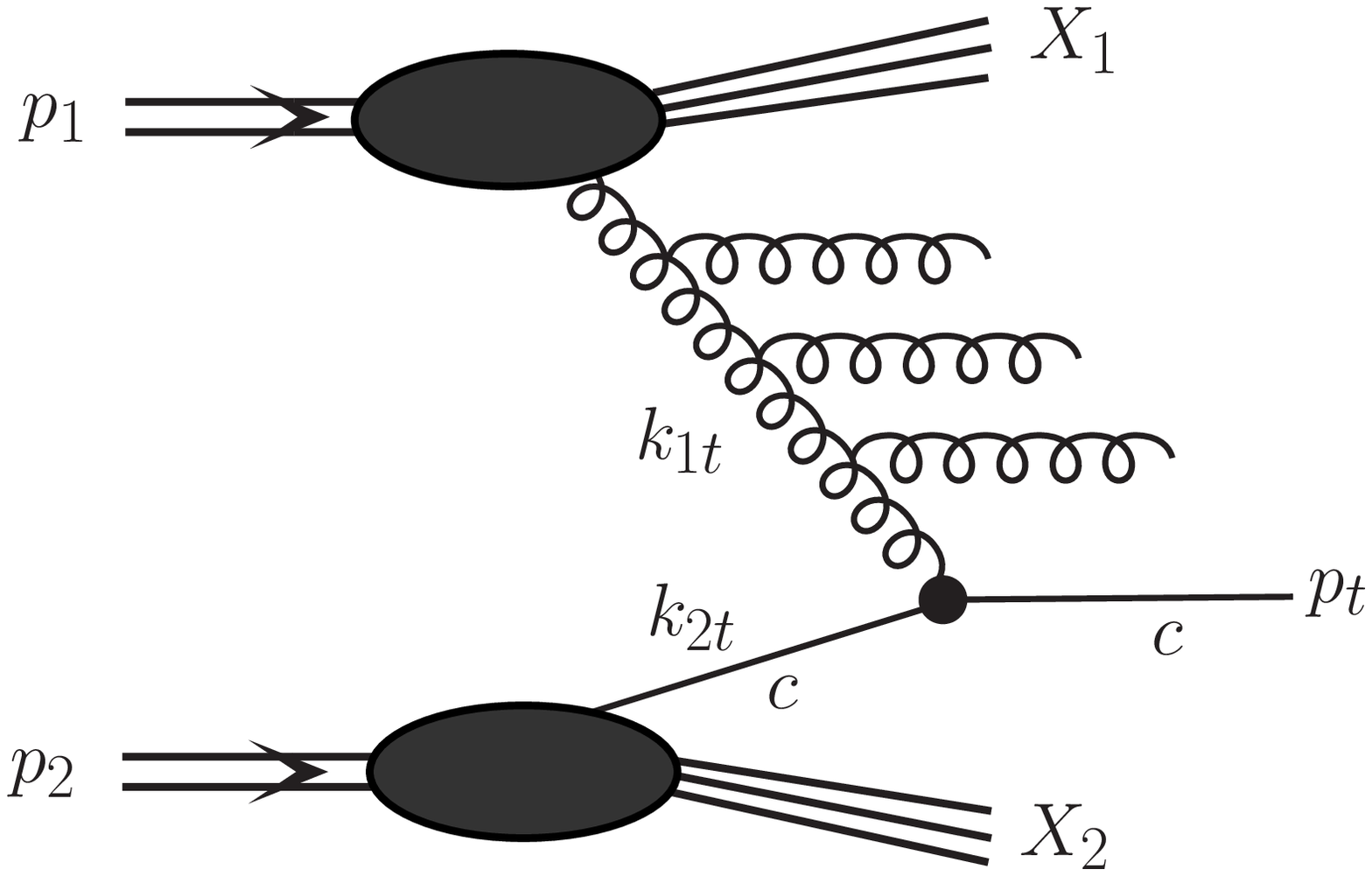}}
\end{minipage}
\hspace{+5mm}
\begin{minipage}{0.35\textwidth}
  \centerline{\includegraphics[width=1.0\textwidth]{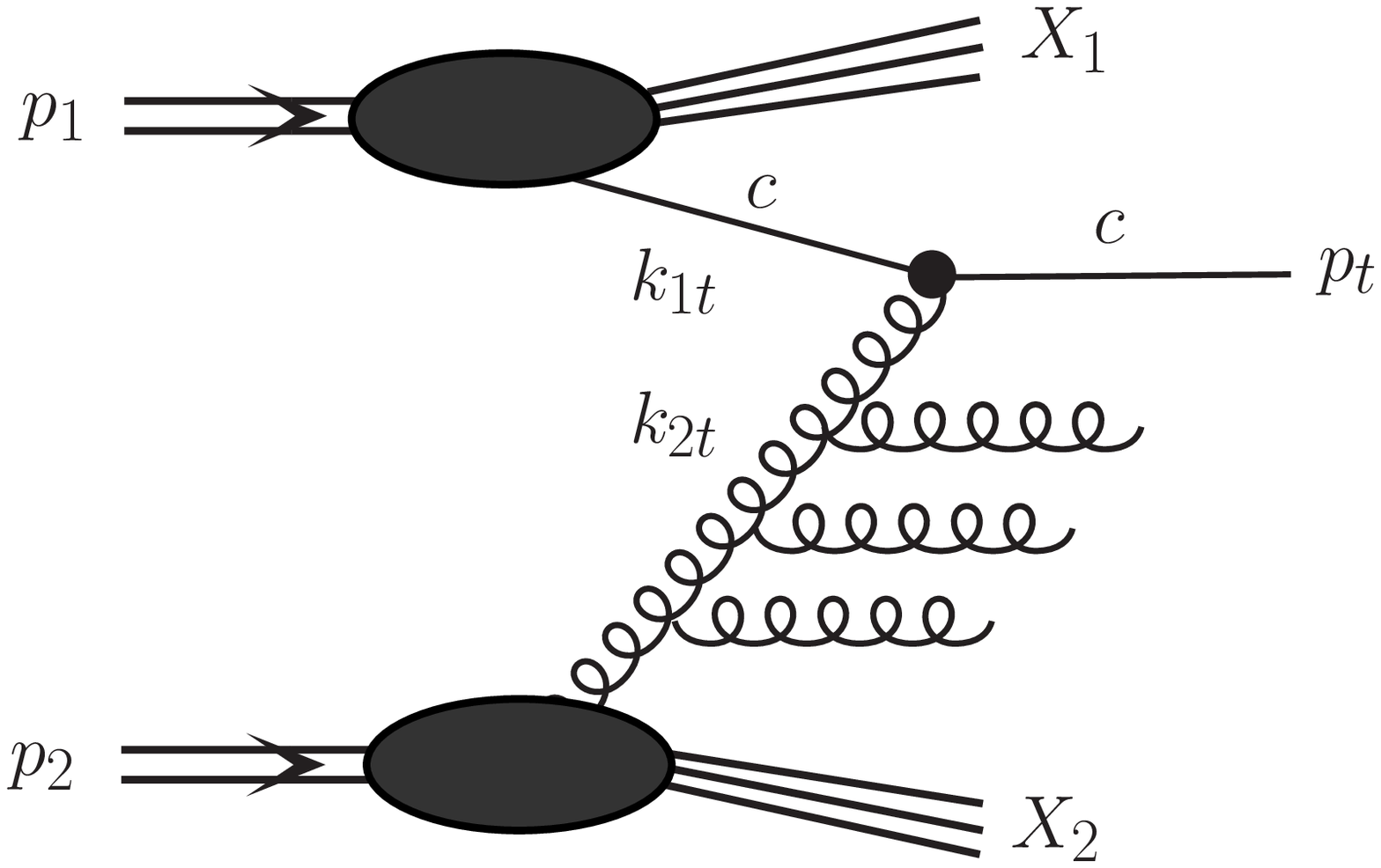}}
\end{minipage}
  \caption{Two leading-order diagrams for charm quark (antiquark)
    production relevant for $k_t$-factorization approach. The extra
    explicit gluonic emissions suggest the use of unintegrated 
   gluon distributions.
\small   
}
\label{fig:kt-factorization_diagrams}
\end{figure}

In the $k_T$-factorization framework the charm quark/antiquark can be
created at one-order higher approach. A relevant formalism was
used previously for production of forward pions in Ref.~\cite{CS2006}. 
In Fig.~\ref{fig:kt-factorization_diagrams} we show basic graphs
for charm quark production within the $2 \to 1$ mechanisms.

The emitted charm-quark (or antiquark) momentum-space distribution can be
written as:
\begin{eqnarray}
\frac{d \sigma(p p \to c \, X)}{d y d^2 p_t}&& = \frac{16 N_c}{N_c^2 - 1} \cdot
\frac{4}{9} \cdot \frac{1}{m_t^2} \times \; \nonumber \\
&&\int 
\alpha_s(\Omega^2) f_g(x_1,k_{1,t}^2,\mu^2) f_c(x_2,k_{2,t},\mu^2)
\delta\left( \vec{k}_{1,t} + \vec{k}_{2,t} - \vec{p}_t \right)
d^2 k_{1,t} d^2 k_{2,t} \; .  \nonumber \\  
\label{kt_factorization_at_LO}
\end{eqnarray}
In the formula above $f$'s are unintegrated gluon or charm
quark/antiquark distributions.
For unintegrated gluon distributions we will take the ones used
recently in the literature in the context of $\eta_c$ or $\chi_c$
production \cite{BPSS2019,BPSS2020} where the kinematics is similar. 
For $\Omega^2$ we can take $\Omega^2 = \min(m_t,k_{1t}^2,k_{2t}^2)$
or just $\Omega^2 = m_t^2$.
The longitudinal momentum fractions are calculated as
\begin{eqnarray}
x_1 &=& \frac{m_t}{\sqrt{s}} \exp(+y) \; , \nonumber \\
x_2 &=& \frac{m_t}{\sqrt{s}} \exp(-y) \; .
\end{eqnarray}
%

\section{Results}

We divide the section with numerical results to four subsections. First three of them are devoted to 
numerical calculations obtained with the collinear-, hybrid- and the $k_T$-factorization approach, respectively.
The last subsection contains explicit predictions for impact of intrinsic charm mechanism on forward production of charm
in different experiments, including low energy LHC experiments like fixed-target LHCb and SHIP, as well as high energy FCC and LHC experiments, like proposed recently LHC-FASER.     

\subsection{The collinear approach}

We start presentation of numerical predictions with the results for $pp
\to gc X$ reaction driven by the $gc \to gc$ leading-order mechanism
calculated in the collinear framework within massive matrix element and
kinematics for the energy $\sqrt{s} = 7$ TeV. Here we take the gluon and
the intrinsic charm distributions as encoded in the CT14nnloIC collinear
PDFs. The three different lines in Fig.~\ref{fig1} correspond to a different choice of the $p_{t0}$ parameter used for the regularization of the cross section. We see that the predictions for charm quark transverse momentum (left panel) and rapidity (right panel) distributions are very sensitive to the choice of this parameter, especially, at small charm quark transverse momenta, which also affects the rapidity spectrum. In the numerical studies below $p_{t0}=1.0$ GeV will be taken as a default choice which leads to a central value of the uncertainty related to the choice of the parameter. 

\begin{figure}[!h]
\begin{minipage}{0.47\textwidth}
  \centerline{\includegraphics[width=1.0\textwidth]{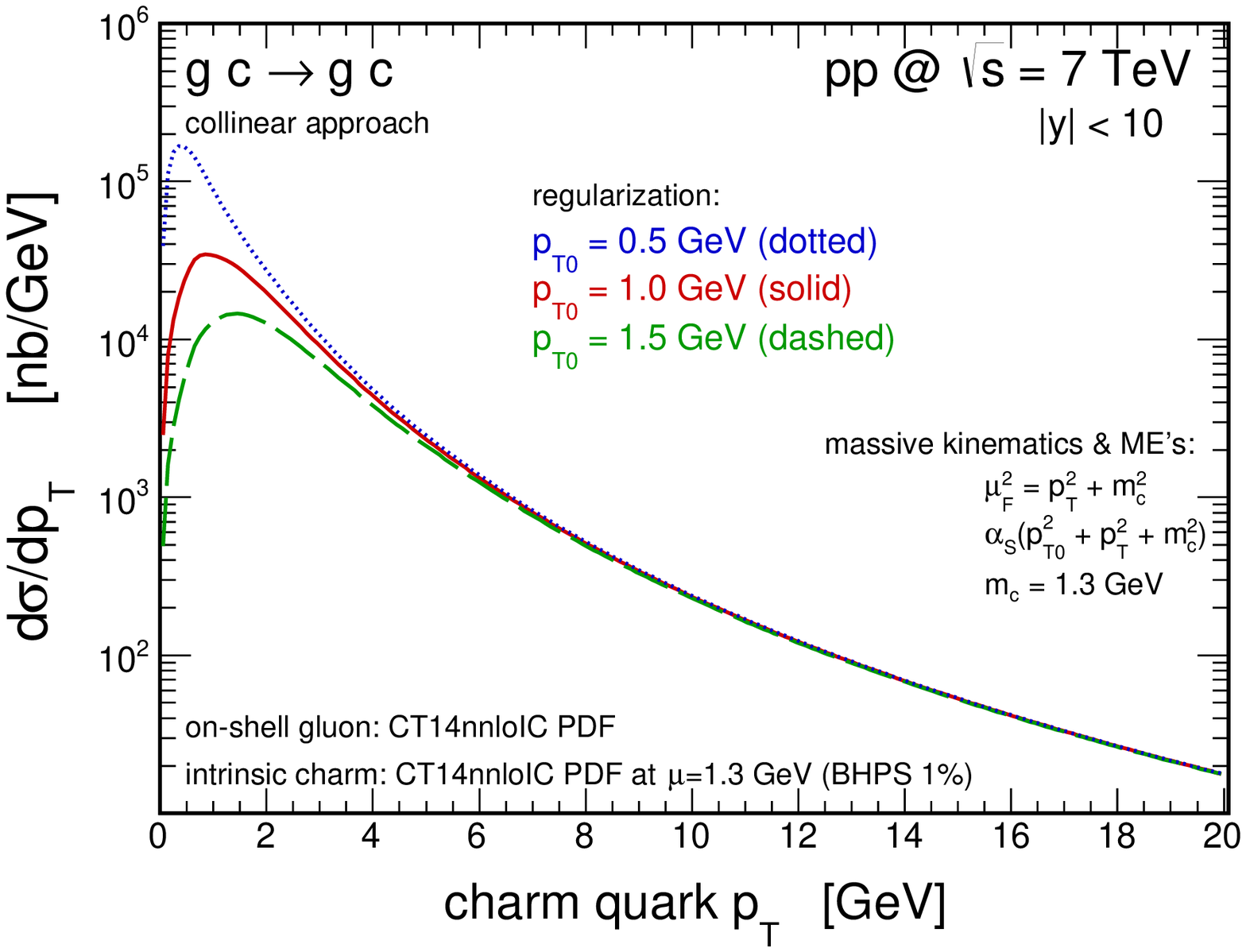}}
\end{minipage}
\begin{minipage}{0.47\textwidth}
  \centerline{\includegraphics[width=1.0\textwidth]{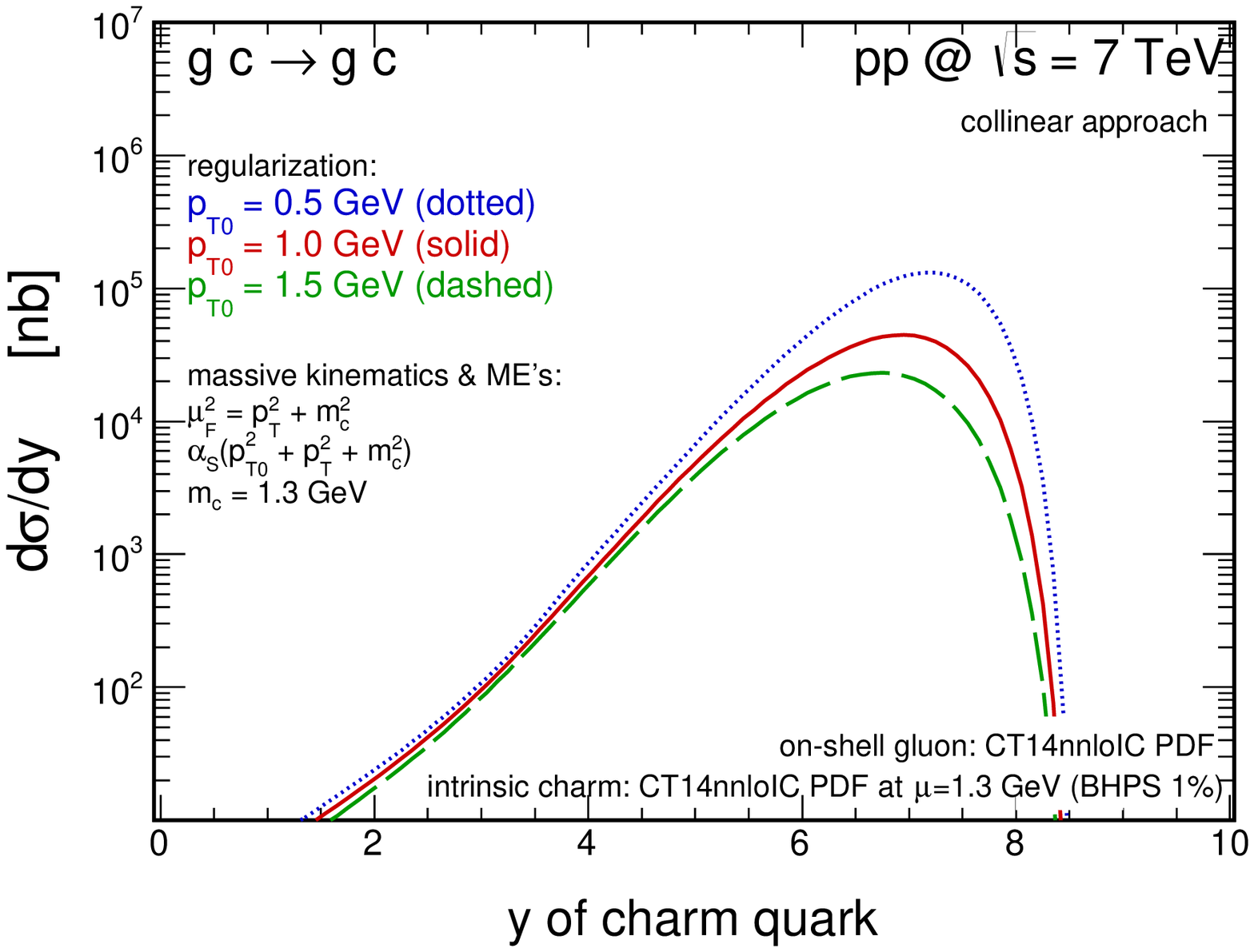}}
\end{minipage}
  \caption{
\small The charm quark transverse momentum (left) and rapidity (right) differential cross sections for $pp$-scattering at $\sqrt{s}=7$ TeV. The results correspond to the $g c \rightarrow g c$ mechanism calculated within the intrinsic charm concept   in the collinear-approach with matrix element and kinematics for massive charm quark. Here three different values of the regularization parameter $p_{T0}$ are used. Details are specified in the figure.   
}
\label{fig1}
\end{figure}

\begin{figure}[!h]
\begin{minipage}{0.47\textwidth}
  \centerline{\includegraphics[width=1.0\textwidth]{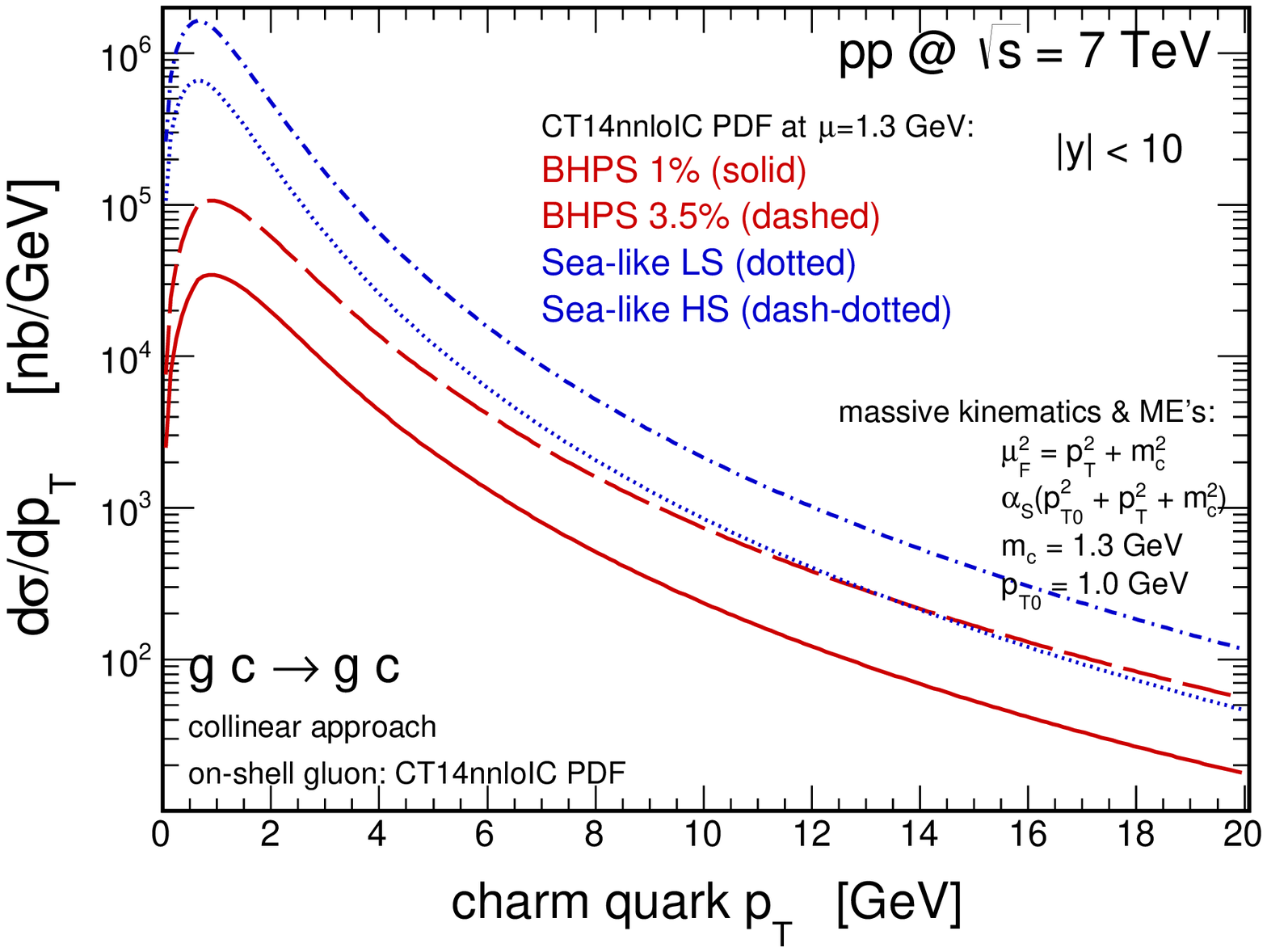}}
\end{minipage}
\begin{minipage}{0.47\textwidth}
  \centerline{\includegraphics[width=1.0\textwidth]{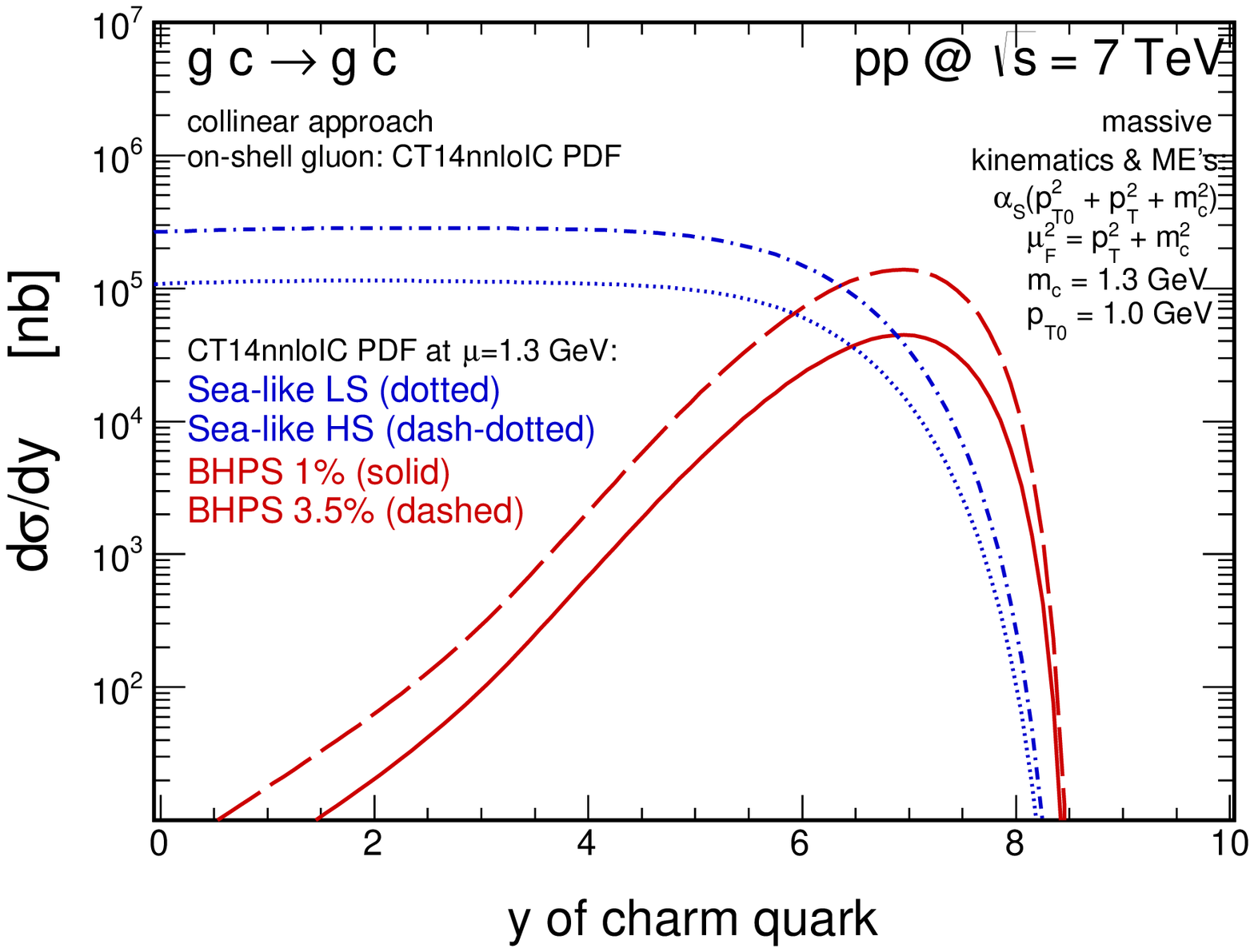}}
\end{minipage}
  \caption{
\small The same as in Fig.~\ref{fig1} but here results obtained with the four different scenarios for intrinsic charm content in a proton are shown. Details are specified in the figure.  
}
\label{fig2}
\end{figure}

In Fig.~\ref{fig2} we present again collinear results for the leading-order $gc \to gc$ mechanism but here we
applied four different sets of the intrinsic charm distribution in a proton at initial scale $\mu = 1.3$ GeV as incorporated in the CT14nnloIC PDFs. Again, we show the differential cross sections as a function of the charm quark transverse momentum (left panel) and rapidity (right panel). The solid, long-dashed, dotted and dash-dotted lines correspond to the BHPS 1\%, BHPS 3.5\%, sea-like LS and sea-like HS models, respectively. The sea-like models lead to a larger cross section than in the case of the BHPS model in the midrapidty region. On the other hand, a larger cross section in the forward direction is obtained within the BHPS models. Clearly, large uncertainties due to the intrinsic charm input are found.  

\begin{figure}[!h]
\begin{minipage}{0.47\textwidth}
  \centerline{\includegraphics[width=1.0\textwidth]{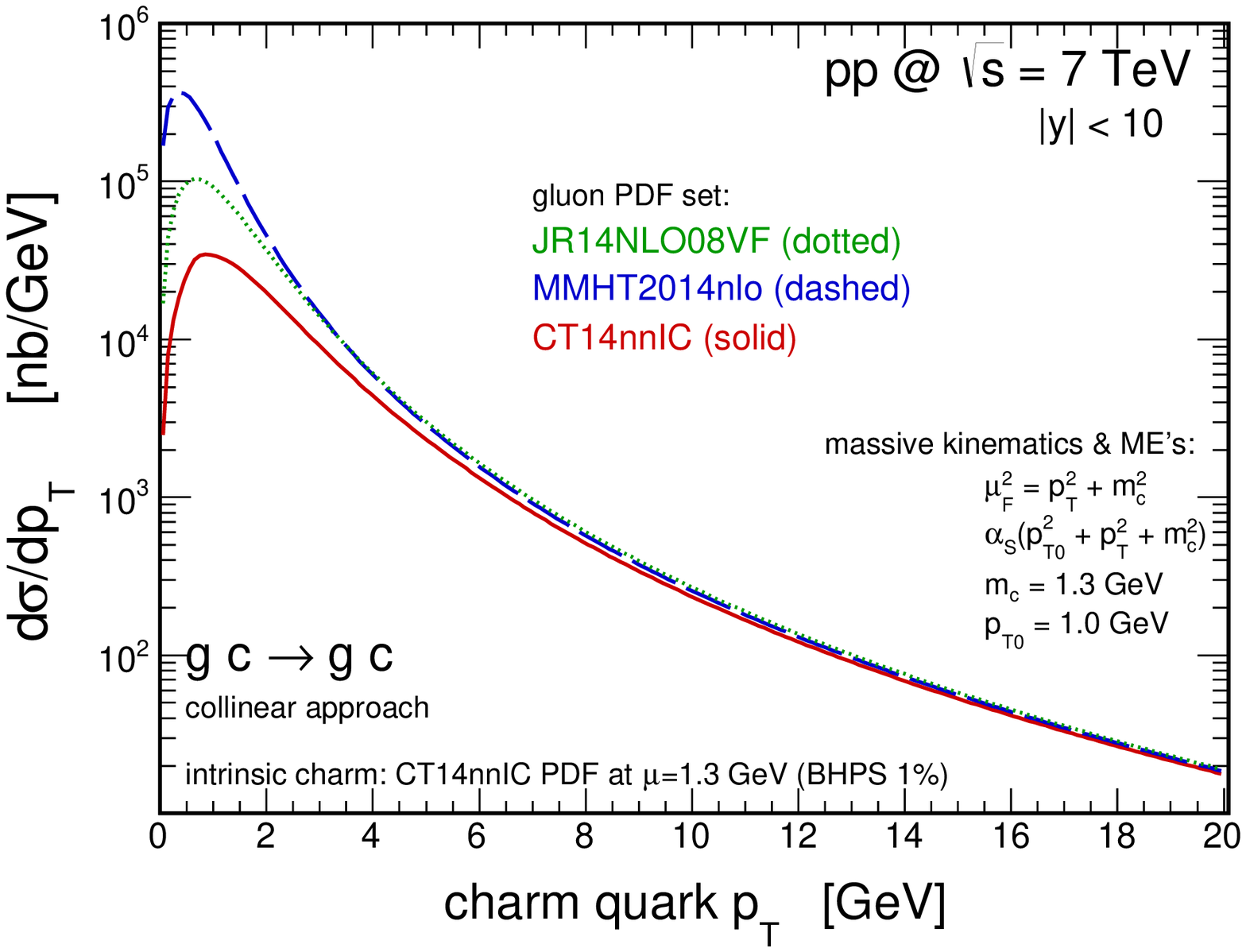}}
\end{minipage}
\begin{minipage}{0.47\textwidth}
  \centerline{\includegraphics[width=1.0\textwidth]{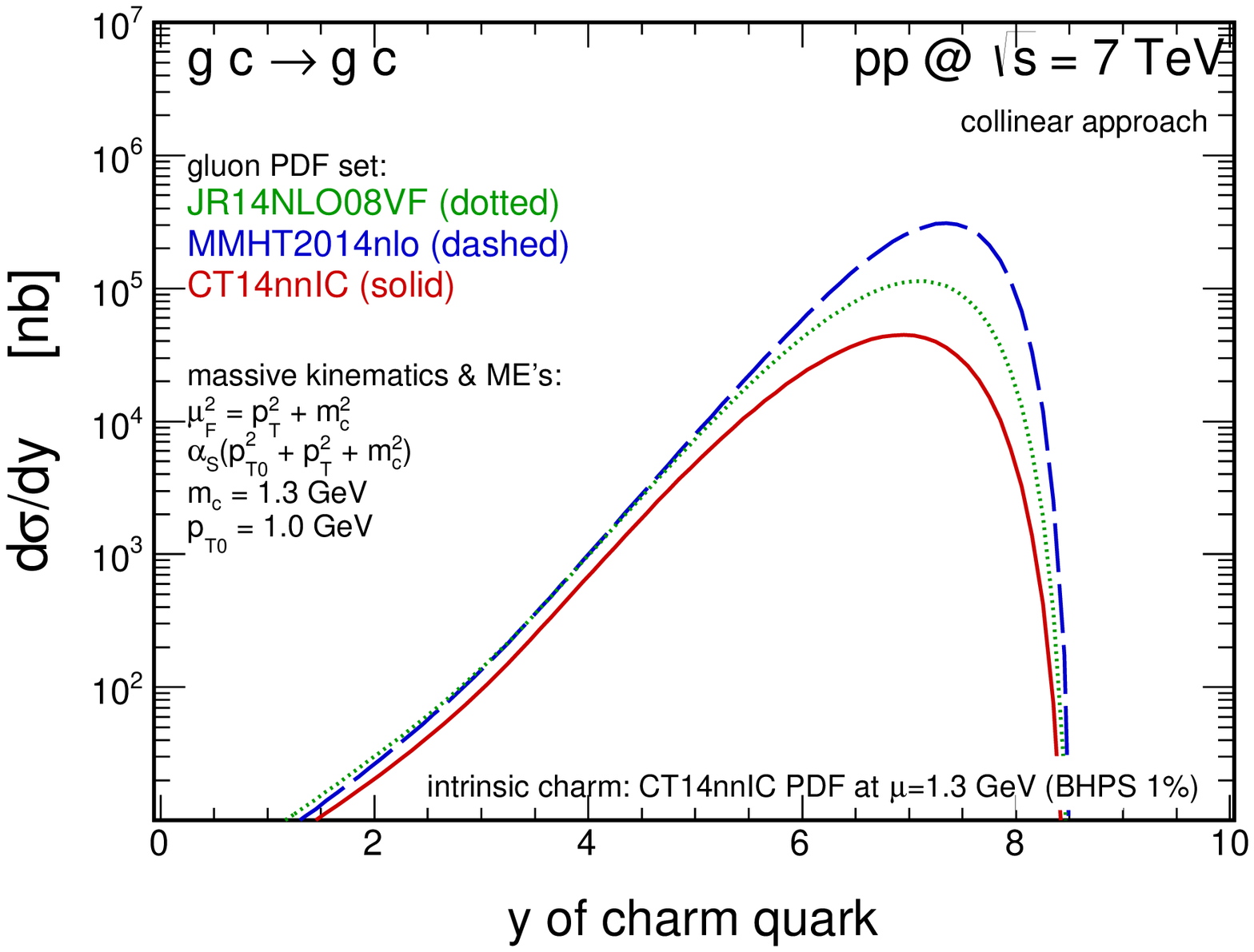}}
\end{minipage}
  \caption{
\small The same as in Fig.~\ref{fig1} but here results obtained with the three different collinear gluon PDFs are shown. Details are specified in the figure.   
}
\label{fig3}
\end{figure}

The intrinsic charm component in the proton is not the only source of uncertainties related to the collinear PDFs. As it is shown in Fig.~\ref{fig3} the gluon PDF also leads to a significant uncertainties of the predictions. Here we show a comparison of the predictions obtained with the default CT14nnloIC (solid lines), the JR14NLO08VF (dotted lines) and the MMHT2014nlo (dashed lines) PDF sets. The gluon PDFs provided by different groups are probed here at small-$x$ and relatively small scales and lead to a quite different results, especially, at small transverse momenta of charm quark.

\begin{figure}[!h]
\begin{minipage}{0.47\textwidth}
  \centerline{\includegraphics[width=1.0\textwidth]{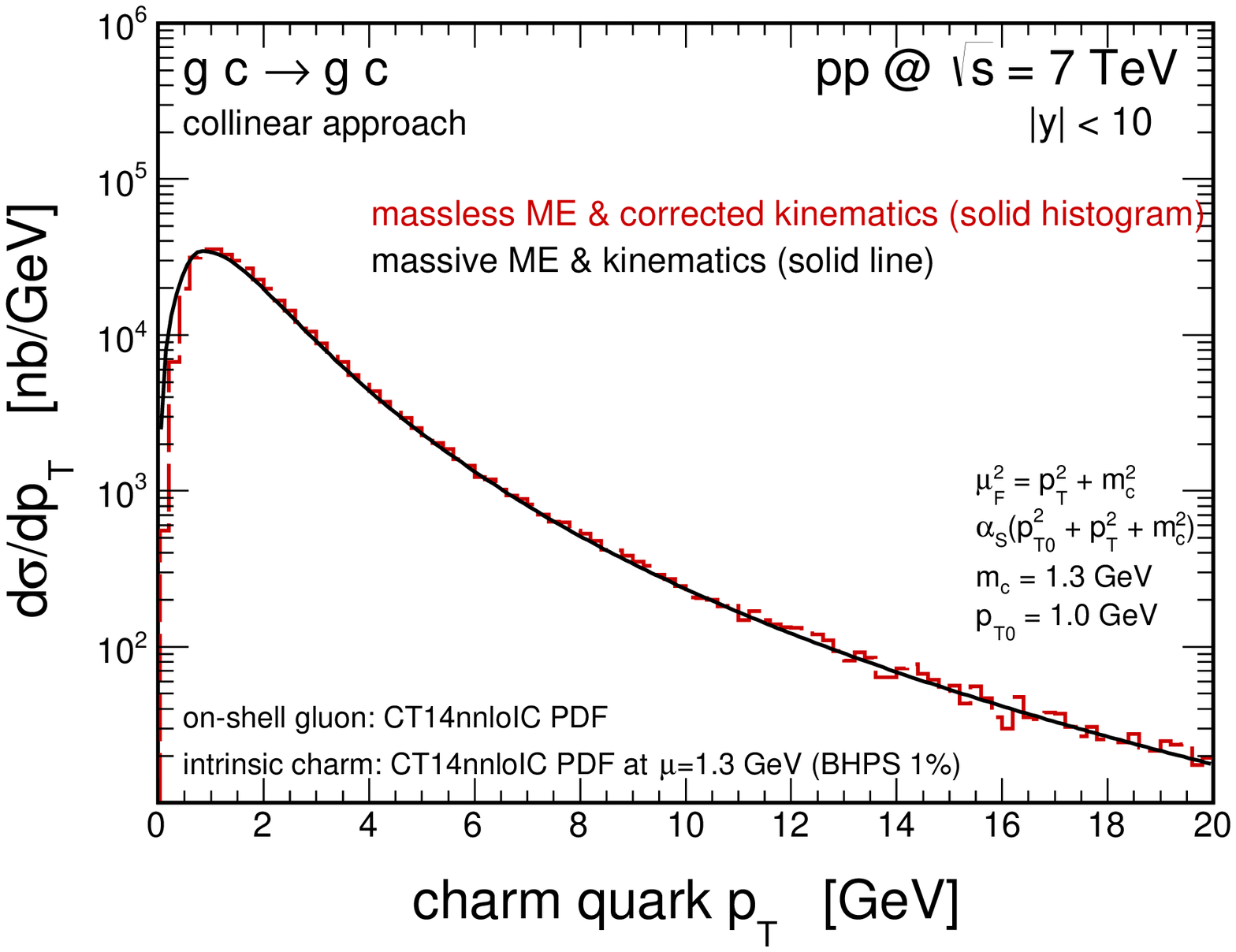}}
\end{minipage}
\begin{minipage}{0.47\textwidth}
  \centerline{\includegraphics[width=1.0\textwidth]{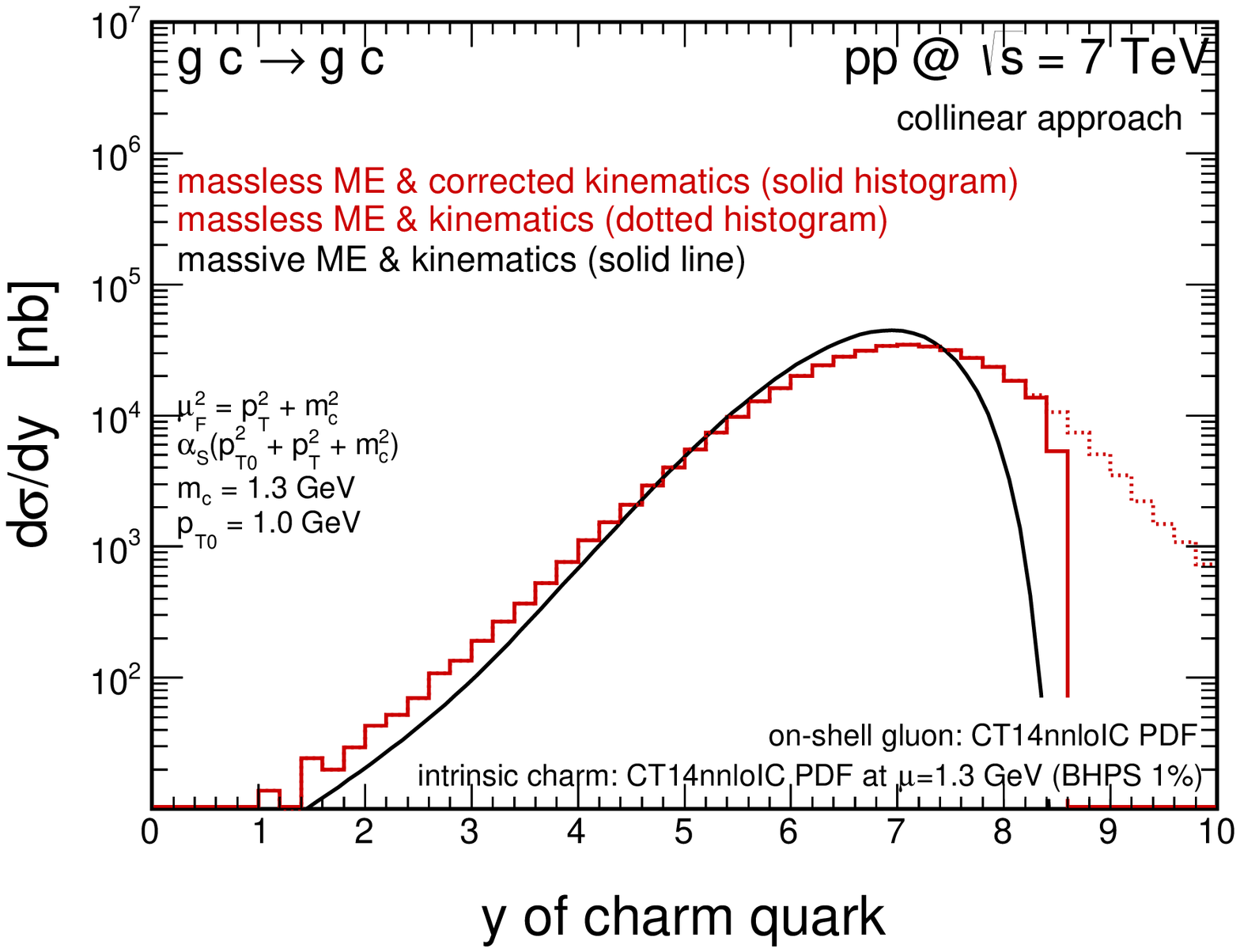}}
\end{minipage}
  \caption{
\small The same as in Fig.~\ref{fig1} but here results of three different schemes of the collinear caclulations are compared. The solid lines correspond to the calculations with massive matrix element and kinematics, the dotted histograms show results for the calculations with massless matrix elements and kinematics, and the solid histograms represent calculations with massless matrix element and kinematics corrected for the charm quark mass.    
}
\label{fig4}
\end{figure}

Now we wish to compare three different schemes for the collinear
calculations of the $pp \to gc X$ reaction via the $gc \to gc$
leading-order partonic subprocess. In Fig.~\ref{fig4} we present
theoretical distributions obtained within the matrix element with
massive quarks (called massive ME for brevity) and kinematics
including quark masses (solid lines, our default choice), within the massless matrix element and massless kinematics (dotted histograms), as well as within the massless matrix element and kinematics corrected for the charm quark mass (solid histograms). In each of the cases, we kept the same choice of the renormalization scale $\mu_{R}^{2} = p_{t0}^{2}+p_{t}^{2}+m_{c}^{2}$ and the factorization scale $\mu_{F}^{2} = p_{t}^{2}+m_{c}^{2}$. The charm quark transverse momentum distributions (left panel) are almost identical and some very small (almost invisible) discrepancies appear only at extremely small transverse momenta. The rapidity distributions (right panel) are found to be very sensitive to the charm quark mass effects. Neglecting the charm quark mass in the kinematics leads to a shift of its rapidity distribution to a far forward direction. Correction of the kinematics by inclusion of the outgoing particles mass in the calculation of $x$-values seems to approximately restore the full massive calculations. This step seems to be necessary in the case of massless calculations, otherwise shapes of the predicted rapidity distributions may not be correct.      

\begin{figure}[!h]
\begin{minipage}{0.47\textwidth}
  \centerline{\includegraphics[width=1.0\textwidth]{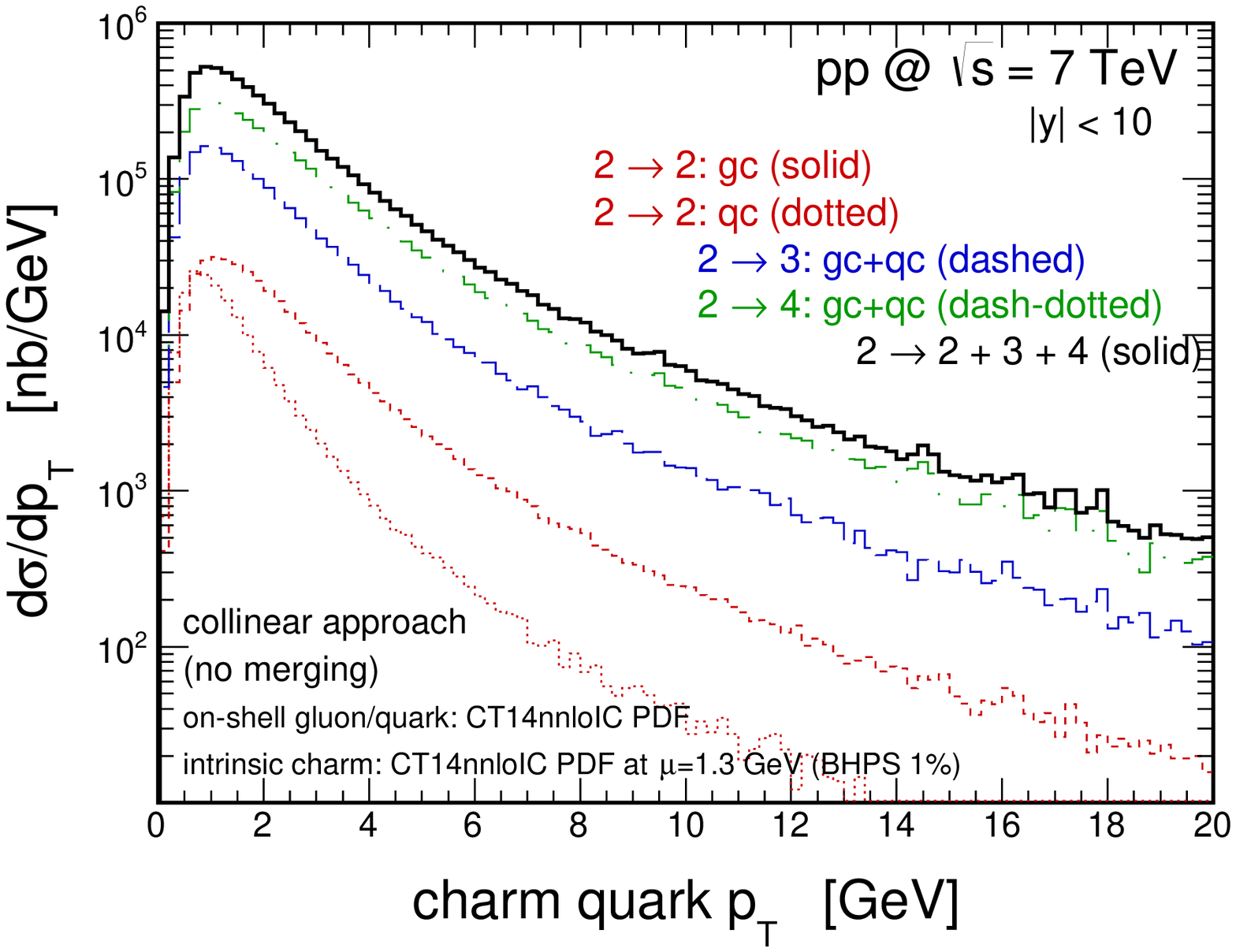}}
\end{minipage}
\begin{minipage}{0.47\textwidth}
  \centerline{\includegraphics[width=1.0\textwidth]{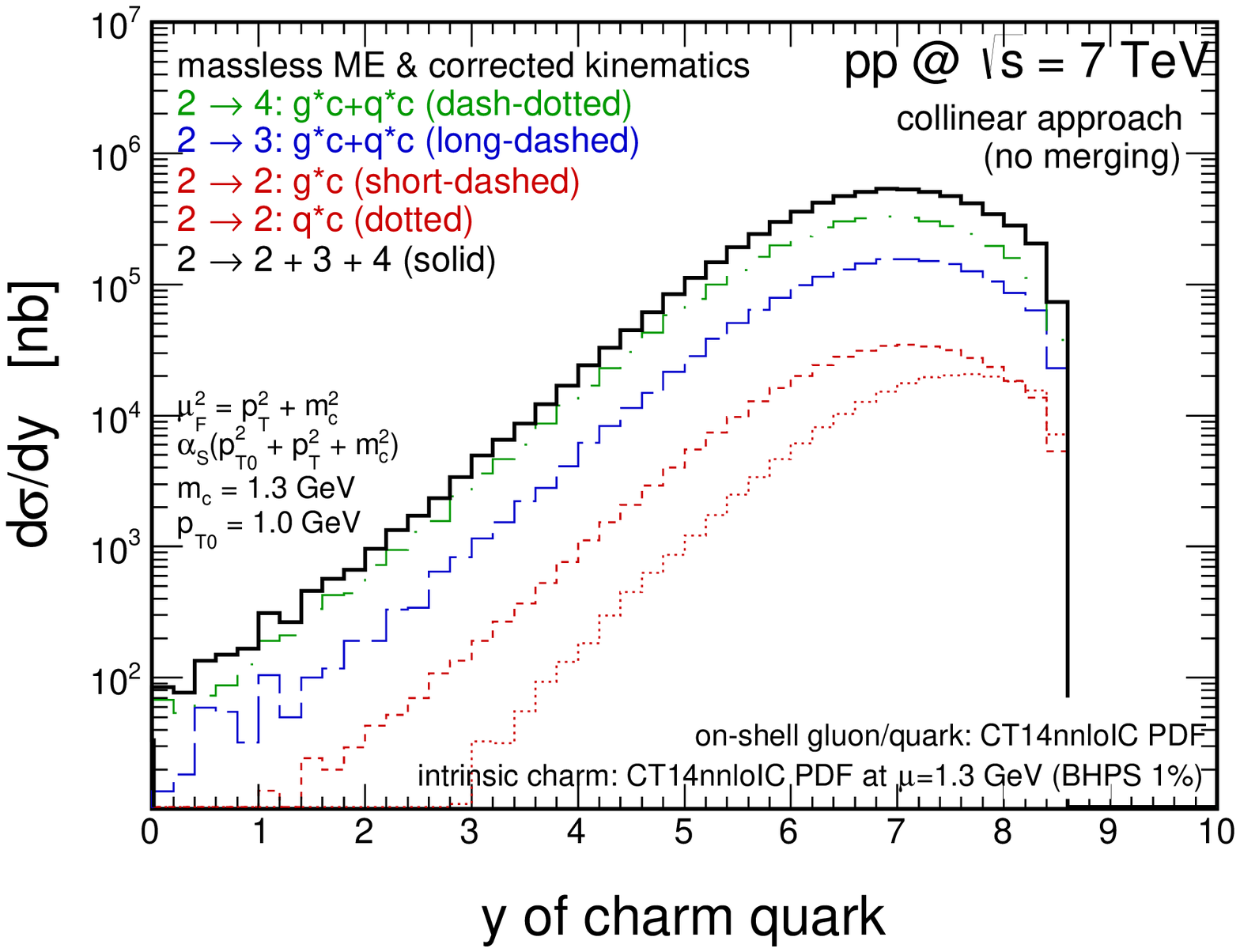}}
\end{minipage}
  \caption{
\small The same as in Fig.~\ref{fig1} but here results of the $2 \to 2$ ($gc$ and $qc$ initial states), $2\to 3$ ($gc + qc$ initial states) and $2\to4$ ($gc + qc$ initial states) mechanisms are shown separately. The calculations are done with massless matrix element and kinematics corrected for the charm quark mass. Details are specified in the figure.     
}
\label{fig5}
\end{figure}

Having discussed the dominant leading-order mechanism we wish to move beyond and consider 
importance of higher-order corrections for the charm quark forward production mechanisms
with intrinsic charm in the initial state. In Fig.~\ref{fig5} we compare our collinear predictions
for the leading-order $2\to2$ mechanisms, both $gc \to gc$ (dotted
histograms) and $qc\to qc$ (short-dashed histograms) shown separately,
and for the higher-order $2\to3$ (long-dashed histograms) and $2\to 4$
(dash-dotted histograms) mechanisms calculated at tree-level. A sum of
the four different components denoted as $2\to 2+3+4$ is also shown but
it does not follow any merging procedure here\footnote{Technically, this
  could be done properly only if the parton level calculations are
  supplemented with a parton shower but it goes beyond the scope of the
  present study.}. For the higher-order contributions the partonic
subprocesses with $gc$ and $qc$ initial states are added together. We
report a huge contribution to the cross section coming from the
higher-order mechanisms (more than order of magnitude). It clearly shows
that the leading-order mechanisms  are not enough in order to get
reasonable predictions for the impact of intrinsic charm concept on
forward charm quark production. Full NLO and even NNLO frameworks are
required for precise studies of the subject within the collinear parton
model. The situation in the case for other approaches, like the hybrid-
and the $k_{T}$-factorization is quite different than in the collinear case what will be
discussed in next two subsections.   

\subsection{The hybrid model}

\begin{figure}[!h]
\begin{minipage}{0.47\textwidth}
  \centerline{\includegraphics[width=1.0\textwidth]{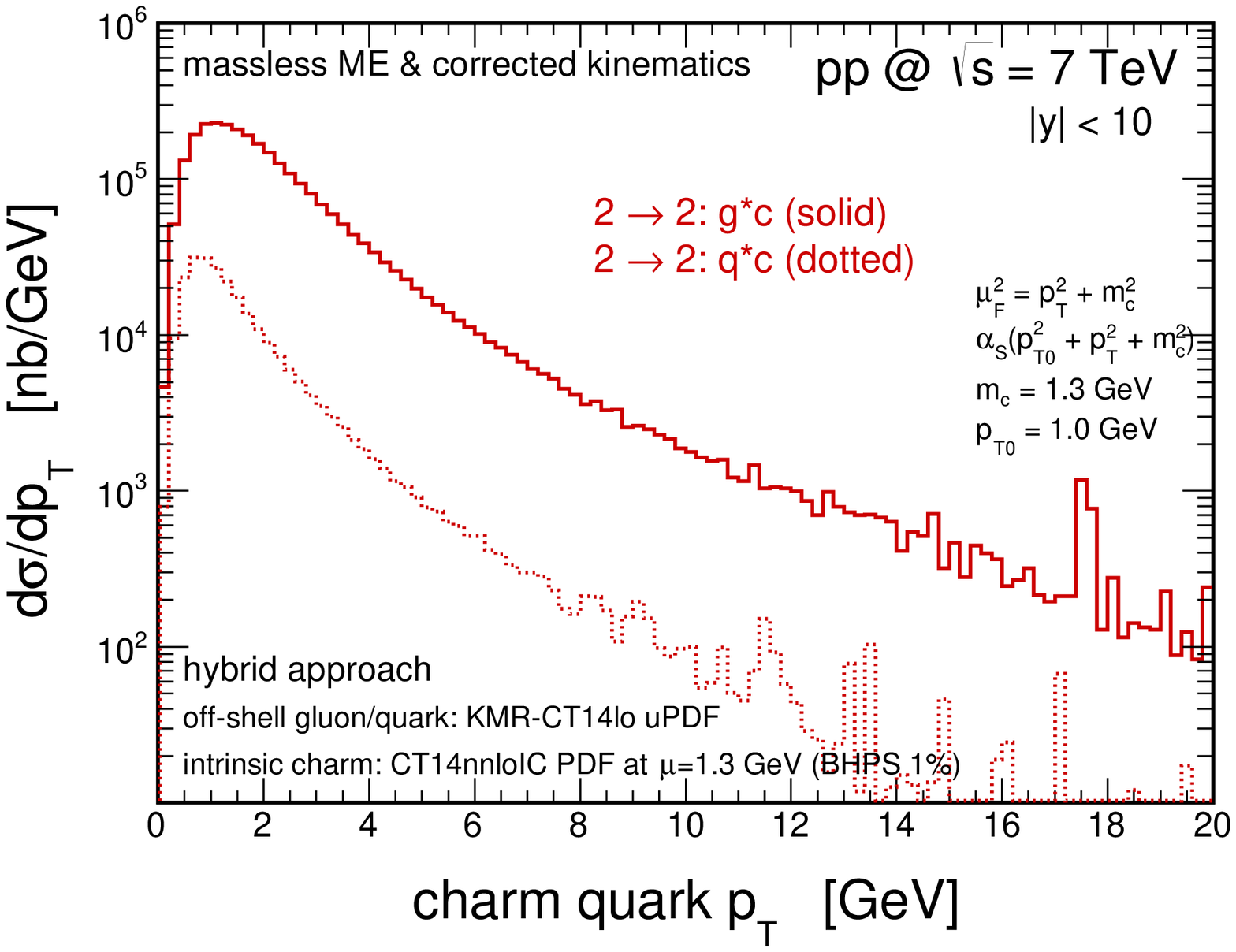}}
\end{minipage}
\begin{minipage}{0.47\textwidth}
  \centerline{\includegraphics[width=1.0\textwidth]{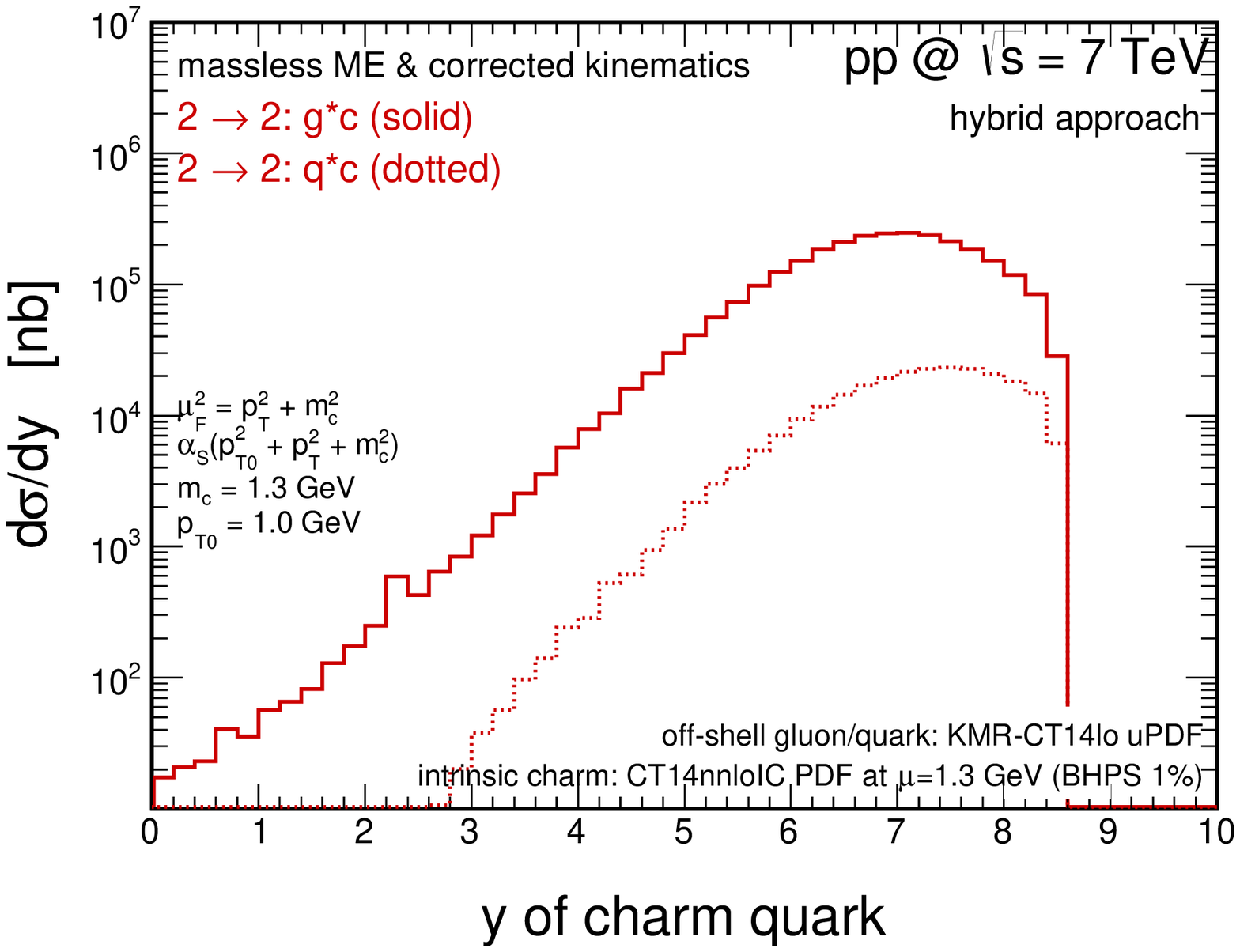}}
\end{minipage}
  \caption{
\small The charm quark transverse momentum (left) and rapidity (right) differential cross sections for $pp$-scattering at $\sqrt{s}=7$ TeV. The results correspond to the $g^* c \rightarrow g c$ and $q^* c \rightarrow q c$ mechanisms calculated within the intrinsic charm concept in the hybrid model with off-shell initial state gluon and/or off-shell light-quark. Here the KMR-CT14lo unintegrated parton densities were used.   
}
\label{fig6}
\end{figure}

Now we wish to start presentation of our numerical results obtained in
the hybrid model. Here the incident small-$x$ parton is assumed 
to be off-mass shell in contrast to the large-$x$ intrinsic charm which
is kept on-shell. In Fig.~\ref{fig6} we show theoretical predictions for
charm quark transverse momentum (left panel) and rapidity (right panel)
distributions for forward charm production within the leading-order
$g^*c \to gc$ and the $q^*c \to qc$ mechanisms. 
Here the KMR-CT14lo gluon and light quark/antiquark uPDFs are used. We
observe that here much larger cross sections are obtained 
than in the analogous calculations done in the collinear framework 
(see two lowest histograms in Fig.~\ref{fig5}). 
Especially, in the hybrid model the gluonic component is much bigger
than its collinear counterpart. Significant effects related to the off-shellness of the incident gluons are found. Considering far forward rapidities of the produced charm quark the transverse momenta (virtualities) of the incident small-$x$ gluons start to play a very important role and lead to a sizeable enhancement of the predicted cross section with respect to the leading-order collinear calculations. 
 
\begin{figure}[!h]
\begin{minipage}{0.47\textwidth}
  \centerline{\includegraphics[width=1.0\textwidth]{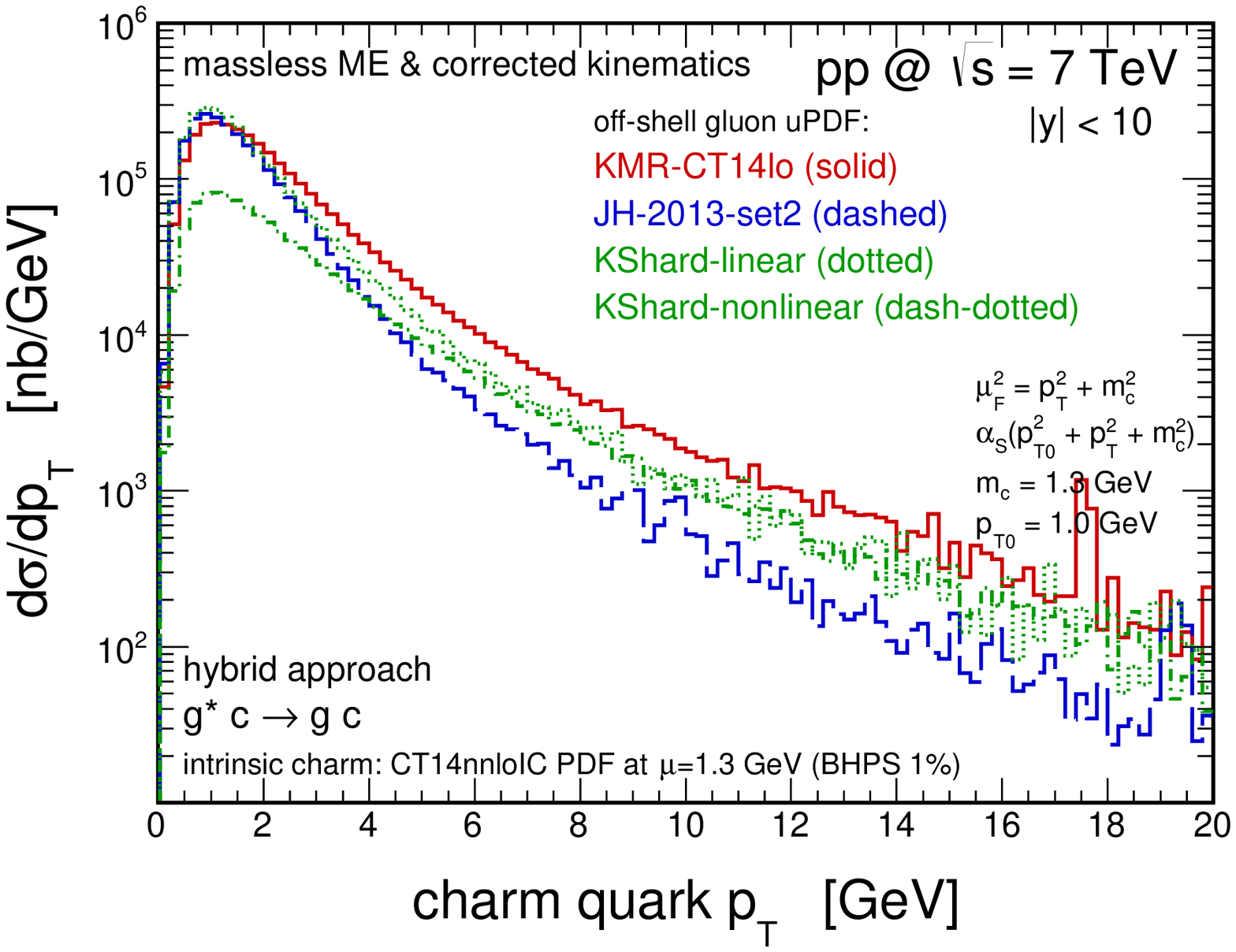}}
\end{minipage}
\begin{minipage}{0.47\textwidth}
  \centerline{\includegraphics[width=1.0\textwidth]{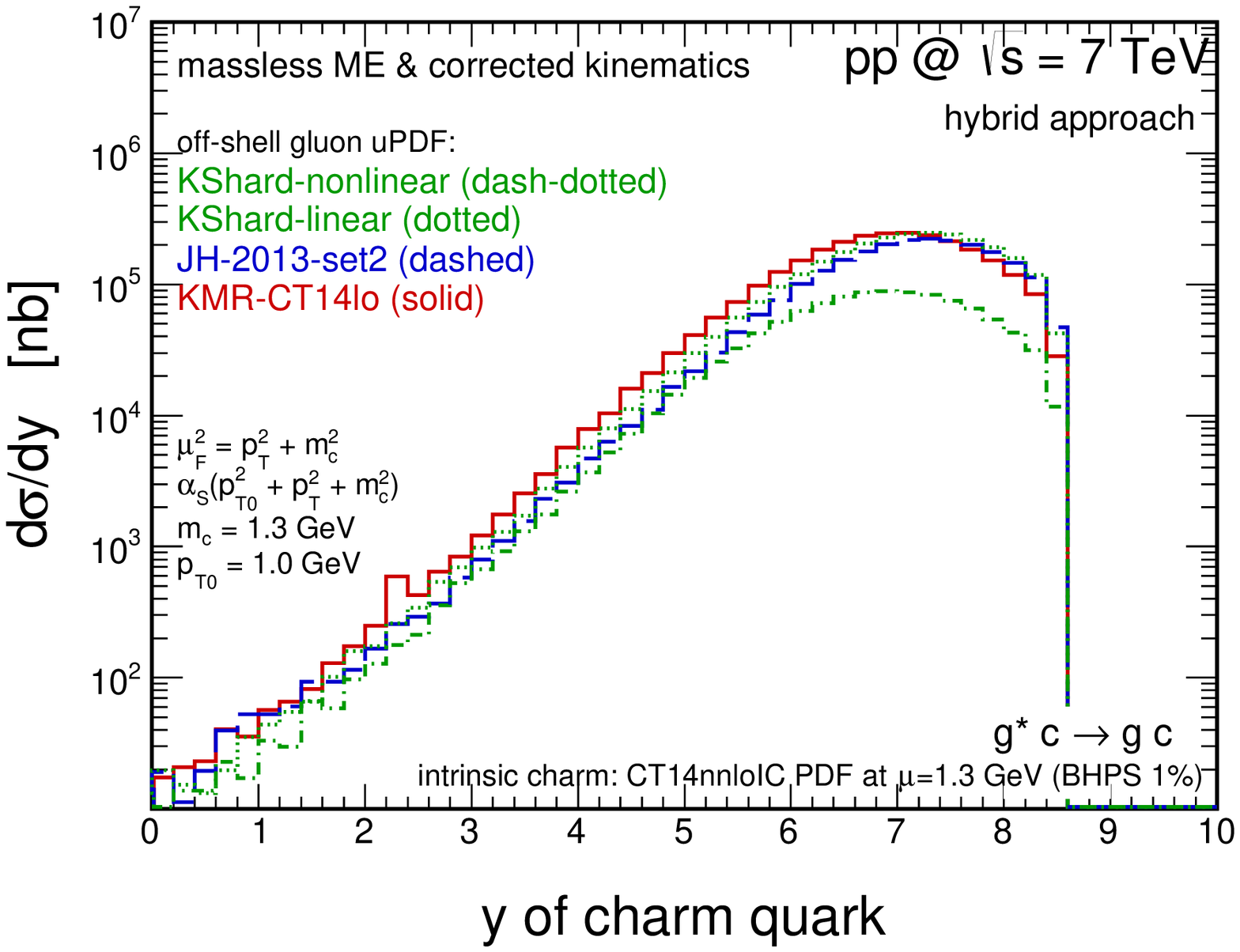}}
\end{minipage}
  \caption{
\small The same as in Fig.~\ref{fig6} but here results for four different unintegrated gluon densities in a proton are shown.
Here only the $g^* c \rightarrow g c$ mechanism is taken into account.   
}
\label{fig7}
\end{figure}

Since in the hybrid model the leading-order quark component $q^* c \rightarrow q c$ is found to be negligible one can safely concentrate on the gluonic $g^* c \rightarrow g c$ channel only. In Fig.~\ref{fig7} we show the relevant predictions for different unintegrated gluon densities from the literature. We compare results obtained with the KMR-CT14lo (solid histograms), the CCFM JH-2013-set2 (dashed histograms) as well as the KS-linear (dotted histograms) and KS-nonlinear (dash-dotted histograms) gluon uPDFs. Different models lead to quite different results, however, they seem to be consistent with each other up to a factor of 5. Main differences appear at larger quark transverse momenta. At small transverse momenta predictions within the KMR-CT14lo, the JH-2013-set2 and the KS-linear uPDFs coincide. It translates also into the rapidity spectrum. Only the KS-nonlinear uPDF leads to a somewhat different behaviour of the cross section at small $p_{T}$'s. We observe that both the transverse momentum and rapidity distributions of charm quark are sensitive to the non-linear evolution effects that lead here to a sizeable damping of the predicted cross section. Thus, the forward production of charm within intrinsic charm concept might be a very good testing ground for studies of the non-linear term in the evolution of unintegrated gluon densities and may shed new light on phenomenon of parton saturation.

\begin{figure}[!h]
\begin{minipage}{0.47\textwidth}
  \centerline{\includegraphics[width=1.0\textwidth]{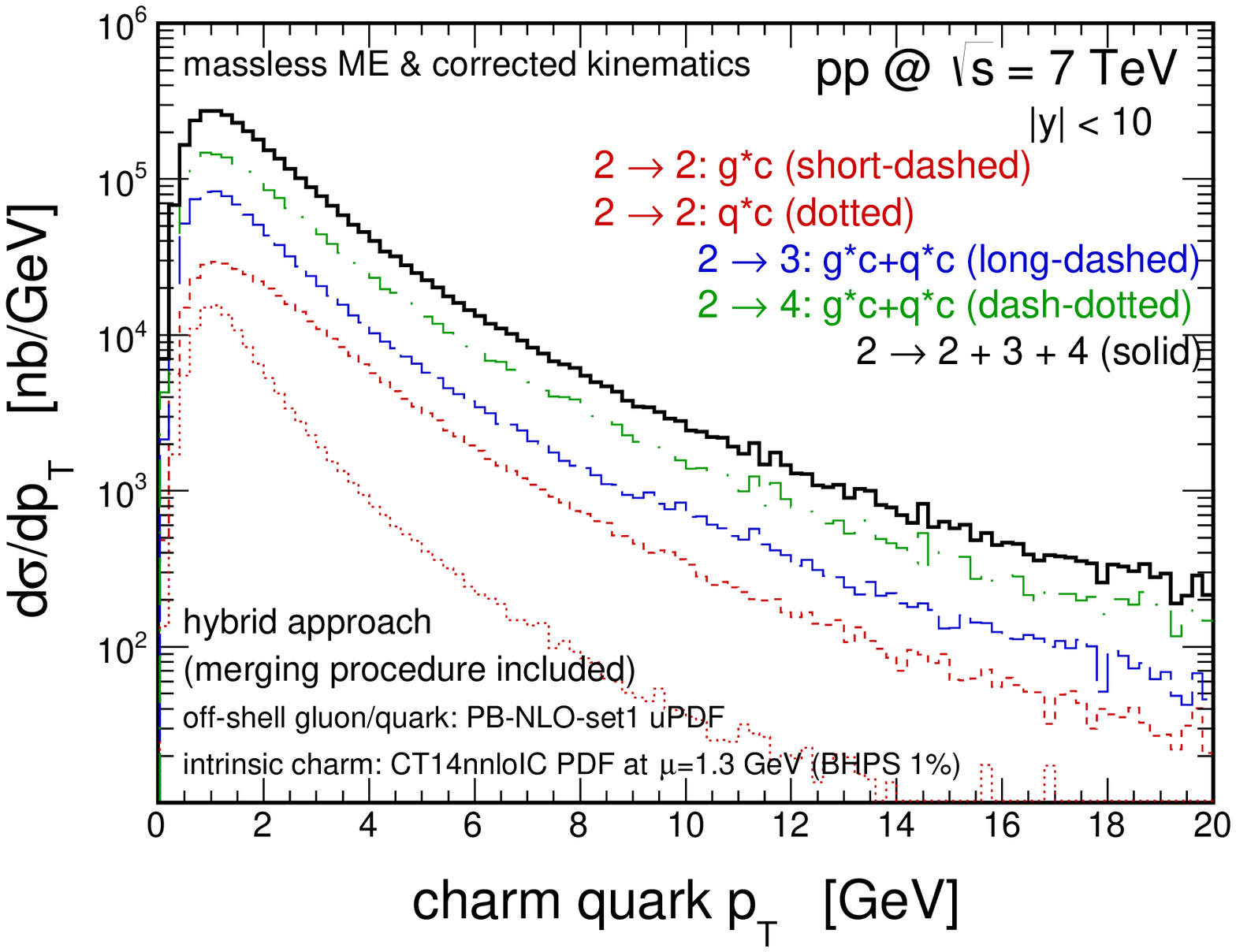}}
\end{minipage}
\begin{minipage}{0.47\textwidth}
  \centerline{\includegraphics[width=1.0\textwidth]{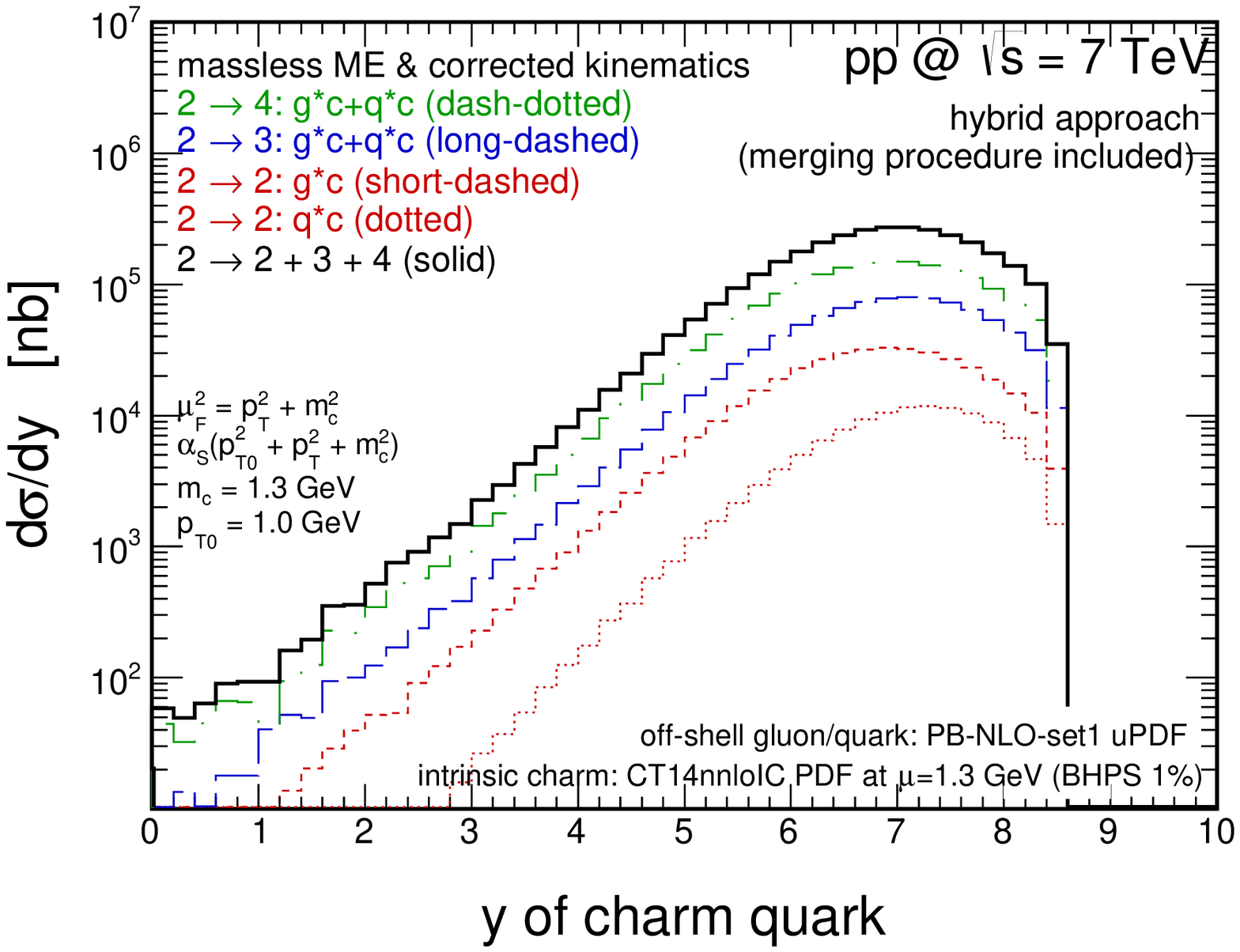}}
\end{minipage}
  \caption{
\small The same as in Fig.~\ref{fig6} but here results for PB-NLO-set1 unintegrated parton densities obtained within the $2 \to 2+3+4$ scheme of the calculation. Here, the $2\to2$, $2\to3$, and $2\to4$ components as well as their sum $2\to 2+3+4$ obtained including merging procedure are shown separately.   
}
\label{fig8}
\end{figure}

Above, we have used those gluon uPDF models that are assumed to allow for an effective resummation of extra real emissions (real higher-order terms). Therefore, their can be successfully used in phenomenological studies based even on leading-order matrix elements (see a discussion in Refs.~\cite{Maciula:2019izq,Maciula:2020cfy}). Here we wish to present results obtained within the scheme of the calculations where the higher-order corrections are not resummed in the uPDF but are taken into account via the hard-matrix elements. This procedure can be tested with the help of the DGLAP-based Parton-Branching uPDFs as was proposed in Ref.~\cite{Maciula:2019izq} and further applied in Refs.~\cite{Lipatov:2019izq,Maciula:2020cfy}. In Fig.~\ref{fig8} we show predictions of the hybrid model for the $2 \to 2$, $2 \to 3$ and $2 \to 4$ mechanisms, as well as for their sum $2 \to 2+3+4$ obtained using a dedicated merging procedure. The results are calculated with the PB-NLO-set1 quark and gluon uPDFs. For the leading-order $2\to 2$ mechanisms we show $g^*c$ and $q^*c$ channels separately while for the higher-order components we plot sum of all possible gluonic and quark channels. As in the collinear case, the higher-order mechanisms are found to be very important also here.    

\begin{figure}[!h]
\begin{minipage}{0.47\textwidth}
  \centerline{\includegraphics[width=1.0\textwidth]{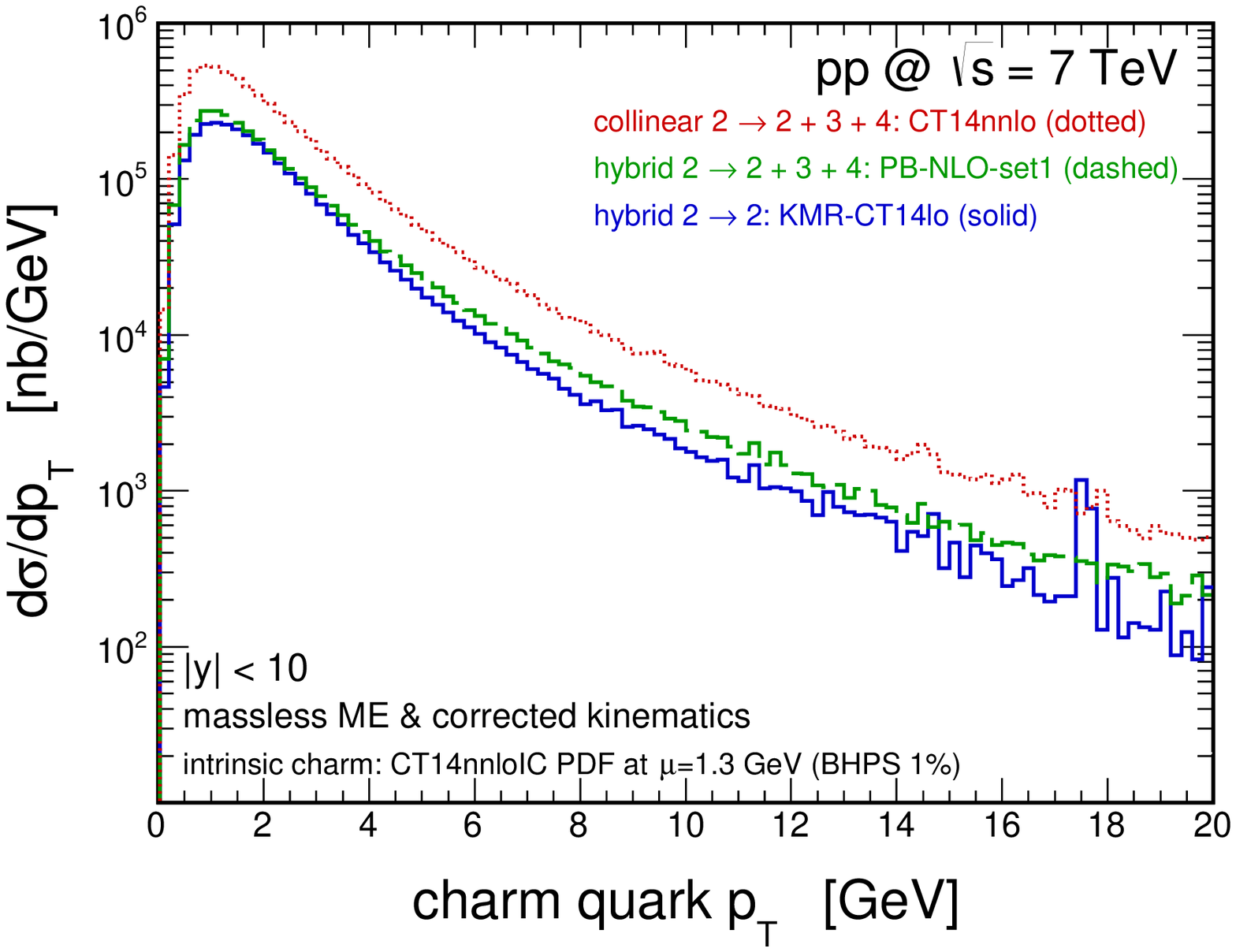}}
\end{minipage}
\begin{minipage}{0.47\textwidth}
  \centerline{\includegraphics[width=1.0\textwidth]{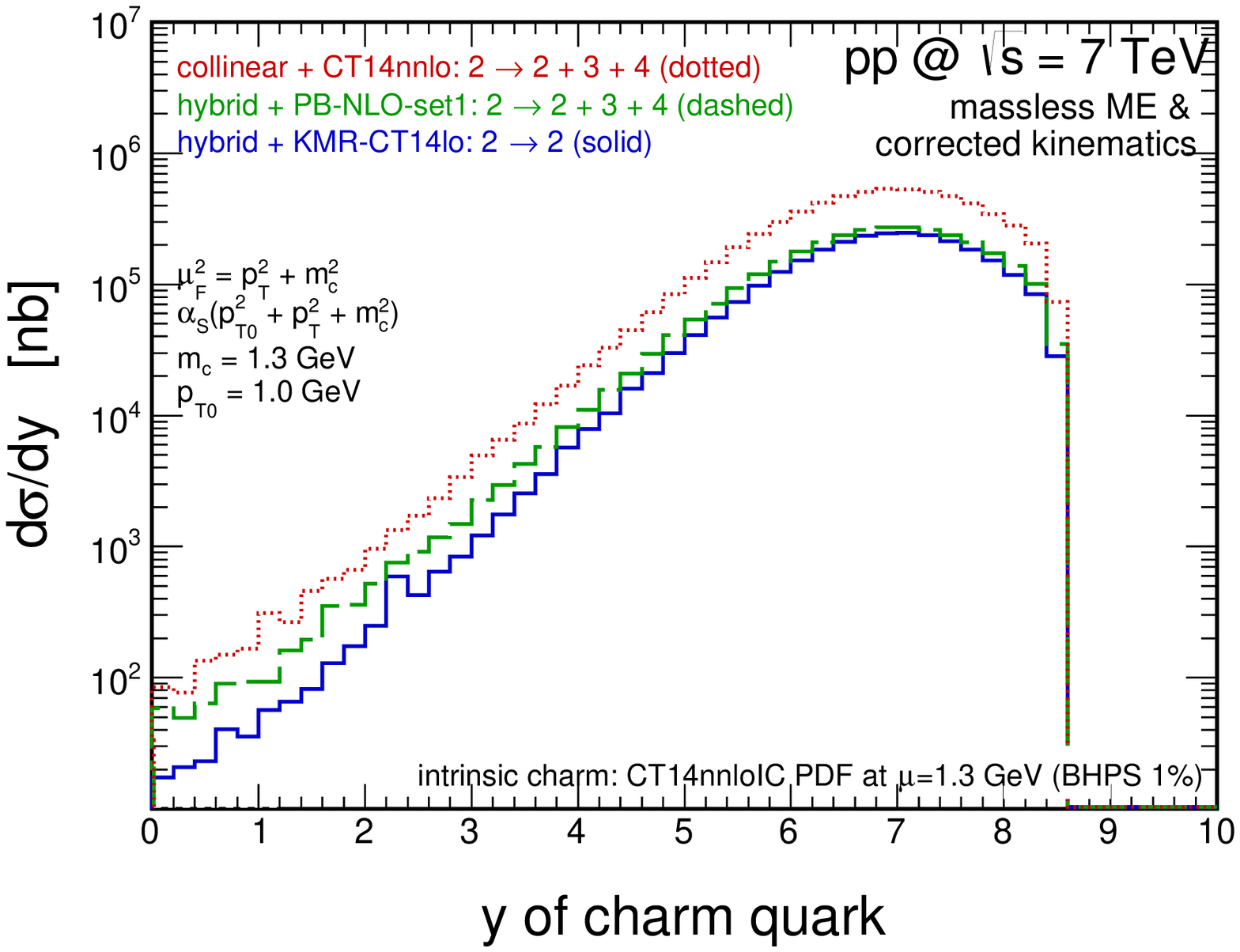}}
\end{minipage}
  \caption{
\small The same as in Fig.~\ref{fig6} but here we compare results for the CT14nnlo collinear PDFs with $2 \to 2+3+4$ collinear model calculations, for the KMR-CT14lo uPDFs with the $2\to 2$ hybrid model calculations and for the PB-NLO-set1 uPDFs with $2 \to 2+3+4$ hybrid model calculations including merging.
}
\label{fig9}
\end{figure}

For a better transparency in Fig.~\ref{fig9} we compare the hybrid model results obtained with the KMR-CT14lo (solid histograms) with the PB-NLO-set1 (dashed histograms) uPDFs, that correspond to the two different hybrid calculation schemes, together with the results obtained in the collinear approach (dotted histograms). Both types of the hybrid model calculations seem to lead to a very similar predictions. It seems to justify the proposed $2 \to 2+3+4$ hybrid calculation scheme with the PB uPDFs and with the applied merging in a qualitative way. On the other hand, the collinear $2 \to 2+3+4$ results seem to be larger by a factor of 2 than their hybrid model counterpart. However, this might be related to a lack of a relevant merging procedure in the collinear case.     

\subsection{The $\bm{k_{T}}$-factorization approach}

Now we wish to present results obtained within the $k_{T}$-factorization
approach. So here we take into account 
also effects related to off-shellness of $c$ quark of the intrinsic
charm in the proton. The transverse momentum dependent intrinsic charm
uPDF is obtained by Gaussian smearing of the collinear PDF. 
Rather small smearing parameter is used that do not allow for a large
transverse momenta of the intrinsic charm. 
It seems to be appropriate for the case of the forward production of
charm. For the unintegrated gluon density the KMR-CT14lo model is used.  

\begin{figure}[!h]
\begin{minipage}{0.47\textwidth}
  \centerline{\includegraphics[width=1.0\textwidth]{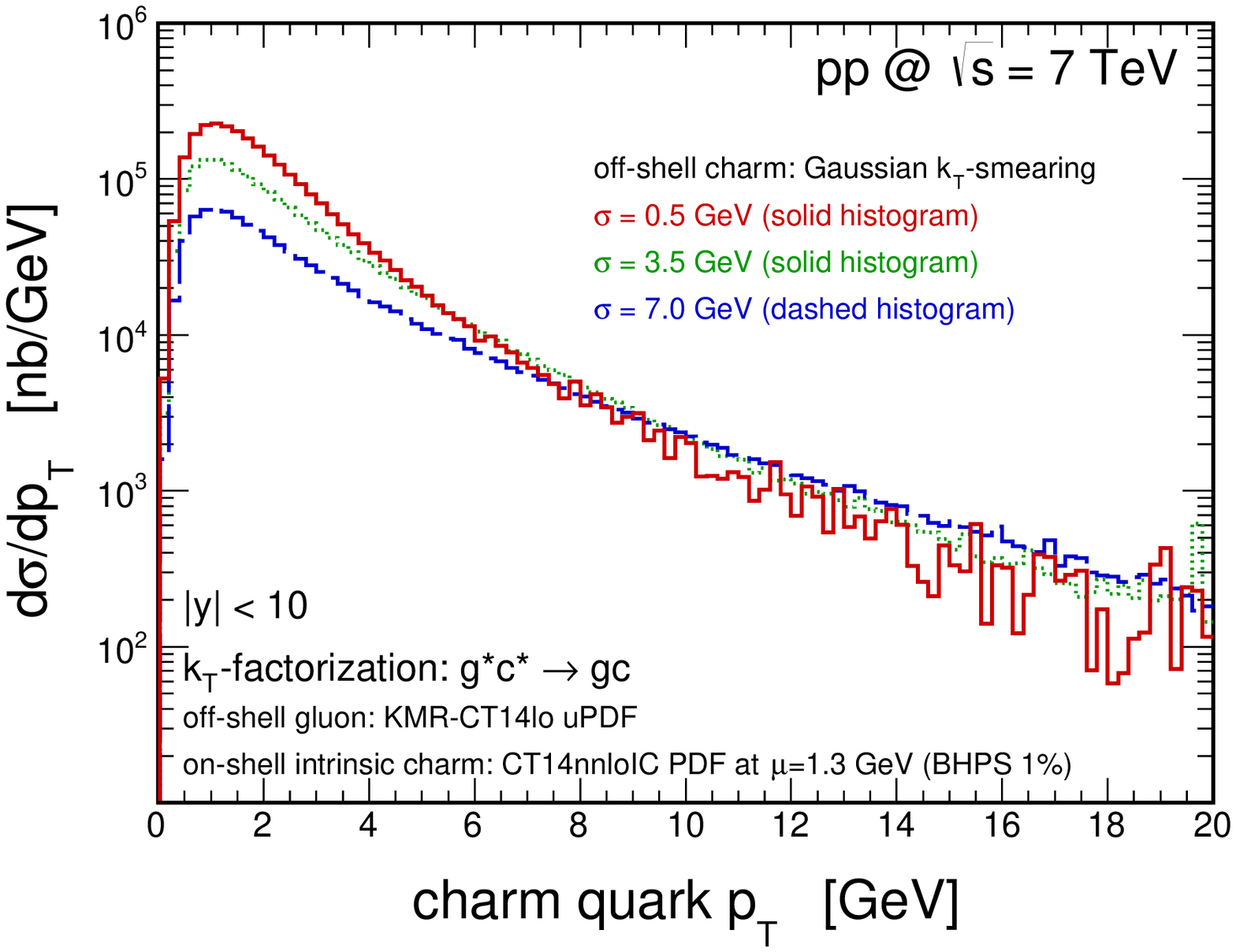}}
\end{minipage}
\begin{minipage}{0.47\textwidth}
  \centerline{\includegraphics[width=1.0\textwidth]{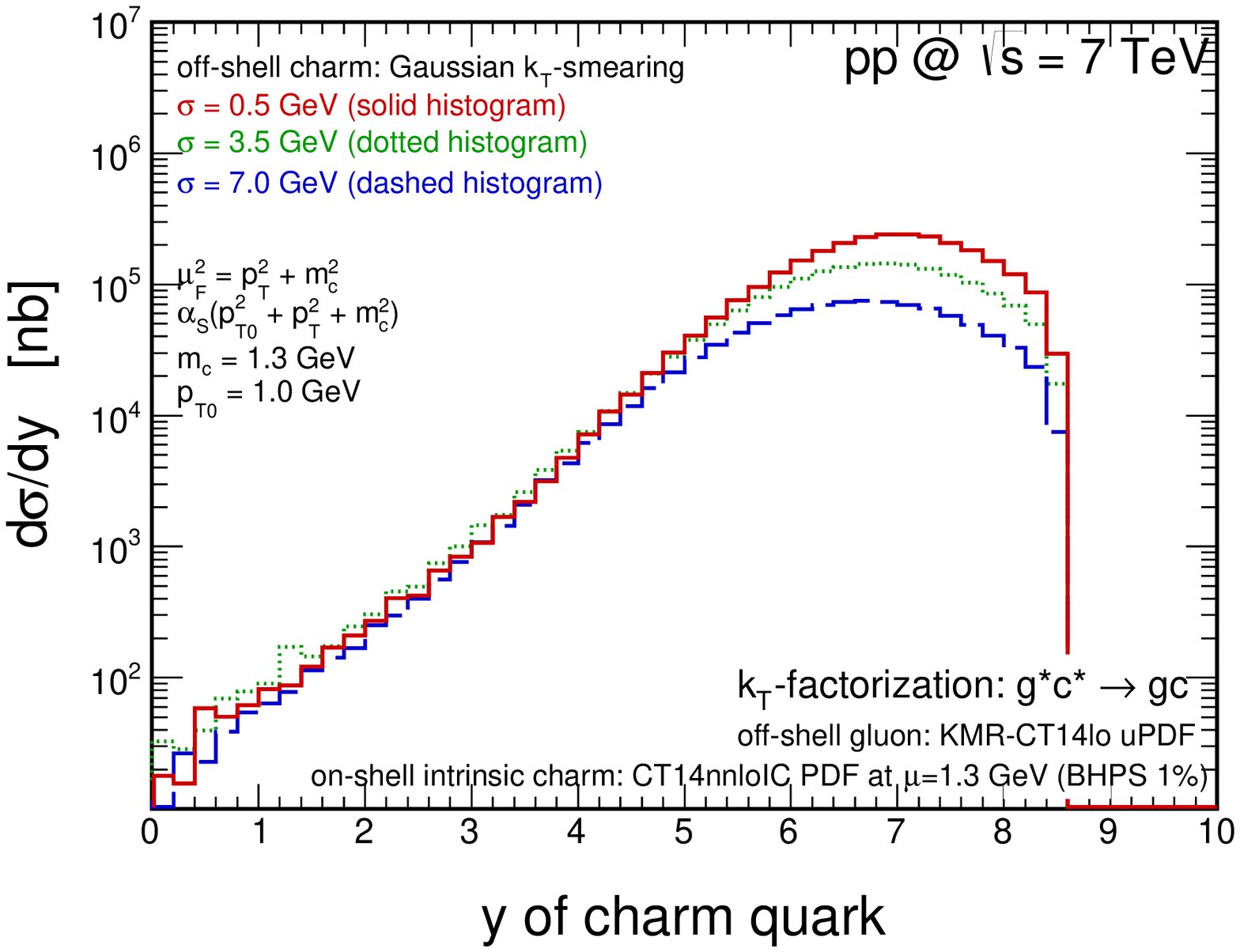}}
\end{minipage}
  \caption{
\small The charm quark transverse momentum (left) and rapidity (right) differential cross sections for $pp$-scattering at $\sqrt{s}=7$ TeV. The results correspond to the $g^* c^* \rightarrow g c$ mechanism calculated within the intrinsic charm concept in the $k_{T}$-factorization approach with both off-shell initial state partons. Here the KMR-CT14lo unintegrated gluon density and Gaussian $k_{t}$-distribution for off-shell charm quark were used. We show results for different values of the smearing parameter $\sigma$.
}
\label{fig10}
\end{figure}

In Fig.~\ref{fig10} we show predictions for the $g^*c^* \to gc$ mechanism with both initial state partons being off-shell.
Here three different values of the smearing parameter in the calculation of the intrinsic charm uPDF are used: $\sigma_{0} = 0.5$ GeV (solid histograms), $3.5$ GeV (dotted histograms) and $7.0$ GeV (dashed histograms). The larger $\sigma_{0}$ is taken the smaller cross section at small outgoing charm quark transverse momenta is obtained (left panel). The same is true for the rapidity spectrum in the forward region (right panel).  

\begin{figure}[!h]
\begin{minipage}{0.47\textwidth}
  \centerline{\includegraphics[width=1.0\textwidth]{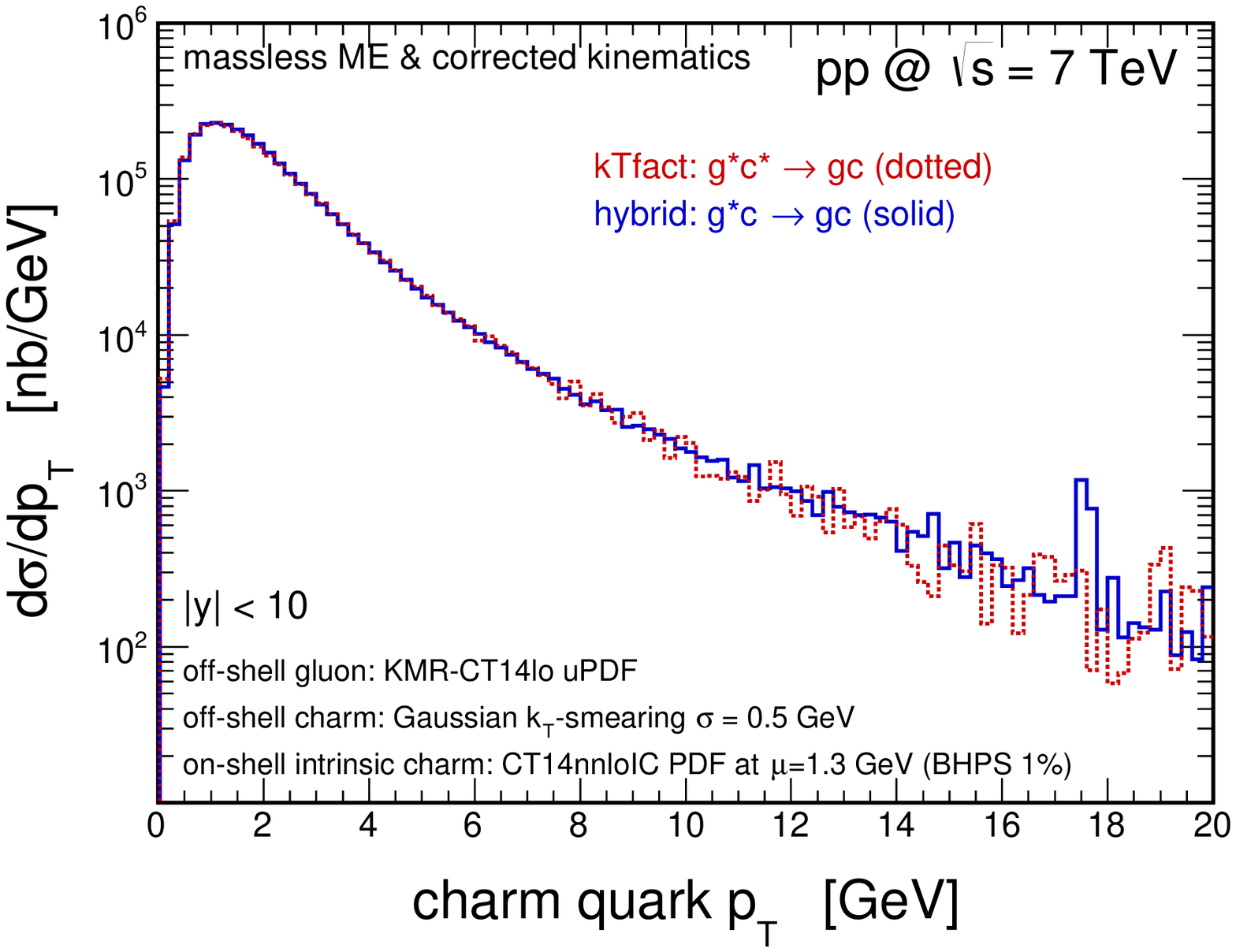}}
\end{minipage}
\begin{minipage}{0.47\textwidth}
  \centerline{\includegraphics[width=1.0\textwidth]{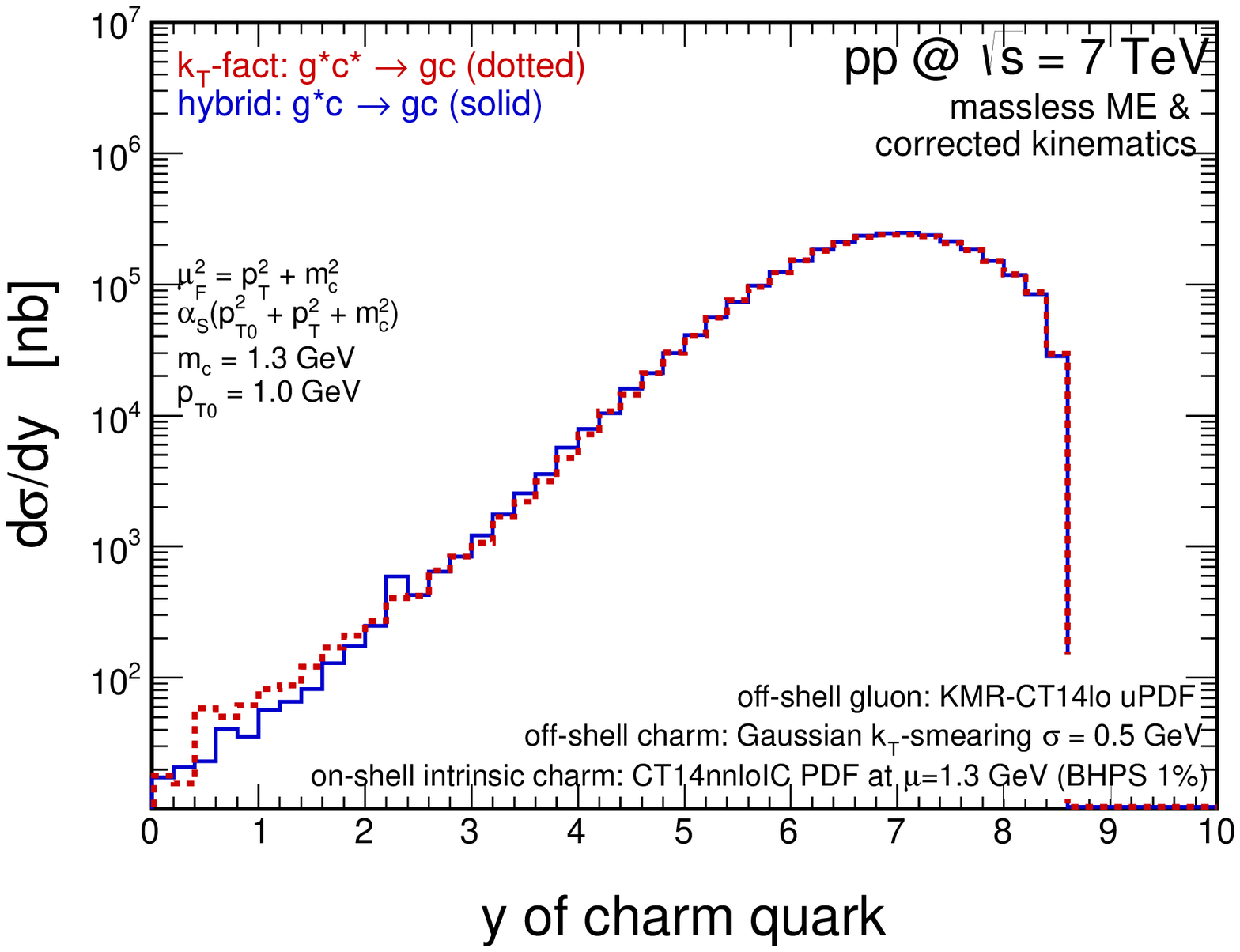}}
\end{minipage}
  \caption{
\small The same as in Fig.~\ref{fig10} but here we compare results for
the hybrid $g^*c\to gc$ and the $k_{T}$-factorization $g^*c^* \to gc$
calculations obtained with the KMR-CT14lo unintegrated gluon densities. 
The off-shell charm quark Gaussian $k_{t}$-distribution is obtained with
the the smearing parameter $\sigma_0 = 0.5$ GeV.
}
\label{fig11}
\end{figure}

In Fig.~\ref{fig11} we illustrate mutual relations between the results obtained with the hybrid and the $k_T$-factorization frameworks. When the smearing parameter in the calculation of the intrinsic charm uPDF is small, e.g. $\sigma_{0} = 0.5$ GeV,
the hybrid model $g^*c \to gc$ results coincide with the $g^*c^* \to gc$ results obtained within the full $k_T$-factorization approach.

\begin{figure}[!h]
\begin{minipage}{0.47\textwidth}
  \centerline{\includegraphics[width=1.0\textwidth]{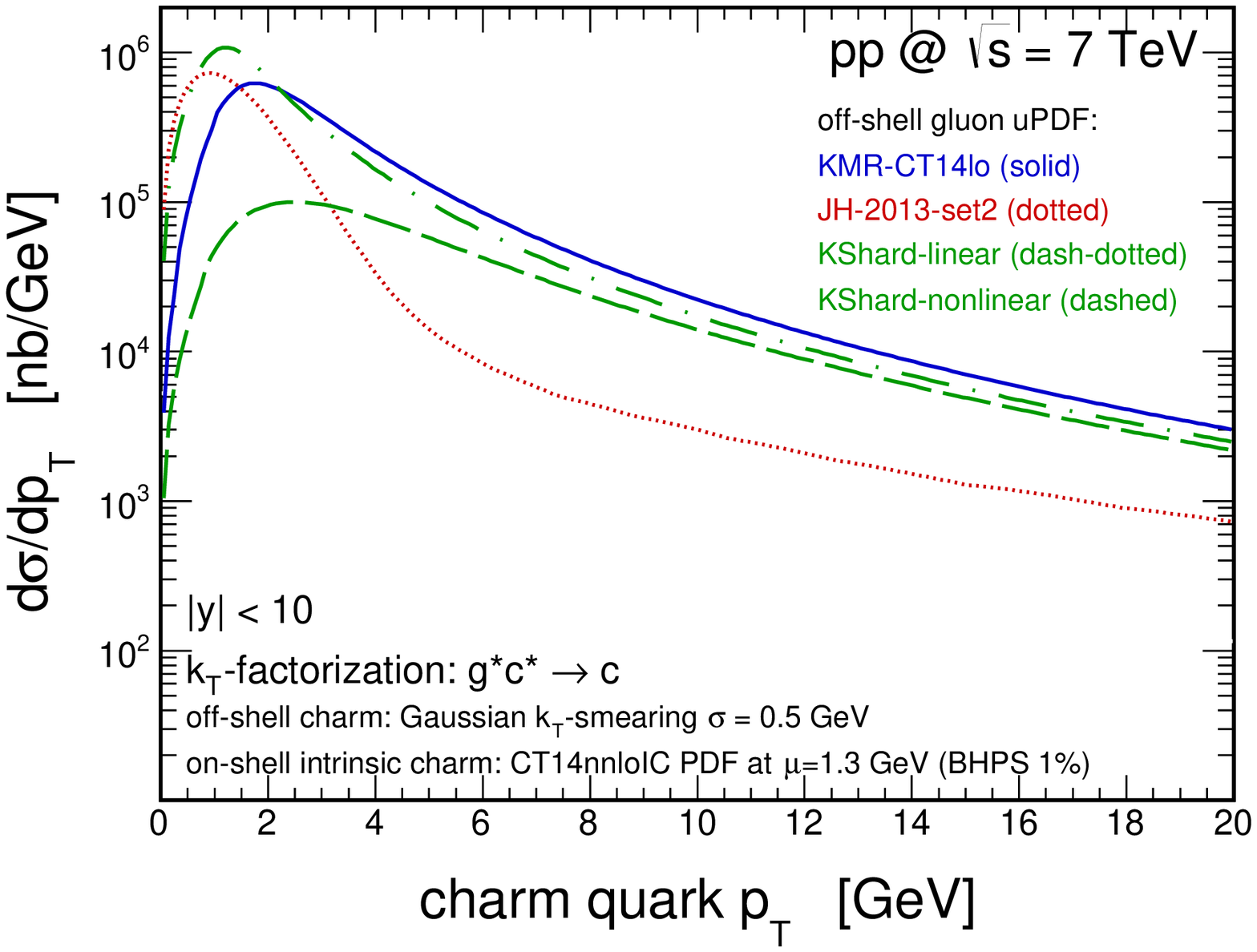}}
\end{minipage}
\begin{minipage}{0.47\textwidth}
  \centerline{\includegraphics[width=1.0\textwidth]{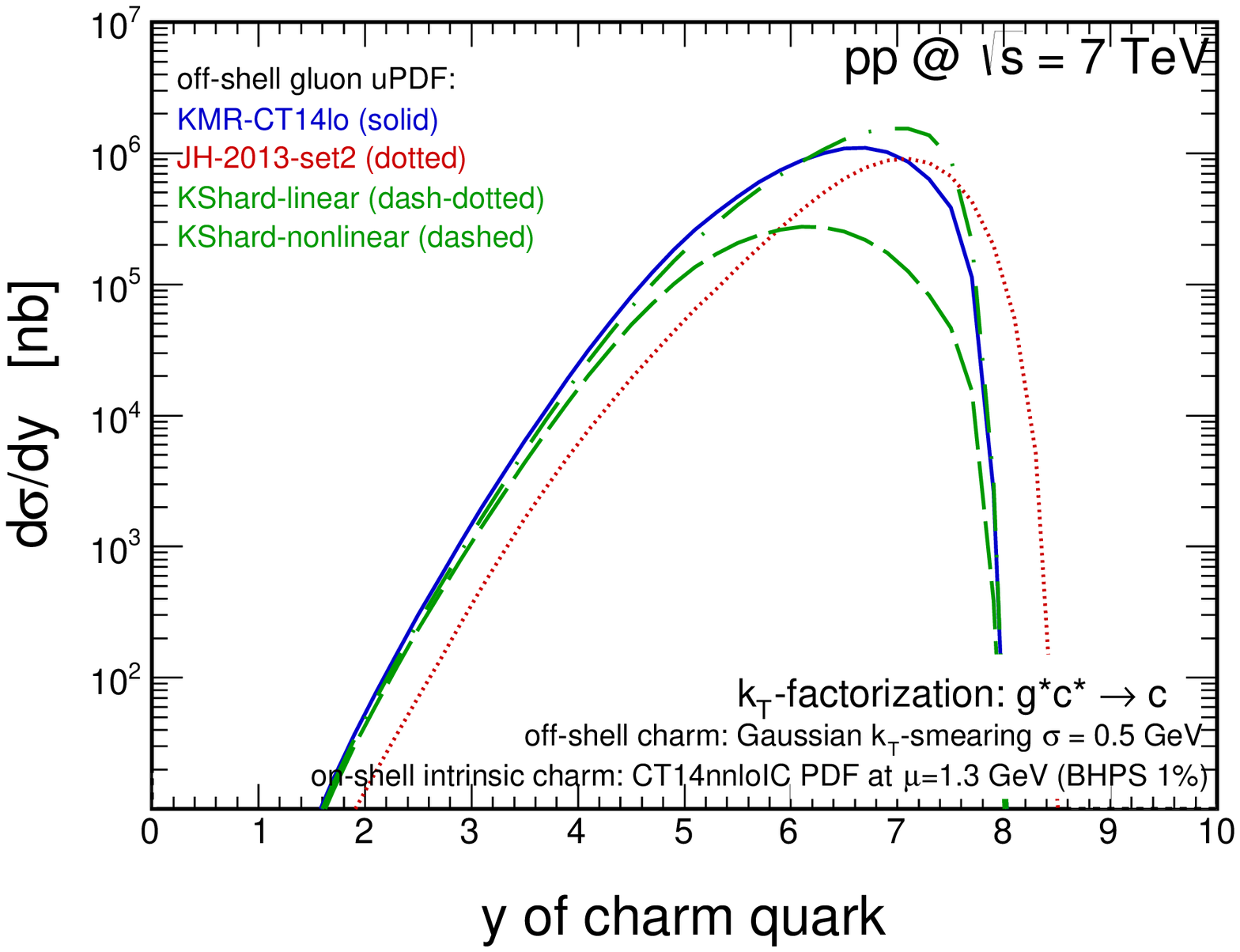}}
\end{minipage}
  \caption{
\small The charm quark transverse momentum (left) and rapidity (right) differential cross sections for $pp$-scattering at $\sqrt{s}=7$ TeV. The results correspond to the $g^* c^* \rightarrow c$ mechanism calculated within the intrinsic charm concept in the $k_{T}$-factorization approach with both off-shell initial state partons. Here the Gaussian $k_{t}$-distribution for off-shell charm quark were used. We show results for different gluon uPDFs.
}
\label{fig12}
\end{figure}

Finally, we wish to present results of the $k_{T}$-factorization approach for the $g^*c^* \to c$ mechanism.
In Fig.~\ref{fig12} we compare the corresponding predictions obtained
with the four different gluon uPDFs: the KMR-CT14lo (solid lines), the
JH-2013-set2 (dotted lines), the KS-linear (dash-dotted lines) 
and the KS-nonlinear (dashed lines).
Different models lead to quite different results. The discrepancies 
between the uPDF models obtained here seem to be larger than in 
the corresponding case of the $g^*c \to gc$ calculations within the hybrid model.  

\subsection{Predictions for future experiments}

Before we go to predictions for different future and present experiments
we wish to summarize the conclusions drawn in the previous subsection 
by a direct comparison of the results corresponding to the approaches discussed above.

\begin{figure}[!h]
\begin{minipage}{0.47\textwidth}
  \centerline{\includegraphics[width=1.0\textwidth]{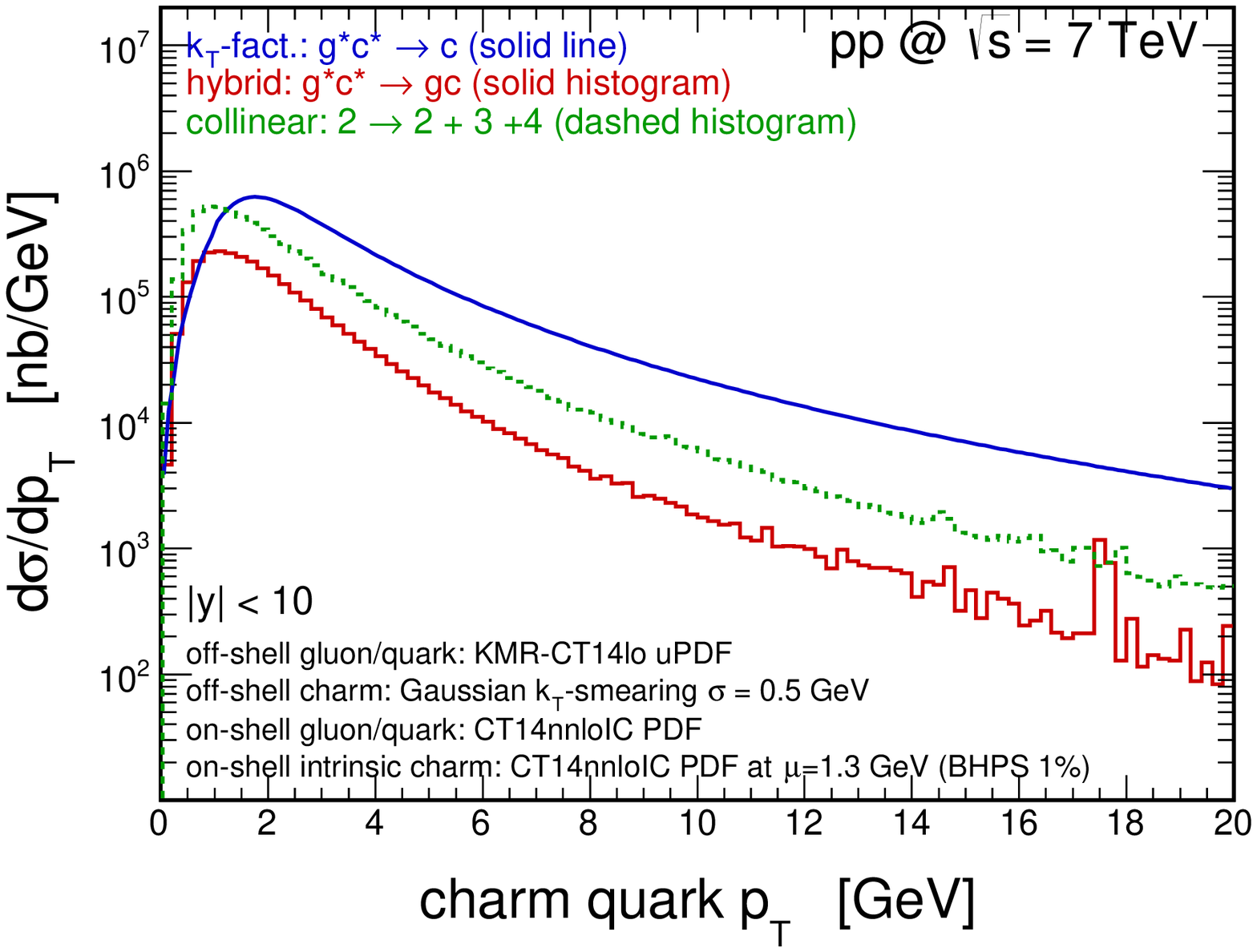}}
\end{minipage}
\begin{minipage}{0.47\textwidth}
  \centerline{\includegraphics[width=1.0\textwidth]{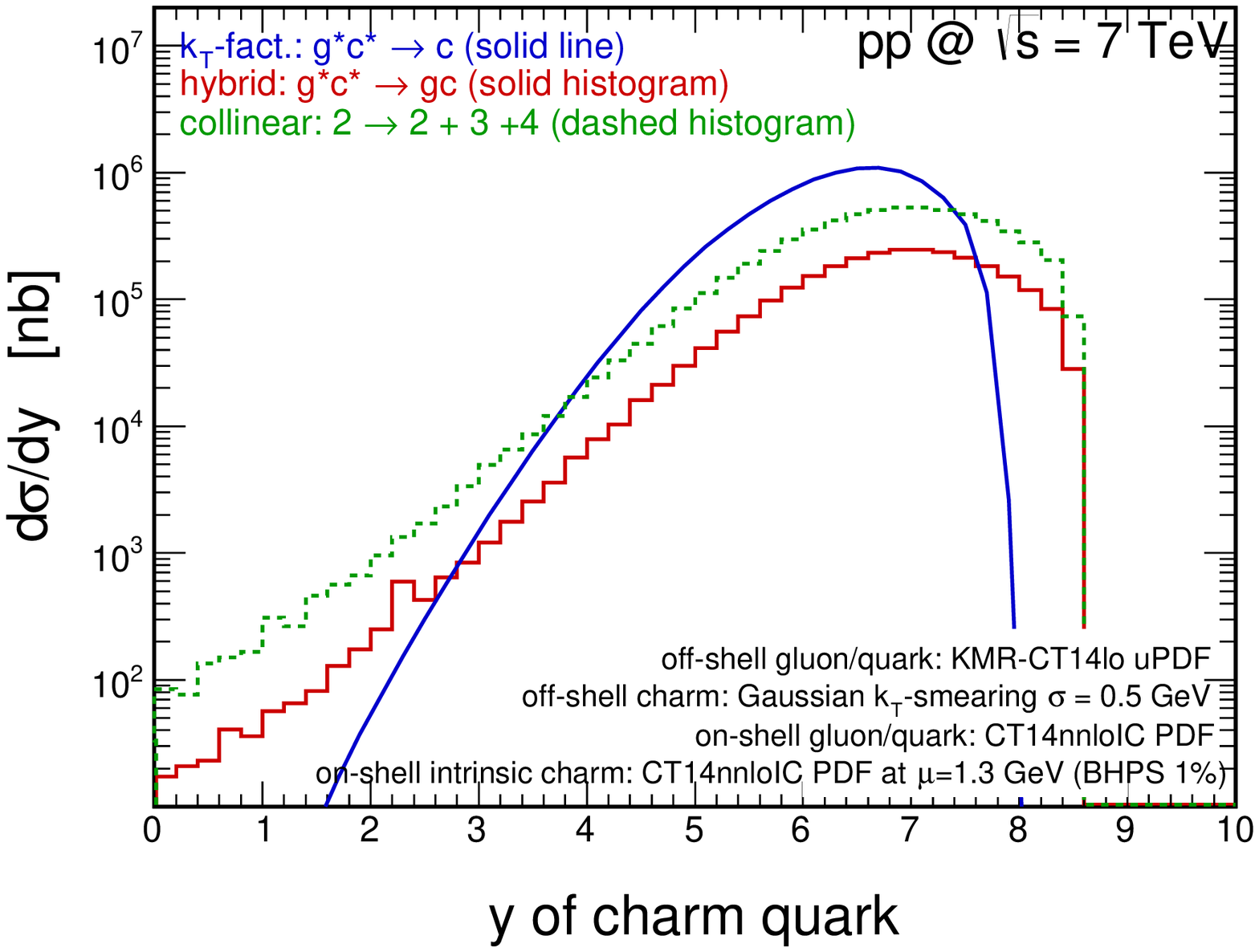}}
\end{minipage}
  \caption{
\small The charm quark transverse momentum (left) and rapidity (right) differential cross sections for $pp$-scattering at $\sqrt{s}=7$ TeV. Here we compare predictions of the three different approaches used in the previous subsections: the $2 \to 2+3+4$ collinear, the hybrid $g^*c \to gc$ and the $k_{T}$-factorization  $g^*c^* \to c$ calculations. 
}
\label{fig13}
\end{figure}

In Fig.~\ref{fig13} we compare predictions of the three different
approaches used in the previous subsections: the $2 \to 2+3+4$ collinear
(dashed histograms), the hybrid $g^*c \to gc$ (solid histograms) and the
$k_{T}$-factorization  $g^*c^* \to c$ (solid lines)
calculations. Different models lead to a very different results with
more than one order of magnitude difference between the lowest and the
highest predicted cross section.
Huge cross section for $g c \to c$ or $c g \to c$ may be partly due to
ignoring other emissions than $c$ or $\bar c$ in the evolution of $x_1$
and $x_2$. 
These large uncertainties of the predictions can be reduced only 
by a forward experiments at forward directions. 
Forward charm production data sets that will be dominated by 
the contribution from intrinsic charm are necessary to draw 
definite conclusions about the level of applicability of 
the different theoretical approaches.
 
\begin{figure}[!h]
\begin{minipage}{0.47\textwidth}
  \centerline{\includegraphics[width=1.0\textwidth]{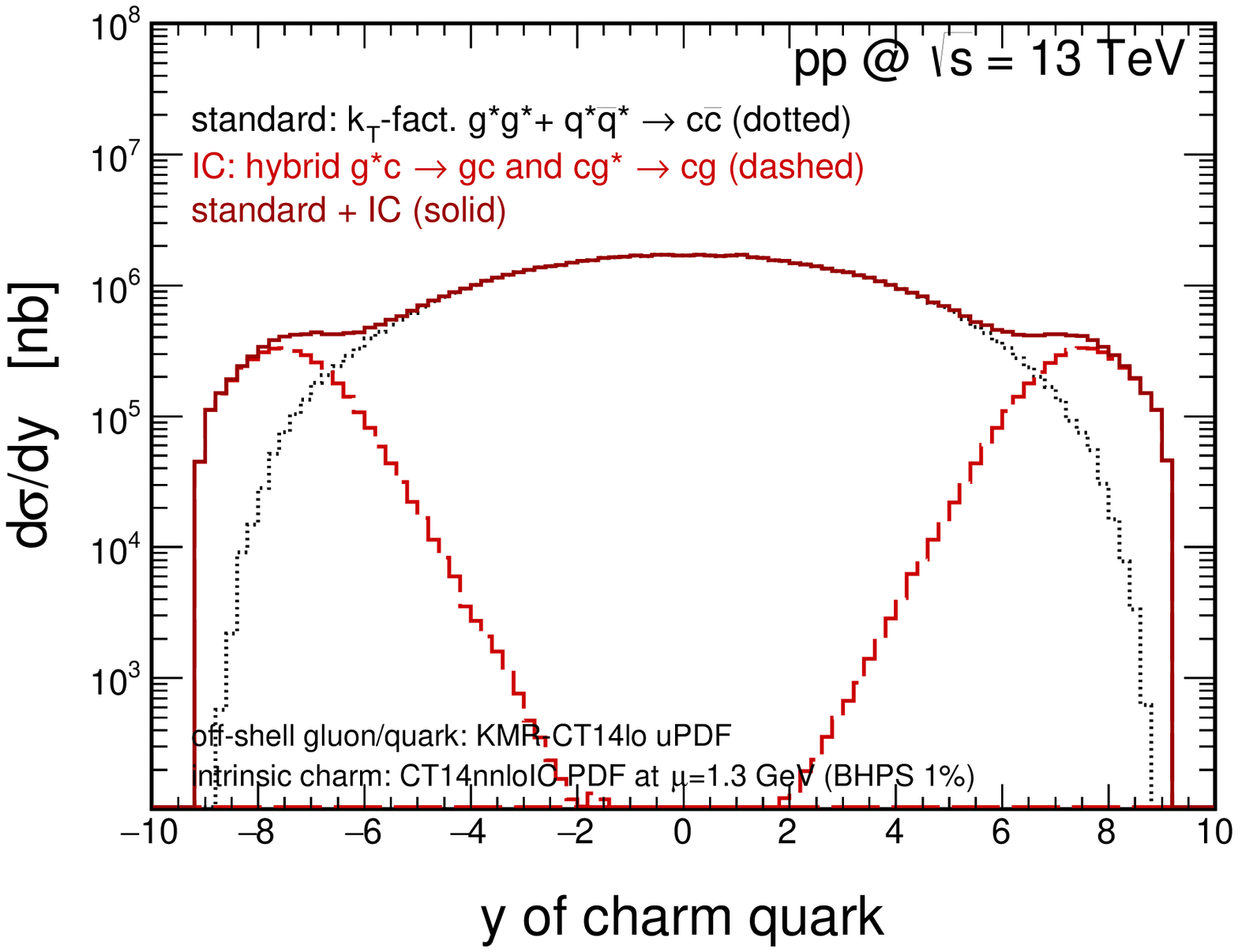}}
\end{minipage}
\begin{minipage}{0.47\textwidth}
  \centerline{\includegraphics[width=1.0\textwidth]{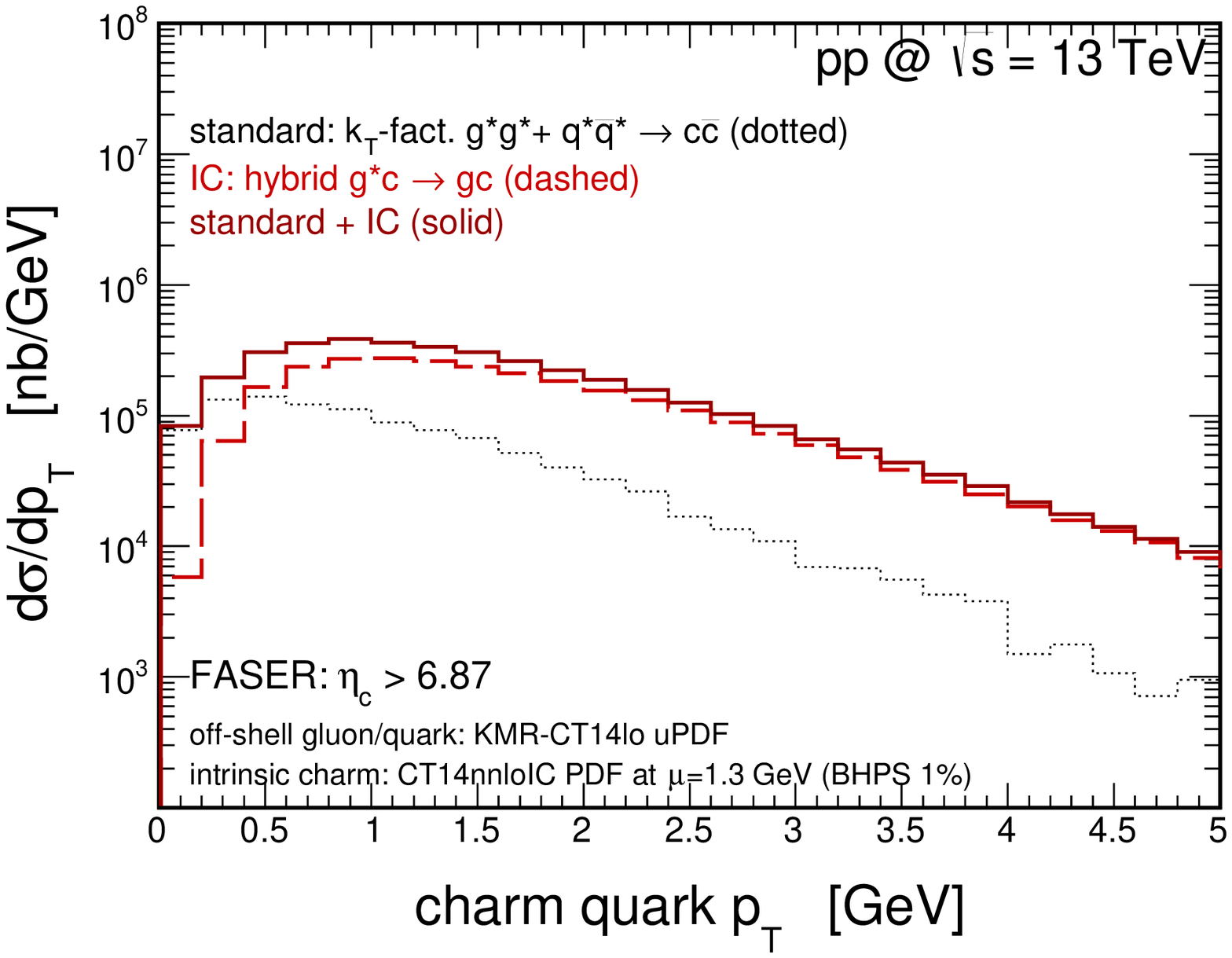}}
\end{minipage}\\
\begin{minipage}{0.47\textwidth}
  \centerline{\includegraphics[width=1.0\textwidth]{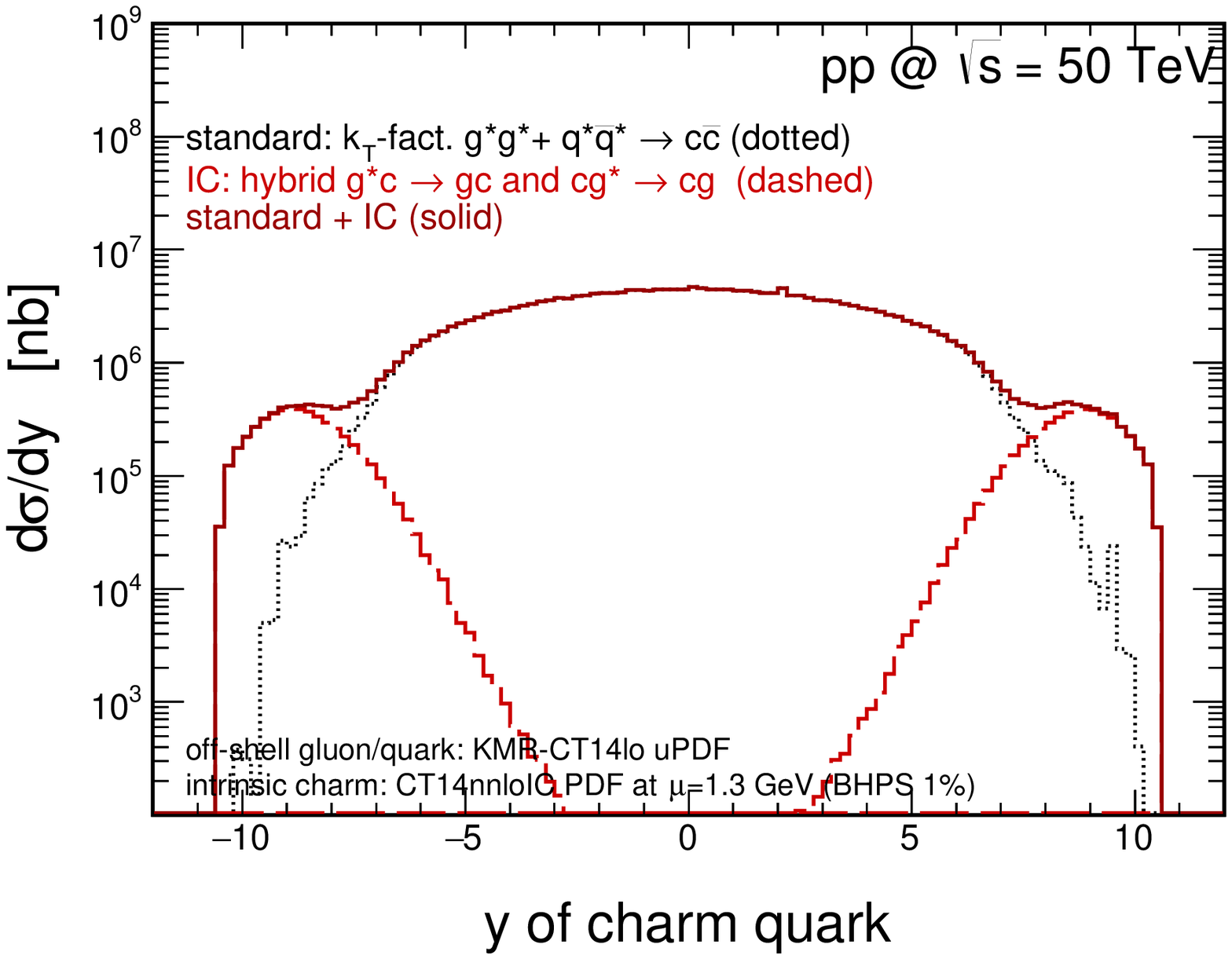}}
\end{minipage}
\begin{minipage}{0.47\textwidth}
  \centerline{\includegraphics[width=1.0\textwidth]{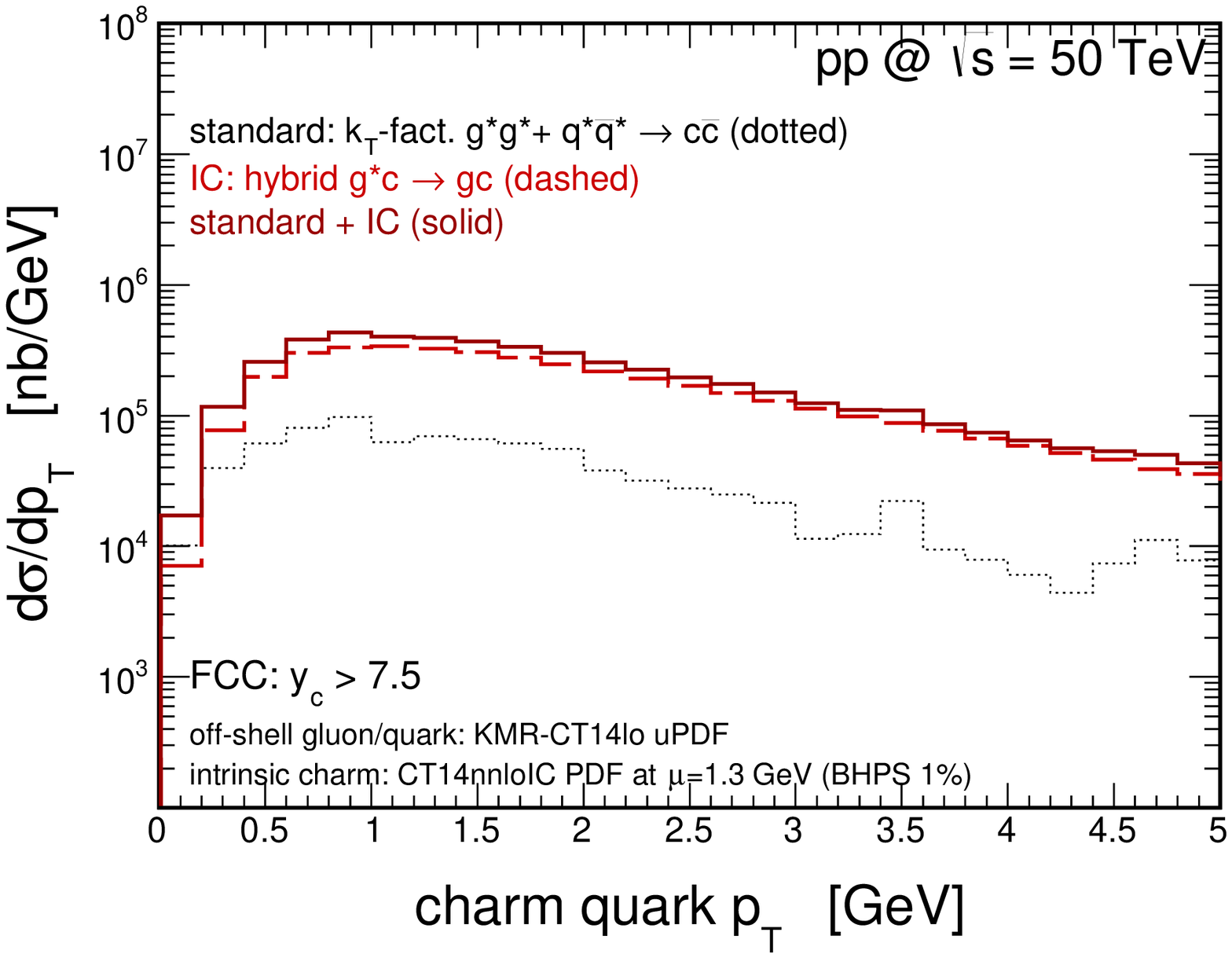}}
\end{minipage}
  \caption{
\small Predictions of the impact of the intrinsic charm component in
charm quark production in different experiments. Here we explore
kinematics relevant for the FASER experiment at the LHC and an exemplary experiment at the FCC.
}
\label{fig14}
\end{figure}

Therefore, now we wish to present results of the study of the impact of
the intrinsic charm component on the forward charm particle production
in already existing or future experiments at different energies. We
start with predictions for the high energy experiments at the LHC and
the FCC, at $\sqrt{s}=13$ and $50$ TeV, respectively (top and bottom
panels in Fig.~\ref{fig14}). In the LHC case the considered kinematics
correspond to the planned FASER experiment. Here we compare predictions
of the $k_{T}$-factorization approach for the $g^*g^* \to c\bar c$
mechanism which is known to give a very good description of the LHC open
charm data \cite{Maciula:2019izq}, and predictions of the $g^*c \to gc$ mechanism
(dashed) within the hybrid model. In both cases the charm production
cross section starts to be dominated by the intrinsic charm component at
very forward rapidities, \textit{i.e.} $y \geq 7$. In this far-forward
region, the transverse momentum distribution of charm quark is also
dominated by the contributions of the intrinsic charm. The predicted
enhancement of the charm cross section could certainly be examined by
the FASER experiment dedicated to a measurement of forward neutrinos
originating from semileptonic decays of $D$ mesons.
The actual predictions for neutrinos will be presented elsewhere.           

\begin{figure}[!h]
\begin{minipage}{0.47\textwidth}
  \centerline{\includegraphics[width=1.0\textwidth]{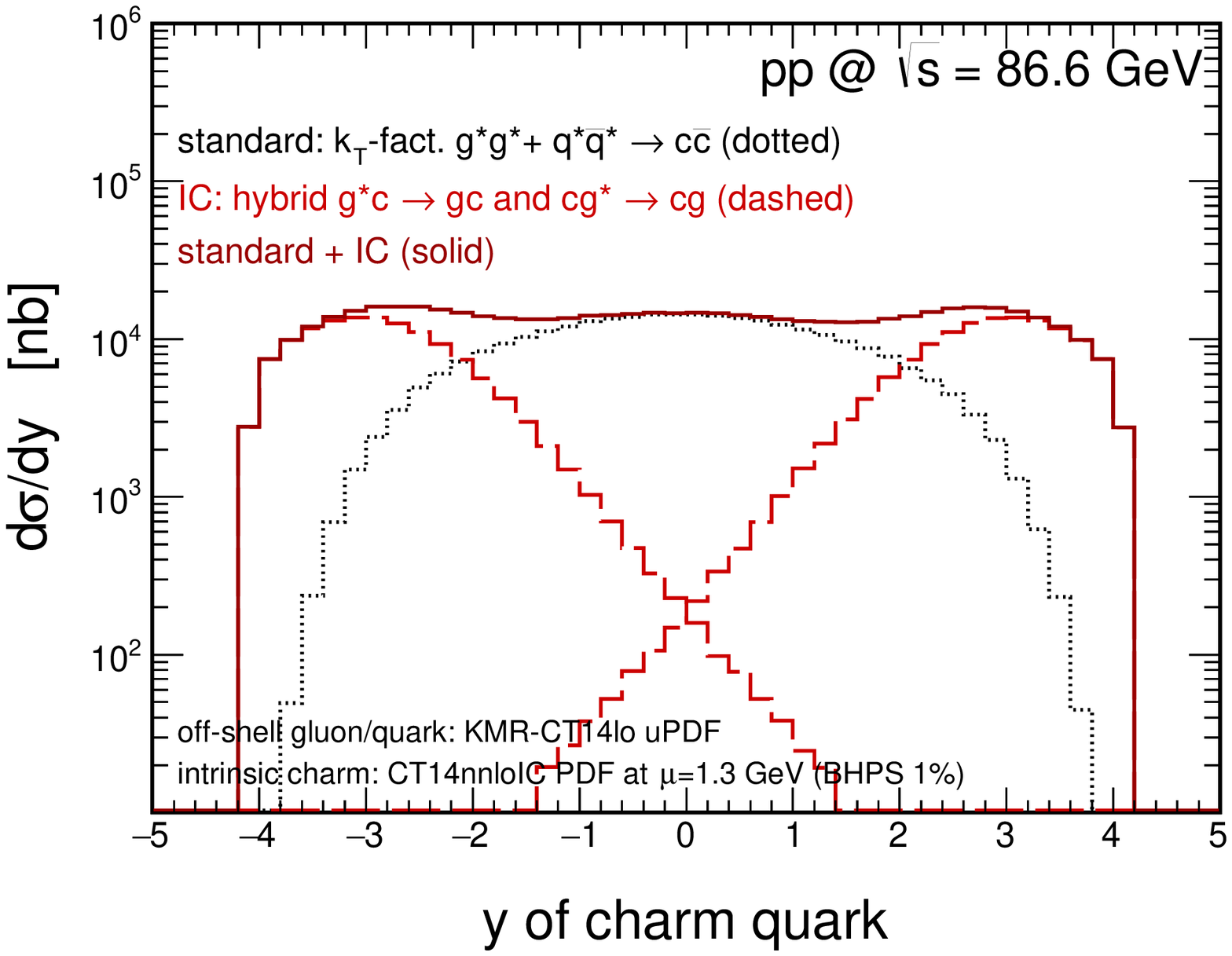}}
\end{minipage}
\begin{minipage}{0.47\textwidth}
  \centerline{\includegraphics[width=1.0\textwidth]{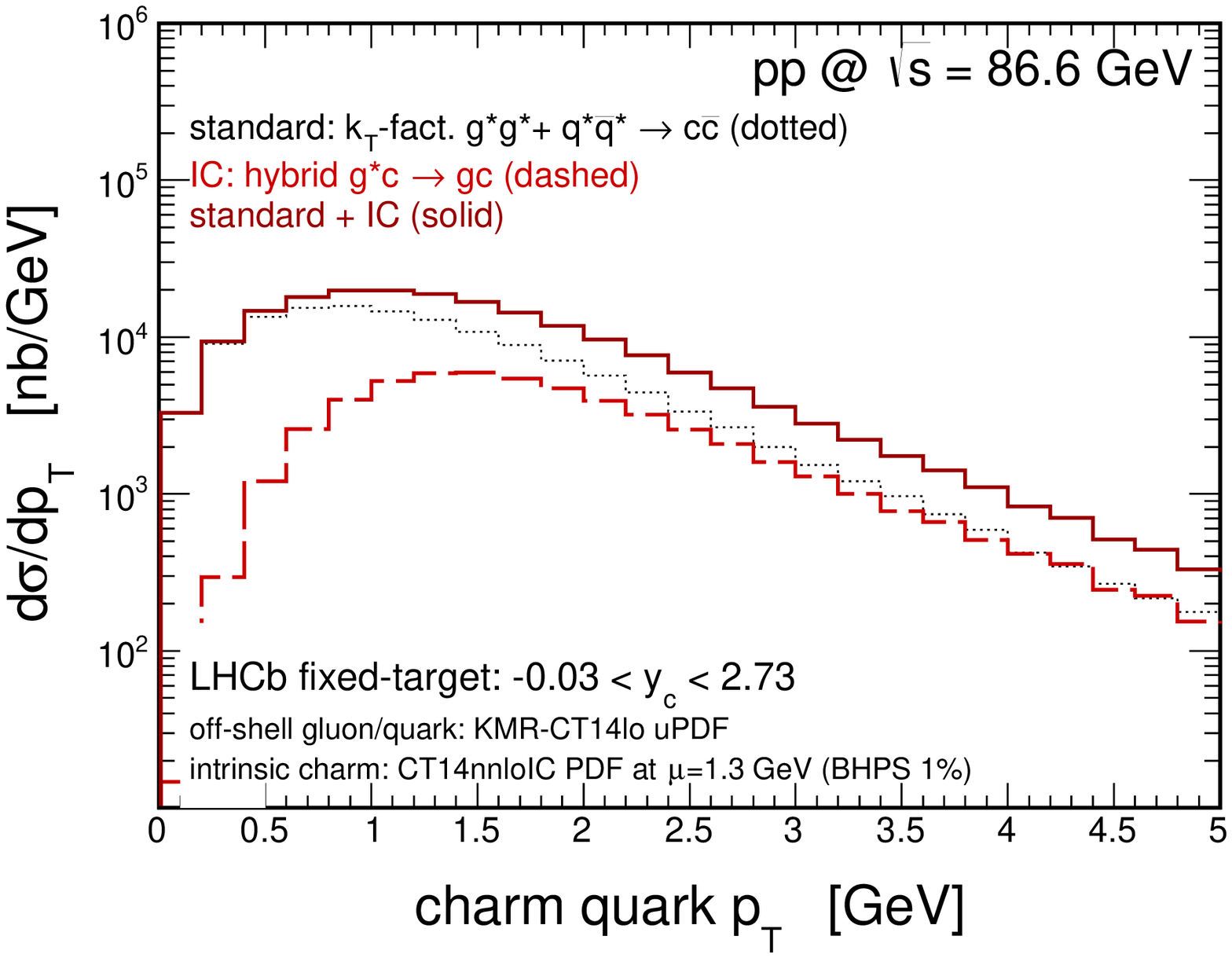}}
\end{minipage}\\
\begin{minipage}{0.47\textwidth}
  \centerline{\includegraphics[width=1.0\textwidth]{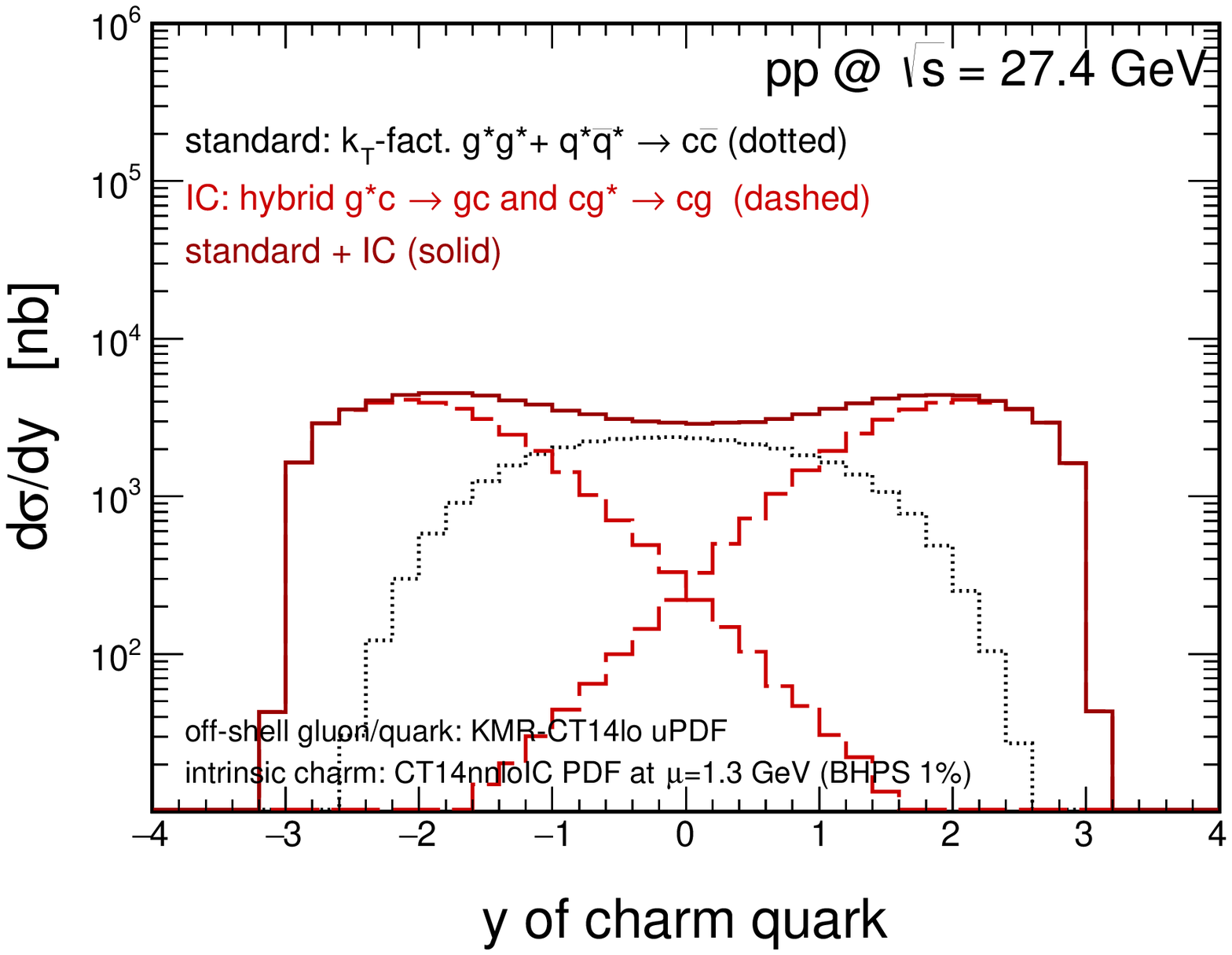}}
\end{minipage}
\begin{minipage}{0.47\textwidth}
  \centerline{\includegraphics[width=1.0\textwidth]{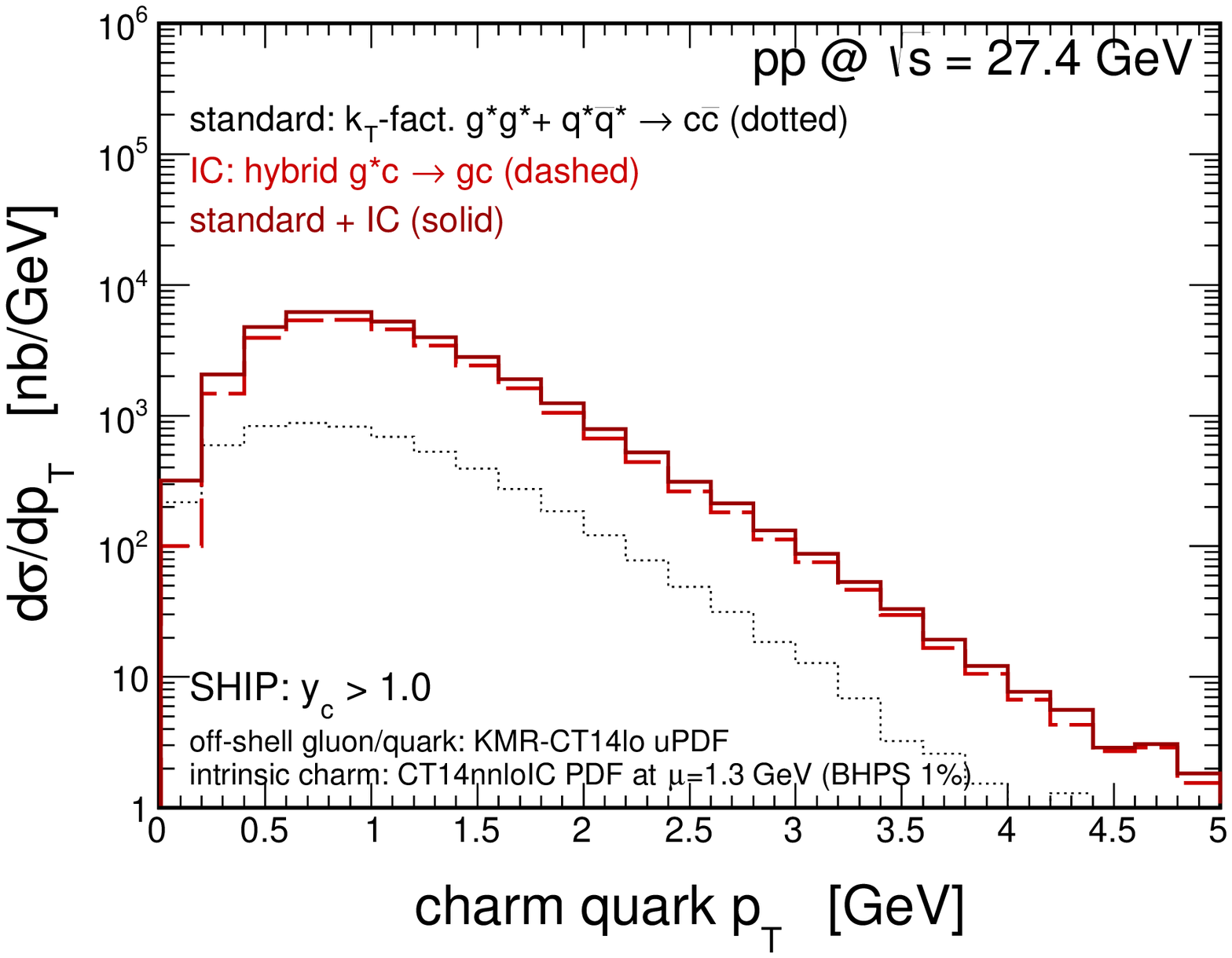}}
\end{minipage}
  \caption{
\small Predictions of the impact of the intrinsic charm component in
charm production in different experiments. 
Here we explore kinematics of the fixed-target mode LHCb and the kinematics relevant for the SHIP experiment.
}
\label{fig15}
\end{figure}

In addition, we also analysed a possibility of experimental study of the
intrinsic charm concept at lower energies.
In Fig.~\ref{fig15} we show predictions for the fixed-target LHC and the
SHIP experiment, at $\sqrt{s}=86.6$ and $27.4$ GeV, respectively (top
and bottom panels). We observe that also at relatively small energies
the intrinsic charm contributions could be identified experimentally. It
seems that the already existing data set on open charm meson production
in the fixed-target LHC mode \cite{Aaij:2018ogq} needs to have the intrinsic charm
component included in the theoretical description. 
Similarly, our results suggests that the predictions of 
the tau-neutrino flux that could be measured in the SHIP experiment 
should include effects related to a possible intrinsic charm content of the proton.   

\section{Conclusions}

In this paper we have discussed the effect of intrinsic charm in the proton
on forward production of $c$ quark or $\bar c$ antiquark at different energies.
Three different approaches: collinear, hybrid and $k_{T}$-factorization have been used with modern
collinear and unintegrated parton distribution functions.

The production mechanism of $c$-quarks and $\bar c$-antiquarks
originating from intrinsic charm in the nucleon is concentrated in forward/backward directions, but details depend on collision energy.
The absolute normalization strongly depends on the approach used. The leading-order (LO) collinear framework leads to the smallest cross section. The cross section becomes much bigger in the $k_{T}$-factorization or in the hybrid model
which effectively include higher-order corrections. The next-to-leading (NLO) and even next-to-next-to-leading (NNLO) tree-level corrections are found to be very important here. Therefore, the $k_{T}$-factorization or the hybrid model will give stringent limits on the intrinsic charm which cannot be constrained at present from first principles.

We have shown that in the collinear approach the LO calculations of the intrinsic charm component are insufficient.
We have included the NLO and NNLO components at tree-level which were found to significantly contribute to the cross section.

Working in the hybrid model or in the $k_{T}$-factorization approach we have shown that the effects related to the off-shellness 
of the incident partons (especially gluons) are large. In both cases higher-order corrections are effectively included already within the basic $gc \to gc$ mechanism. We have used different models for gluon unintegrated parton distribution functions (uPDFs) from the literature. We obtained different results for different gluon uPDFs. The forward charm production was recognized as a useful testing ground for the small-$x$ behaviour of the gluon uPDFs. We have shown in addition that the final results are also sensitive to the concept of gluon saturation in a proton. Unintegrated gluon densities derived from linear and non-linear evolution equations lead to a quite different results. We have performed also leading-order calculations within
$k_T$-factorization approach where the basic process is either
$g + c \to c$ or $c + g \to c$ as done for forward
production of charm quarks. 

We have shown that the intrinsic charm component dominates over the standard pQCD (extrinsic) mechanism of $c\bar c$-pair production
at forward (or far-forward) rapidities starting from low energy fixed-target experiment at $\sqrt{s}=27.4$ and $86.6$ GeV, through the LHC Run II nominal energy $\sqrt{s}=13$ TeV, and up to the energies relevant for the IceCube experiment ($\sqrt{s}=50$ TeV).
The LHC experiments at low energies (fixed-target experiments) can provide valuable information already now. Future LHC experiments on $\nu_{\tau}$ neutrino production such as SHIP sand FASER are an interesting alternative in next few years.   

In the present study we intentionally limited to the production of charm
quarks/antiquarks.
The production of charmed mesons or baryons is currently uder discussion
and a new fragmentation scheme was proposed \cite{szczurek2020} very
recently.
We leave the predictions for production of charmed hadrons and their
semileptonic decays for a separate study.
However, the consequences for high-energy neutrino production 
have been discussed shortly in the context of the IceCube experiment 
and experiments proposed at the LHC (SHIP and FASER).

\section*{Acknowledgements}
We would like to thank Victor Goncalves for a discussion on IC. 
This study was partially supported by the Polish National
Science Center grant UMO-2018/31/B/ST2/03537 and by the Center for
Innovation and Transfer of Natural Sciences and Engineering Knowledge 
in Rzesz\'ow.

\end{document}